\documentclass[aps,prd,floats,superscriptaddress,showpacs,nofootinbib,twocolumn,preprintnumbers]{revtex4-1}

\usepackage{amssymb}
\usepackage{amsmath}

\usepackage{graphicx}
\usepackage{graphics}
\usepackage{dcolumn}
\usepackage{color}
\usepackage{rotate}
\usepackage{fancyhdr} 
\usepackage{hyperref}
\usepackage{indentfirst}

\usepackage{umoline}
\usepackage{ulem}

\def\be{\begin{eqnarray}}
\def\ee{\end{eqnarray}}
\def\ba{\begin{eqnarray}}
\def\ea{\end{eqnarray}}
\def\no{\nonumber}

\definecolor{darkred}{rgb}{.743,0,0}

\begin{document}
\preprint{CERN-TH-2018-102, INR-TH-2018-008, KCL-PH-TH/2018-16}
\title{Galactic Rotation Curves vs. Ultra-Light Dark Matter:\\
Implications of the Soliton -- Host Halo Relation}

\author{Nitsan Bar}\affiliation{Weizmann Institute of Science,
  Rehovot, Israel} 

\author{Diego Blas}\affiliation{Theory department, CERN, CH-1211
  Geneve 23, Switzerland}
\affiliation{Theoretical Particle Physics and Cosmology Group,
  Department of Physics,\\ 
King's College London, Strand, London WC2R 2LS, UK}

\author{Kfir Blum}\affiliation{Weizmann Institute of Science,
  Rehovot, Israel}
\affiliation{Theory department, CERN, CH-1211
  Geneve 23, Switzerland} 
 
\author{Sergey Sibiryakov}\affiliation{Theory department, CERN,
  CH-1211 Geneve 23, Switzerland}
\affiliation{LPPC, IPHYS, Ecole Politechnique F\'ed\'erale de
  Lausanne, CH-1015 Lausanne, Switzerland}
\affiliation{Institute for Nuclear Research of the Russian
  Academy of Sciences, 
60th October Anniversary Prospect, 7a, 117312 Moscow, Russia} 

\date{\today}

\begin{abstract}
Bosonic ultra-light dark matter (ULDM) would form cored density
distributions at the center of galaxies. These cores, seen in
numerical simulations, admit analytic description as the
lowest energy bound state
solution (``soliton'') of the Schroedinger-Poisson equations.
Numerical simulations of ULDM galactic halos found empirical scaling
relations between the mass of the large-scale host halo and the mass
of the central soliton. 
We discuss how the simulation results of different groups can be
understood in terms of the basic properties of the
soliton. Importantly, simulations imply that the energy per unit
mass in the soliton and in the virialised host halo should be
approximately equal. This relation lends itself
to observational tests, because it predicts that the 
peak circular velocity, measured for the host halo in the outskirts of the galaxy, should
approximately repeat itself in the central region. Contrasting this prediction to the measured rotation curves of well-resolved near-by galaxies, we show that
ULDM in the mass range $m\sim (10^{-22}\div 10^{-21})$~eV, which has
been invoked as a possible solution to the small-scale puzzles of
$\Lambda$CDM, is in tension with the data. 
We suggest that a dedicated analysis of the Milky Way inner
gravitational potential 
could probe ULDM up to $m\lesssim 10^{-19}$~eV.  
\end{abstract}

\maketitle


\section{Introduction}
Two forms of matter could provide the cosmological equation of state,
required for dark matter (DM):  non-relativistic massive particles
(e.g. WIMPs), or the classical background of an ultra-light bosonic
field oscillating around a minimum of its potential (e.g. axions or
axion-like particles).  
On cosmological scales, WIMPs and ultra-light dark matter (ULDM)
behave similarly. However, on scales comparable to its de Broglie
wavelength, ULDM behaves markedly different from 
WIMPs~\cite{Hu:2000ke,Arbey:2001qi,Lesgourgues:2002hk,Chavanis:2011zi,Chavanis:2011zm,Schive:2014dra,Schive:2014hza,Marsh:2015wka,Chen:2016unw,Schwabe:2016rze,Veltmaat:2016rxo,Hui:2016ltb,Mocz:2017wlg}. 

Axion-like particles with exponentially suppressed masses arise in the
context of string
theory~\cite{Svrcek:2006yi,Arvanitaki:2009fg,Marsh:2015xka,Hui:2016ltb}. They
are produced by misalignment 
and for mass in the range $m\sim (10^{-22}\div 10^{-18})$~eV their
cosmic abundance could naturally match the observed DM density.

ULDM with mass $m\sim (10^{-22} \div 10^{-21})$~eV is particularly
motivated due to several  puzzles, facing the standard WIMP paradigm
on galactic scales (see, e.g.~\cite{DelPopolo:2016emo,Hui:2016ltb} for
recent reviews). This range of $m$ is in some tension with the matter power spectrum, inferred from Ly-$\alpha$
forest
analyses~\cite{Bozek:2014uqa,Armengaud:2017nkf,Irsic:2017yje,Zhang:2017chj,Kobayashi:2017jcf}, which 
yields a bound $m\gtrsim10^{-21}$~eV. Nevertheless, since the
strongest Ly-$\alpha$ bound constraints come from the smallest and 
most non-linear
scales, where systematic effects are challenging, we believe it 
prudent to seek additional methods to constrain the model for
$m\gtrsim10^{-22}$~eV.   

Numerical simulations~\cite{Schive:2014dra,Schive:2014hza,Mocz:2017wlg,Veltmaat:2016rxo,Schwabe:2016rze} show that ULDM forms cored density profiles in the inner region of galactic halos, roughly within the de Broglie wavelength, 
\be\label{eq:ldB}x_
{dB}&\approx&190\left(\frac{v}{100~{\rm km/s}}\right)^{-1}\left(\frac{m}{10^{-22}~{\rm eV}}\right)^{-1}~{\rm pc}.\ee
The core, referred to as ``soliton'' in the
literature~\cite{Chavanis:2011zi,Chavanis:2011zm,Marsh:2015wka,Chen:2016unw},
corresponds to a coherent quasi-stationary solution of the ULDM
equations of motion and its density profile can be derived
analytically\footnote{Strictly speaking, the soliton solution is only known numerically. With some abuse of the word, we refer to it here as analytical to emphasize that the solution can be found by integrating a simple differential equation in a procedure that takes $<1$~sec on a standard laptop.}. This match between analytic and numerical simulation results offers
a unique opportunity to understand and extend the simulations,
making the ULDM model potentially more predictive than the WIMP
paradigm.  
For near-by dwarf galaxies ($v\sim10$~km/s) or Milky Way (MW)-like galaxies
($v\sim100$~km/s), the soliton core could be resolved with current
observational tools, offering a test of the model.  

Attempts to detect, constrain, or fit ULDM soliton cores using galactic rotation velocity and velocity dispersion data were discussed in the literature~\cite{Schive:2014dra,Schive:2014hza,Marsh:2015wka,Calabrese:2016hmp,Gonzales-Morales:2016mkl,Robles:2012uy,Bernal:2017oih,Lesgourgues:2002hk,Arbey:2001qi}. 
Some of these analyses fitted a soliton profile to describe entire
galaxies~\cite{Lesgourgues:2002hk,Arbey:2001qi,Calabrese:2016hmp}. This
exercise leads to small ULDM particle mass, $m\lesssim10^{-23}$~eV, in
strong tension with cosmological
bounds~\cite{Bozek:2014uqa,Armengaud:2017nkf,Irsic:2017yje,Zhang:2017chj,Kobayashi:2017jcf}. Furthermore,
numerical
simulations~\cite{Schive:2014dra,Schive:2014hza,Mocz:2017wlg,Veltmaat:2016rxo,Schwabe:2016rze}
have shown that galactic halos above a certain mass
would exhibit a more complex structure, with the central soliton
transiting into a large-scale host halo composed of an incoherent
superposition of multiple scalar field wavepackets.
Consequently, other
works~\cite{Marsh:2015wka,Gonzales-Morales:2016mkl,Robles:2012uy,Bernal:2017oih}
analysed galactic profiles using a soliton+host halo description.  

A key question in comparing the soliton+host halo model to kinematical
data, is how to model the transition between the central soliton and
the host halo. Previous phenomenological analyses~\cite{Marsh:2015wka,Gonzales-Morales:2016mkl,Robles:2012uy,Bernal:2017oih}
defined separate free parameters for the host halo and for the
soliton, for each individual galaxy in the sample. On the numerical
simulation side, several studies focused on finding a soliton--host
halo
relation~\cite{Schive:2014dra,Mocz:2017wlg,Veltmaat:2016rxo,Schwabe:2016rze}.  
The point in finding a soliton--host halo relation is, of course, that
it could tie together the behaviour of ULDM in the large-scale host
halo to predict the central core with fewer free parameters. 

In this work we consider the soliton--host halo relations, reported by
different numerical simulation
groups~\cite{Schive:2014dra,Schive:2014hza,Mocz:2017wlg,Veltmaat:2016rxo,Schwabe:2016rze}. Our
first observation is that properties of the analytic soliton solution
can provide important insight on the numerical results. We show that:  
(i) the soliton--host halo relation, reported in~\cite{Mocz:2017wlg},
essentially attributes the total energy (kinetic+gravitational) of the
halo to the dominant soliton. This energetic dominance of the soliton
is unlikely to hold for realistic galaxies above a certain size; 
(ii) the soliton--host halo relation, reported
in~\cite{Schive:2014hza}, essentially equates the energy per unit mass
in the soliton to the energy per unit mass in the virialised halo. This
relation can apply to real galaxies.  

Assuming the soliton--host halo relation
of~\cite{Schive:2014hza}, we show that it 
leads to a prediction: the peak circular velocity
characterising the host halo on large scales (few kpc for typical
$(10^9\div 10^{10})$~M$_\odot$ galaxies) should repeat itself in
the core on small scales ($\lesssim1$~kpc), insensitive to the details
of the host halo density profile. This implies an observational
constraint that can be tested without free parameters. Applying this
test to high resolution rotation curves of late-type galaxies
from~\cite{deBlok:2002vgq,Lelli:2016zqa}, we find that ULDM
in the mass range $m\sim (10^{-22}\div 10^{-21})$~eV is disfavoured
by the data. Baryonic physics is unlikely to cure the tension: the discrepancy between the predictions of ULDM (with the
soliton--host halo relation) and the data is too large. In many galaxies in the sample we analyse, baryonic feedback would need to overcome a dark matter-to-baryon mass ratio of $\gtrsim10:1$, to destroy the
soliton. 
As a result, if the soliton--host halo relation
of~\cite{Schive:2014dra,Schive:2014hza} holds for real systems, 
ULDM is disfavoured 
below $m\sim10^{-21}$~eV and is unlikely to play a role in solving the small
scale puzzles of $\Lambda$CDM. 

We also discuss observational imprints of the soliton in big
galaxies, such as the MW. In this case the shape of the
soliton is modified due to the gravitational potential 
of baryonic matter and supermassive black
hole (SMBH). Preliminary
numerical simulations of ULDM halos with
stars~\cite{2017arXiv171201947C} 
indicate that
in the presence of baryons, the soliton mass stays the same or
increases compared to the soliton--host halo prediction of pure ULDM. If these results
are confirmed, solitons formed by ULDM with masses $m\lesssim
10^{-19}$~eV will produce order-one contribution into the mass
budget of the inner MW. A dedicated analysis
of inner MW kinematics -- including simultaneous
modeling of the baryonic mass and the soliton -- could potentially
test ULDM up to $m\lesssim10^{-19}$~eV. 

The outline of this paper is as follows. In Sec.~\ref{ssec:solprop} we
review some basic properties of the soliton.  
In Sec.~\ref{s:sim} we discuss the soliton--host halo relations found
in simulations, and show that these relations can be understood in
terms of fundamental properties of the soliton and the halo. The
simulations of~\cite{Schive:2014dra,Schive:2014hza} are considered in
Sec.~\ref{ss:Schive}; they can be summarised by the statement
$(E/M)|_{\rm soliton}=(E/M)|_{\rm halo}$, where $E$ is the total
energy (kinetic+gravitational, within the virial radius, for the halo)
and $M$ is the mass (again within the virial radius, for the
halo). The simulations of~\cite{Mocz:2017wlg} are considered in
Sec.~\ref{ss:Mocz}; their soliton--halo result can be summarised by
the statement $E|_{\rm soliton}=E|_{\rm halo}$. We argue that this
result is unlikely to represent realistic galaxies above a certain
size.  

In Sec.~\ref{s:bhconsp}, adopting the soliton--host halo relation
of~\cite{Schive:2014dra,Schive:2014hza}, we work out its
observational consequences. We show that in ULDM galaxies satisfying
this relation, the rotation velocity in the inner core should be
approximately as high as the peak rotation velocity in the outer part
of the galaxy. In Sec.~\ref{ss:compsim} we compare this analysis to
numerical profiles taken directly from the published simulation
results, finding good agreement. In Sec.~\ref{ss:compdat} we compare
this prediction to real galaxies from~\cite{deBlok:2002vgq,Lelli:2016zqa},
finding tension  
for $m\sim (10^{-22}\div 10^{-21})$~eV. We show that the intrinsic
scatter in the soliton--host halo relation does not resolve the
discrepancy. 

In
Sec.~\ref{s:bareff} we calculate how baryonic effects could modify the
soliton solution. In Sec.~\ref{s:fixbar} we consider a smooth fixed
distribution of baryonic mass, deferring the case of a super-massive
black hole to App.~\ref{ss:smbh}. 
These results are applied to the MW in 
Sec.~\ref{ss:MW}.

In Sec.~\ref{sec:discussion} we compare our results to previous literature and
discuss some caveats to our analysis.
Sec.~\ref{s:sum} contains a summary of our results and open questions.

\section{Soliton properties: analytic considerations}\label{ssec:solprop}
In this section we review the relevant properties of the soliton that
help to understand results from numerical simulations.  

We consider a real, massive, free\footnote{Analyses of interacting
  fields can be found in,
  e.g.~\cite{Chavanis:2011zi,Chavanis:2011zm,RindlerDaller:2012vj,Desjacques:2017fmf}.}
scalar field $\phi$, satisfying the Klein-Gordon equation of motion
and minimally coupled to gravity.   
In the non-relativistic regime it is convenient to decompose $\phi$ as
\be\label{eq:schroedfield}\phi(x,t)=\frac{1}{\sqrt{2}m}e^{-imt}\psi(x,t)+c.c.,\ee
with complex field $\psi$ that varies slowly in space and time, such
that $|\nabla\psi|\ll m|\psi|$ and $|\dot\psi|\ll m|\psi|$. The field
$\psi$
satisfies the Schroedinger-Poisson (SP) equations~\cite{Ruffini:1969qy}
\be\label{eq:SP1} i\partial_t\psi&=&-\frac{1}{2m}\nabla^2\psi+m\Phi\psi,\\
\label{eq:SP2}\nabla^2\Phi&=&4\pi G|\psi|^2.\ee
We look for a quasi-stationary phase-coherent solution, described by the ansatz\footnote{$M_{pl}=1/\sqrt{G}$.}
\be\label{eq:schroedfield2}\psi(x,t)&=&\left(\frac{mM_{pl}}{\sqrt{4\pi}}\right)e^{-i\gamma mt}\chi(x).\ee
The ULDM mass density is
\be\rho&=&\frac{\left(mM_{pl}\right)^2}{4\pi}\chi^2\\
&\approx&4.1\times10^{14}\left(\frac{m}{10^{-22}~\rm eV}\right)^2\chi^2~{\rm M_\odot/pc^3}.\no\ee
The parameter $\gamma$ is proportional to the ULDM energy per unit
mass.
Validity of the non-relativistic regime requires $|\gamma|\ll 1$. 
Since we are looking for gravitationally bound configurations,
$\gamma<0$.  

Assuming spherical symmetry and defining $r=mx$, the SP equations for $\chi$ and $\Phi$ are given by
\be\partial_r^2\left(r\chi\right)&=&2r\left(\Phi-\gamma\right)\chi,\label{eq:SPselfX}\\
\partial_r^2\left(r\Phi\right)&=&r\chi^2\label{eq:SPselfV}.\ee
Finding the ground state solution amounts to solving Eqs.~(\ref{eq:SPselfX}-\ref{eq:SPselfV}) subject to $\chi(r\to0)={\rm const}$, $\chi(r\to\infty)=0$, with no nodes. Given the boundary value of $\chi$ at $r\to0$, the solution is found for a unique value of $\gamma$. 

\begin{figure}[htbp!]
\centering
\includegraphics[width=0.45\textwidth]{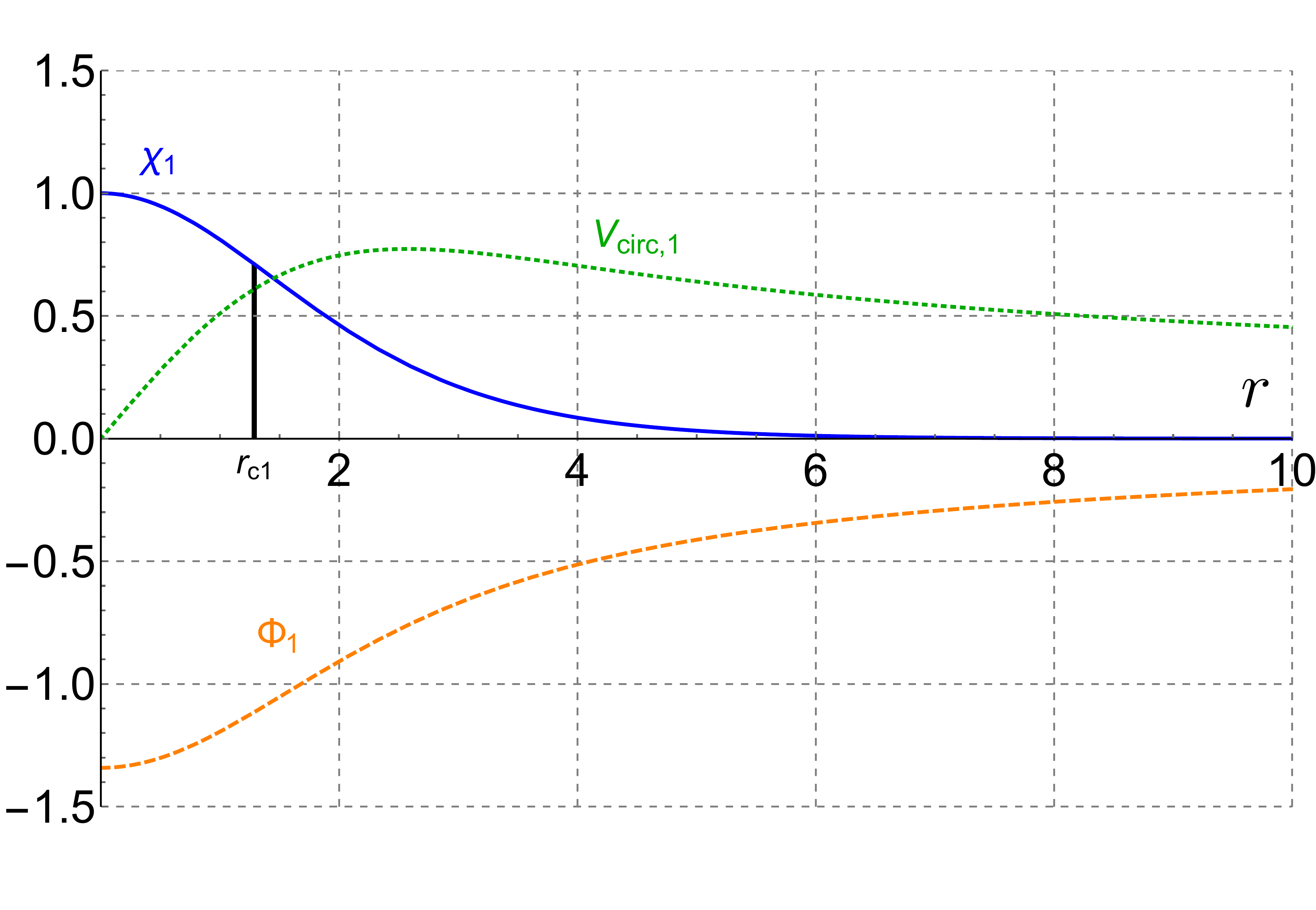}
\caption{Profile of the ``standard'' $\chi_1$ soliton with $\lambda=1$ (blue
  solid). We also show the corresponding gravitational potential
  (orange dashed) and circular velocity of a test particle (dotted green).
}\label{fig:chi1}
\end{figure}

It is convenient to first solve
Eqs.~(\ref{eq:SPselfX}-\ref{eq:SPselfV}) with the boundary condition
$\chi(0)=1$. Let us call this auxiliary solution $\chi_1(r)$, with
$\gamma_1$. A numerical calculation
gives~\cite{Chavanis:2011zi,Chavanis:2011zm,Marsh:2015wka} 
\be\label{eq:g1}\gamma_1&\approx&-0.69,\ee
and the solution is plotted in Fig.~\ref{fig:chi1}.
The mass of the $\chi_1$ soliton is
\be \label{eq:M1}M_1&=&\frac{M_{pl}^2}{m}\int_0^\infty drr^2\chi_1^2(r)\\
&\approx&2.79\times10^{12}\left(\frac{m}{10^{-22}~\rm eV}\right)^{-1}~{\rm M_\odot}.\no\ee
Its core radius, defined as the radius where the mass density drops by a factor of 2 from its value at the origin, is
\be x_{c1}&\approx&0.082\left(\frac{m}{10^{-22}~\rm eV}\right)^{-1}~{\rm pc}.\ee

Other solutions of Eqs.~(\ref{eq:SPselfX}-\ref{eq:SPselfV}) can be obtained from $\chi_1(r),\,\Phi_1(r)$ by a scale transformation. That is, the functions $\chi_\lambda(r),\Phi_\lambda(r)$, together with the eigenvalue $\gamma_\lambda$, given by
\be\chi_\lambda(r)&=&\lambda^2\chi_1(\lambda r),\label{eq:chilam}\\
\Phi_\lambda(r)&=&\lambda^2\Phi_1(\lambda r),\label{eq:Vlam}\\
\gamma_\lambda&=&\lambda^2\gamma_1,\label{eq:glam}\ee
also satisfy Eqs.~(\ref{eq:SPselfX}-\ref{eq:SPselfV}) with correct boundary conditions for any $\lambda>0$. The soliton mass and core radius for $\chi_\lambda$ are 
\be\label{eq:Mlambda} M_\lambda&=&\lambda M_1,\\
x_{c\lambda}&=&\lambda^{-1}x_{c1}.\ee
A mnemonic for the numerical value of $\lambda$ is given by
\be\label{eq:lamM}\lambda&=&3.6\times10^{-4}\left(\frac{m}{10^{-22}~\rm eV}\right)\left(\frac{M_\lambda}{10^9~{\rm M_{\odot}}}\right).\ee
%
The product of the soliton mass and core radius is independent of $\lambda$,
\be\label{eq:MXsol} M_{\lambda}
x_{c\lambda}&\approx&2.27\times10^{8}\left(\frac{m}{10^{-22}~\rm
    eV}\right)^{-2}~{\rm kpc\,M_\odot}.
\ee 

Formally, solutions exist for any positive value of $\lambda$ and
hence for any soliton mass. However, if we select $\lambda\gtrsim1$ we
reach $|\gamma_\lambda|>1$, outside of the regime of validity of the
non-relativistic approximation. Thus, self-consistent solutions are
limited to $\lambda\ll1$  
and their eigenvalue $|\gamma_\lambda|=\lambda^2|\gamma_1|\ll1$,
consistent with the non-relativistic approximation.  

The energy in an arbitrary non-relativistic ULDM configuration is
\be \label{eq:Egen}E&=&\int d^3x\left(
\frac{\left|\nabla\psi\right|^2}{2m^2}+\frac{\Phi\left|\psi\right|^2}{2}\right)\,=\,E_k+E_p,\ee
with kinetic (potential) energy $E_k$ ($E_p$).  
For the ansatz (\ref{eq:schroedfield2}), integrating by parts and using Eqs.~(\ref{eq:SP1}-\ref{eq:schroedfield2}) we have
\be \label{eq:EMgam}E&=&\frac{1}{3}M\,\gamma.\ee
Note that spherical symmetry is not needed for Eq.~(\ref{eq:EMgam}) to hold. 

Considering the $\chi_\lambda$ solitons, we find $E_{p,\lambda}=-2E_{k,\lambda}=2E_{\lambda}$ with
\be E_\lambda&\approx&-0.476\,\lambda^3\frac{M_{pl}^2}{m},\\
M_\lambda&\approx&2.06\,\lambda\frac{M_{pl}^2}{m}.\ee
This leads to a relation for an isolated soliton~\cite{Chavanis:2011zi,Chavanis:2011zm},
\be\label{eq:MEsol} \frac{M_\lambda}{(M_{pl}^2/m)}&\approx&2.64\left|\frac{E_\lambda}{(M_{pl}^2/m)}\right|^{\frac{1}{3}}.\ee
Another useful relation gives the energy per unit mass from the scaling parameter $\lambda$,
\be\label{eq:E2M}\frac{|E_\lambda|}{M_\lambda}&\approx&0.23\,\lambda^2,\ee
which can also be written as
\be\label{eq:MlamSchieve} M_\lambda&\approx&4.3\left(\frac{|E_\lambda|}{M_\lambda}\right)^{\frac{1}{2}}\frac{M_{pl}^2}{m}.
\ee

The circular velocity curve for a test particle in the soliton gravitational potential is given by
\be V_{\rm circ,\lambda}^2(r)&=&r\partial_r\Phi_{\lambda}(r).\ee
The circular velocity rises as $V_{\rm circ,\lambda}\propto r$ at
small $r$ and decreases as $V_{\rm circ,\lambda}\propto
r^{-\frac{1}{2}}$ at large $r$, see Fig.~\ref{fig:chi1}. 
The peak of $V_{\rm circ}$ is obtained at
\be\label{eq:xpeakVc} x_{\rm peak,\lambda}&\approx&0.16\,\lambda^{-1}\left(\frac{m}{10^{-22}~\rm eV}\right)^{-1}~{\rm pc}\\
&\approx&460\left(\frac{m}{10^{-22}~\rm eV}\right)^{-2}\left(\frac{M_\lambda}{10^9~{\rm M_{\odot}}}\right)^{-1}~{\rm pc},\no\ee
and the peak velocity is
\be\label{eq:maxVc} {\rm max}V_{\rm circ,\lambda}&\approx&2.3\times10^5\,\lambda~{\rm km/s}\\
&\approx&83\left(\frac{m}{10^{-22}~\rm eV}\right)\left(\frac{M_\lambda}{10^9~{\rm M_{\odot}}}\right)~{\rm km/s}.\no\ee

\section{Making contact with numerical simulations}\label{s:sim}
We now discuss results from the numerical simulations of three different groups, Refs.~\cite{Schive:2014hza,Schive:2014dra}, Ref.~\cite{Mocz:2017wlg}, and Refs.~\cite{Veltmaat:2016rxo,Schwabe:2016rze}.%

\begin{figure}[hbp!]
\centering
\includegraphics[width=0.45\textwidth]{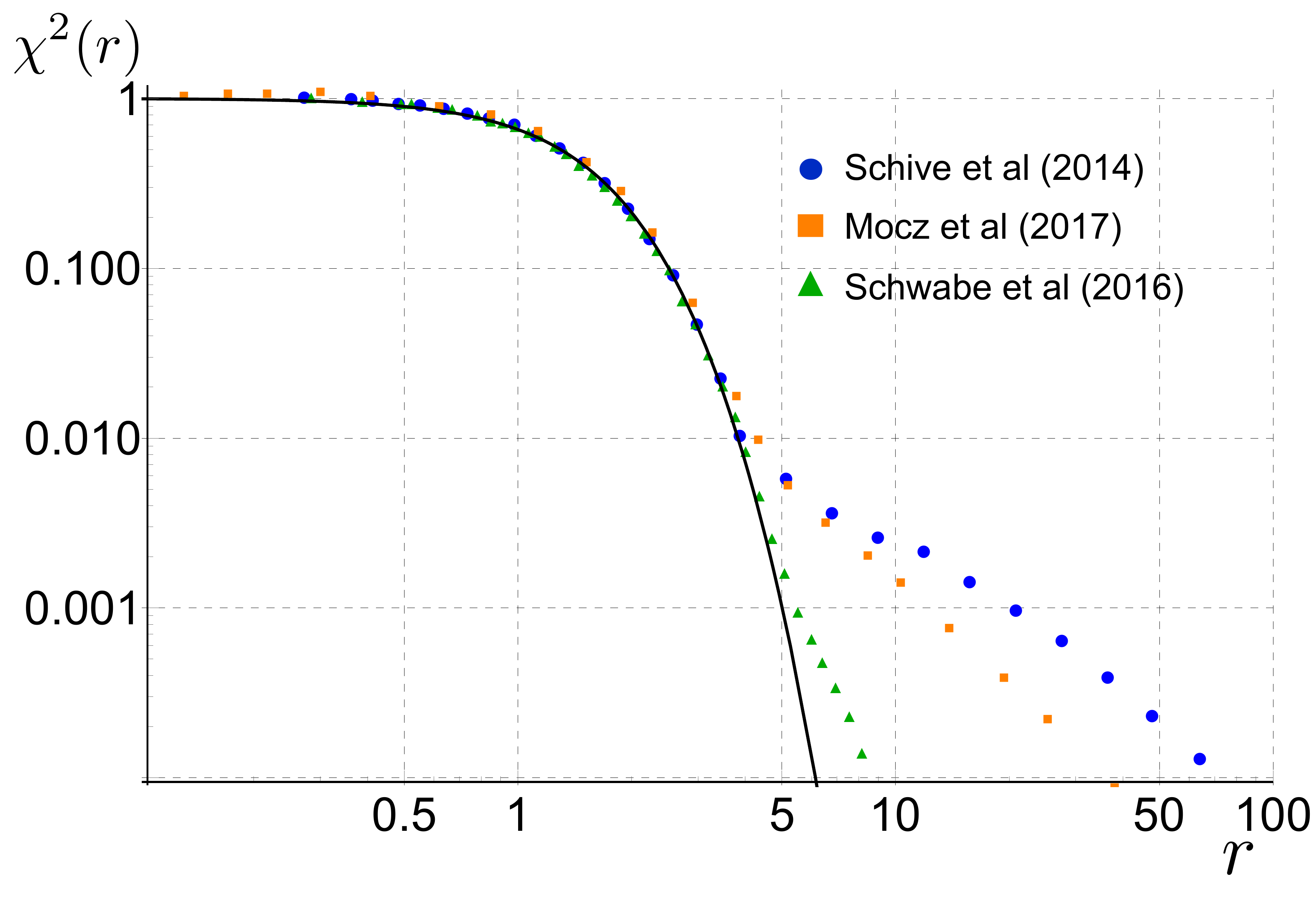}
\caption{
Review of results from numerical simulations by different
groups. Markers show density profiles of simulated halos 
from Schive et al.~\cite{Schive:2014dra}
(blue circles), Mocz et al.~\cite{Mocz:2017wlg} (orange squares), and
Schwabe et al.~\cite{Schwabe:2016rze} (green triangles). The central regions of
the halos are
described by the soliton (solid line). 
}\label{fig:soliton_simulations}
\end{figure}

The first point to note is that soliton configurations, in a form
close to the idealised form discussed in Sec.~\ref{ssec:solprop},
actually occur dynamically in the central region of the halo in the
numerical simulations\footnote{The first simulations of cosmological
  ULDM galaxies~\cite{Woo:2008nn} did not have sufficient resolution
  to resolve the central core.}. In Fig.~\ref{fig:soliton_simulations}
we collect representative density profiles from
Ref.~\cite{Schive:2014dra} (blue), Ref.~\cite{Mocz:2017wlg} (orange),
and Ref.~\cite{Schwabe:2016rze} (green). We refer to those papers for
more details on the specific set-ups in each simulation. To make
Fig.~\ref{fig:soliton_simulations}, in each case, we find the
$\lambda$ parameter that takes the numerical result into the $\chi_1$
soliton, rescale the numerical result accordingly and present it in
comparison with the analytic $\chi_1^2(r)$ profile. 

While different groups agree that solitons form in the centers of
halos, they do not appear to agree on the matching between the inner
soliton profile and the host halo.  
Refs.~\cite{Schive:2014hza,Schive:2014dra} and
Ref.~\cite{Mocz:2017wlg} reported scaling relations between the
central soliton and the host halo. 
As we show below, the scaling relations found by
both groups are connected to properties of a single, isolated,
self-gravitating soliton (part of these observations were made
in~\cite{Veltmaat:2016rxo,Schwabe:2016rze}).

\subsection{Soliton vs. host halo: the simulations of Ref.~\cite{Schive:2014hza,Schive:2014dra}}\label{ss:Schive}
%
%
At cosmological redshift $z=0$, the numerical simulations of~\cite{Schive:2014hza,Schive:2014dra} yield approximately NFW-like halos which transit, in the central region, into a core with core radius and mass density
\be x_c&\approx&160\left(\frac{M_h}{10^{12}~{\rm M_\odot}}\right)^{-\frac{1}{3}}\left(\frac{m}{\rm10^{-22}~eV}\right)^{-1}~{\rm pc},\label{eq:rc}\\
\rho(x)&\approx&\frac{190\left(\frac{m}{10^{-22}~\rm eV}\right)^{-2}\left(\frac{x_c}{100~\rm pc}\right)^{-4}}{\left(1+0.091\left(\frac{x}{x_c}\right)^2\right)^8}~{\rm M_{\odot}\,pc^{-3}},\label{eq:rhosol}
\ee
where $M_h$ is the virial mass of the host halo. 
As noted in~\cite{Schive:2014hza,Schive:2014dra,Marsh:2015wka}, Eqs.~(\ref{eq:rc}-\ref{eq:rhosol}) are an excellent numerical fit for a soliton $\chi_\lambda$. The mass of this soliton is 
\be\label{eq:Mc} M&\approx&1.4\times10^9\!
\left(\frac{m}{10^{-22}\,\rm eV}\right)^{-1}
\!\left(\frac{M_h}{10^{12}~{\rm M_\odot}}\right)^{\frac{1}{3}}
\!{\rm M_{\odot}},\ee
so its $\lambda$ parameter is
\be\label{eq:lamsim}\lambda&\approx&4.9\times10^{-4}\left(\frac{M_h}{10^{12}~{\rm M_\odot}}\right)^{\frac{1}{3}}.\ee
Note that Eq.~(\ref{eq:Mc}) is applicable only as long as the halo
exceeds a minimal mass,
\be
\label{eq:minMh}
M_{h,{\rm min}}\sim 5.2\times
10^7\bigg(\frac{m}{10^{-22}{\rm eV}}\bigg)^{-3/2}
{\rm M}_\odot\;.
\ee
Smaller mass halos would be dominated by the soliton. 

Ref.~\cite{Schive:2014hza} showed that Eq.~(\ref{eq:Mc}) is consistent
with the relation, 
\be\label{eq:McEh}
M_c&\approx&\alpha\,\left(\frac{|E_h|}{M_h}\right)^{\frac{1}{2}}\frac{M_{pl}^2}{m},\ee 
where $M_{c}$ is the core mass (mass within $x<x_c$); $M_h,\,E_h$ are
the virial mass and energy of the simulated halo; and $\alpha=1$
provides a good fit to the data. Ref.~\cite{Schive:2014hza} gave a
heuristic argument, pointing out that Eq.~(\ref{eq:McEh}) relates
the soliton scale radius (chosen as the core radius $x_c$
in~\cite{Schive:2014hza}) with the velocity dispersion of particles in the
host halo, in qualitative agreement with a wave-like ``uncertainty
principle''. 

However, there is another way to express Eq.~(\ref{eq:McEh}). The core
mass of a $\chi_\lambda$ soliton is related to its total mass via
$M_{c\lambda}\approx0.236M_\lambda$. Thus, using
Eq.~(\ref{eq:MlamSchieve}) we have an analytic relation
$M_{c\lambda}\approx1.02\left(\frac{|E_\lambda|}{M_\lambda}\right)^{\frac{1}{2}}\frac{M_{pl}^2}{m}$.  
This allows us to rephrase the empirical
Eq.~(\ref{eq:McEh}) by a more intuitive (though equally empirical)
expression: 
\be\label{eq:E2MsolE2Mh} \frac{E}{M}\bigg|_{\rm
  soliton}&\approx&\frac{E}{M}\bigg|_{\rm halo}.
\ee 
Therefore, the soliton--host halo relation in the simulations of
Ref.~\cite{Schive:2014hza,Schive:2014dra} can be summarised by the
statement that the energy per unit mass of the soliton matches the
energy per unit mass of the host halo.

\subsection{Soliton vs. host halo: the simulations of Ref.~\cite{Mocz:2017wlg}}\label{ss:Mocz}
The simulations of Ref.~\cite{Mocz:2017wlg} pointed to an empirical scaling relation between the soliton mass $M$ and the total energy of the ULDM distribution in the simulation box, $E_h$, 
\be \label{eq:Mocz}
\frac{M}{(M_{pl}^2/m)}&\approx&2.6\left|\frac{E_h}{(M_{pl}^2/m)}\right|^{\frac{1}{3}}.
\ee
However, this is just Eq.~(\ref{eq:MEsol}), if we replace the halo
energy $E_h$ by the energy of the  soliton. Because the central
density profile found in~\cite{Mocz:2017wlg} was a $\chi_\lambda$
soliton, to a good approximation, it must be the case that the total
energy of the halo in the simulations of~\cite{Mocz:2017wlg} was
dominated by the central soliton contribution. This situation is
unlikely to hold for realistic cosmological host halos with $M_h$ significantly above $M_{h,\rm min}$. 

How could this have happened? The initial conditions in the
simulations of~\cite{Mocz:2017wlg} were a collection of N solitons,
which were then allowed to merge. It appears that these initial
conditions were constructed such that one initial state soliton -- the
soliton of initially largest mass -- grew to absorb the
entire energy of the system. Differently from 
Ref.~\cite{Schive:2014hza} that considered initial conditions of
N identical initial solitons, the simulations
of~\cite{Mocz:2017wlg} initiated their N solitons with a random flat
distribution in soliton radius. Such distribution would be skewed
towards large soliton energy because $E_\lambda\propto
x_{c\lambda}^{-3}$. Considering the initial condition set-up as
explained in~\cite{Mocz:2017wlg}, we find that the most massive
initial state soliton typically needed to grow in mass by only a
factor of $1.5\div 2$, to absorb the entire energy of the halo.  

Note that energy dominance of the central soliton over the host halo,
implied by Eq.~(\ref{eq:Mocz}), is not the same, of course, as
equating the energy per unit mass of the soliton and the halo, implied
by Eq.~(\ref{eq:E2MsolE2Mh}). Halos
in~\cite{Schive:2014hza,Schive:2014dra} attained masses up to two
orders of magnitude larger than the central soliton mass, meaning
their halo energy was two orders of magnitude larger than the energy
of the soliton.

\subsection{Comments}\label{ss:disc}
As far as we can currently determine, Eq.~(\ref{eq:E2MsolE2Mh}) may
indeed reflect a realistic soliton--host halo relation for large
enough cosmological halos.  
In the following sections, we take a leap of faith and assume that the
simulations of~\cite{Schive:2014hza,Schive:2014dra} produced the
correct scaling relation.  
We stress that Eq.~(\ref{eq:E2MsolE2Mh}) is an empirical result, and
was only tested in~\cite{Schive:2014hza,Schive:2014dra} for host halo
masses ranging from $\sim10^8$~M$_\odot$ to
$\sim10^{11}$~M$_\odot$. Our key numerical analysis will concern
systems in this range of mass.  

We defer a theoretical study of the origin of
Eq.~(\ref{eq:E2MsolE2Mh}) to future work. Here
we give only a few comments. We stress that the discussion in the rest of this section does not affect any of our results. 

For a soliton, $E/M=\gamma/3$. On the other hand, $\gamma m$ can be
associated with the chemical potential of ULDM particles in the
soliton (see e.g.~~\cite{Eby:2017zyx} and references therein). This may appear to 
hint that Eq.~(\ref{eq:E2MsolE2Mh}) corresponds to 
thermodynamic equilibrium between the ULDM particles in the host halo
and in the soliton. However, there is some evidence to the contrary
from simulations.

Ref.~\cite{2017arXiv171201947C} simulated ULDM, adding collisionless point particles (``stars''). 
The stars aggregated dynamically in a cuspy profile,
resulting in a more massive soliton compared to the pure ULDM
simulations~\cite{Schive:2014hza,Schive:2014dra} with a given host
halo mass.  
Testing the reversibility of the system,
Ref.~\cite{2017arXiv171201947C} adiabatically ``turned off'' the stars
after the initial system virialised. When eliminating the stars, the
soliton+halo system did not relax back to
Eq.~(\ref{eq:E2MsolE2Mh}). Instead, the excess ULDM mass that was
contained in the soliton in the presence of stars remained captured in
the soliton, and did not return to the host halo. 
The final state of the system was not described by
Eq.~(\ref{eq:E2MsolE2Mh}): the soliton ended up containing larger
(negative) $E/M$ than the halo, and larger mass compared with
Eq.~(\ref{eq:Mc}).

\section{Soliton-host halo relation and galactic rotation
  curves}\label{s:bhconsp} 
As we have seen, the soliton--host halo relation found in the
simulations of~\cite{Schive:2014hza,Schive:2014dra} can be summarised
by Eq.~(\ref{eq:E2MsolE2Mh}), equating the energy per unit mass of the
virialised host halo to that in the soliton component. For a
virialised system, the energy per unit mass maps to kinetic energy
density: in particular, the characteristic circular velocity (or, up
to an $\mathcal{O}(1)$ geometrical factor, the velocity dispersion) of
test particles in the halo and in the soliton should match. The peak
circular velocity of the soliton, given by
Eqs.~(\ref{eq:xpeakVc}-\ref{eq:maxVc}), occurs deep in the inner part,
$x<1$~kpc, of the galaxy; while the peak circular velocity of an
NFW-like halo occurs far out at $x\sim 2\,R_s$, with $R_s$ the NFW
characteristic radius, of order 10~kpc for a MW-like galaxy. Thus, if
the scaling derived from the simulations
of~\cite{Schive:2014hza,Schive:2014dra} is correct, ULDM predicts that
the peak rotation velocity in the outskirts of a halo should
approximately repeat itself in the deep inner region.  
We now discuss this result quantitatively.

Consider a halo with an NFW density profile 
\be\rho_{NFW}(x)&=&\frac{\rho_c\delta_c}{\frac{x}{R_s}\left(1+\frac{x}{R_s}\right)^2},\ee
where 
\be\rho_c(z)=\frac{3H^2(z)}{8\pi G},\,\,\,\,\delta_c&=&\frac{200}{3}\frac{c^3}{\ln(1+c)-\frac{c}{1+c}}.\ee
The profile has two parameters: the radius $R_s$ and the concentration
parameter $c=R_{200}/R_s$, where $R_{200}$ is the radius where the
average density of the halo equals 200 times the cosmological critical
density, roughly indicating the virial radius of the halo.  
The gravitational potential of the halo is 
\be\Phi_{NFW}(x)&=&-\frac{4\pi G\rho_c\delta_c R_s^3}{x}\ln\left(1+\frac{x}{R_s}\right).\ee
Near the origin, $x\ll R_s$, $\Phi_{NFW}$ is approximately constant, $\Phi_{NFW}\left(x\ll R_s\right)\approx\Phi_h$, and is related to the mass of the halo, $M_{200}=200\rho_c\frac{4\pi}{3} c^3R_s^3$, via
\be\label{eq:PhiNFW}\Phi_h&=&-G\left(\frac{4\pi\delta_c}{\left(\ln(1+c)-\frac{c}{1+c}\right)^2}\right)^{\frac{1}{3}}\rho_c^{\frac{1}{3}}\,M_{200}^{\frac{2}{3}}.
\ee 

We can estimate the energy per unit mass of the virialised halo by
\be \frac{E}{M}\bigg|_{\rm halo}&\approx&\pi\frac{\int_0^{R_{200}} dxx^2\rho_{NFW}(x)\Phi_{NFW}(x)}{M_{200}}.\ee
This gives
\be\label{eq:E2MNFW} \frac{E}{M}\bigg|_{\rm halo}&\approx&\frac{\tilde c}{4}\,\Phi_h,
\ee
where
\be
\tilde c&=&\frac{c-\ln(1+c)}{(1+c)\ln(1+c)-c}.\ee
Typical values of the concentration parameter are in the range
$c\sim5\div 30$~\cite{BoylanKolchin:2009an}. In this range, $\tilde c$
varies between $\tilde c\sim0.55 \div 0.35$, respectively. (For reference,
fits of the MW outer rotation curve give
$c\sim10\div 20$~\cite{Piffl:2014mfa}.) 

Plugging Eq.~(\ref{eq:E2MNFW}) into the soliton--host halo relation
Eq.~(\ref{eq:E2MsolE2Mh}), the scaling parameter $\lambda$ is fixed as 
\be\label{eq:E2Mmatch} -0.23\,\lambda^2&\approx&\frac{E_\lambda}{M_\lambda}\,\approx\,\frac{\tilde c}{4}\Phi_h,\ee
which implies\footnote{In the numerical estimates we use
  $H_0=70$~km/s/Mpc for the present-day Hubble constant.}
\be \label{eq:Mhostan}M_\lambda&\approx&2.1\,\sqrt{-\tilde c\,\Phi_h}\,\frac{M_{pl}^2}{m}\nonumber\\
&\approx&2.4\times10^9\left(\frac{m}{10^{-22}~\rm eV}\right)^{-1}\nonumber\\
&&\times\left(\frac{H(z)}{H_0}\right)^{\frac{1}{3}}\left(\frac{M_{200}}{10^{12}~\rm M_\odot}\right)^{\frac{1}{3}}f(c)~{\rm M_\odot},
\ee
with
\be
f(c)&=&0.54\sqrt{\left(\frac{c}{1+c}\right)\frac{c-{\ln(1+c)}}{\left(\ln(1+c)-\frac{c}{1+c}\right)^2}}.\no
\ee
Eq.~(\ref{eq:Mhostan}) depends weakly on the NFW concentration
parameter, via the factor $f(c)$ that varies in the range $0.9\div1.1$
for $c=5\div 30$.   
It agrees parametrically with the simulation result, Eq.~(\ref{eq:Mc})
(including the redshift dependence, which we have suppressed in
Eq.~(\ref{eq:Mc})). It also agrees quantitatively to about $20\%$; to
see this, we need to account for the slightly different definition of
the halo mass $M_h$, used in~\cite{Schive:2014hza}, and our
$M_{200}$. We do this comparison in App.~\ref{app:MhM200}. 

Consider the rotation velocity curve of an ULDM galaxy satisfying Eq.~(\ref{eq:E2MsolE2Mh}). The NFW rotation curve is given by
\be \frac{V^2_{\rm circ,h}(x)}{V^2_{\rm circ,h}(R_s)}&
=&\frac{2(1+\xi)\ln(1+\xi)-2\xi}{\xi(1+\xi)(\ln(4)-1)},\,\,\,\,\,
\xi\equiv \frac{x}{R_s}.\ee
This halo rotation curve peaks at $x\approx2.16\,R_s$ with a peak value
\be\label{eq:nfwVmax}{\max}V_{\rm
  circ,h}&\approx&1.37\times10^5(-\Phi_h)^{\frac{1}{2}}~{\rm km/s}. 
\ee
On the other hand, in the inner galaxy $x\ll R_s$,  
the circular velocity due to the soliton peaks to a local maximum of 
\be\label{eq:maxVch} {\max}V_{\rm circ,\lambda}&\approx&1.51\times10^5\left(\frac{\tilde c}{0.4}\right)^{\frac{1}{2}}(-\Phi_h)^{\frac{1}{2}}~{\rm km/s},
\ee
where we used Eq.~(\ref{eq:E2Mmatch}) to fix $\lambda$ and Eq.~(\ref{eq:maxVc}) to relate it to ${\max}V_{\rm circ,\lambda}$.

\begin{figure}[hbp!]
\centering
\includegraphics[width=0.45\textwidth]{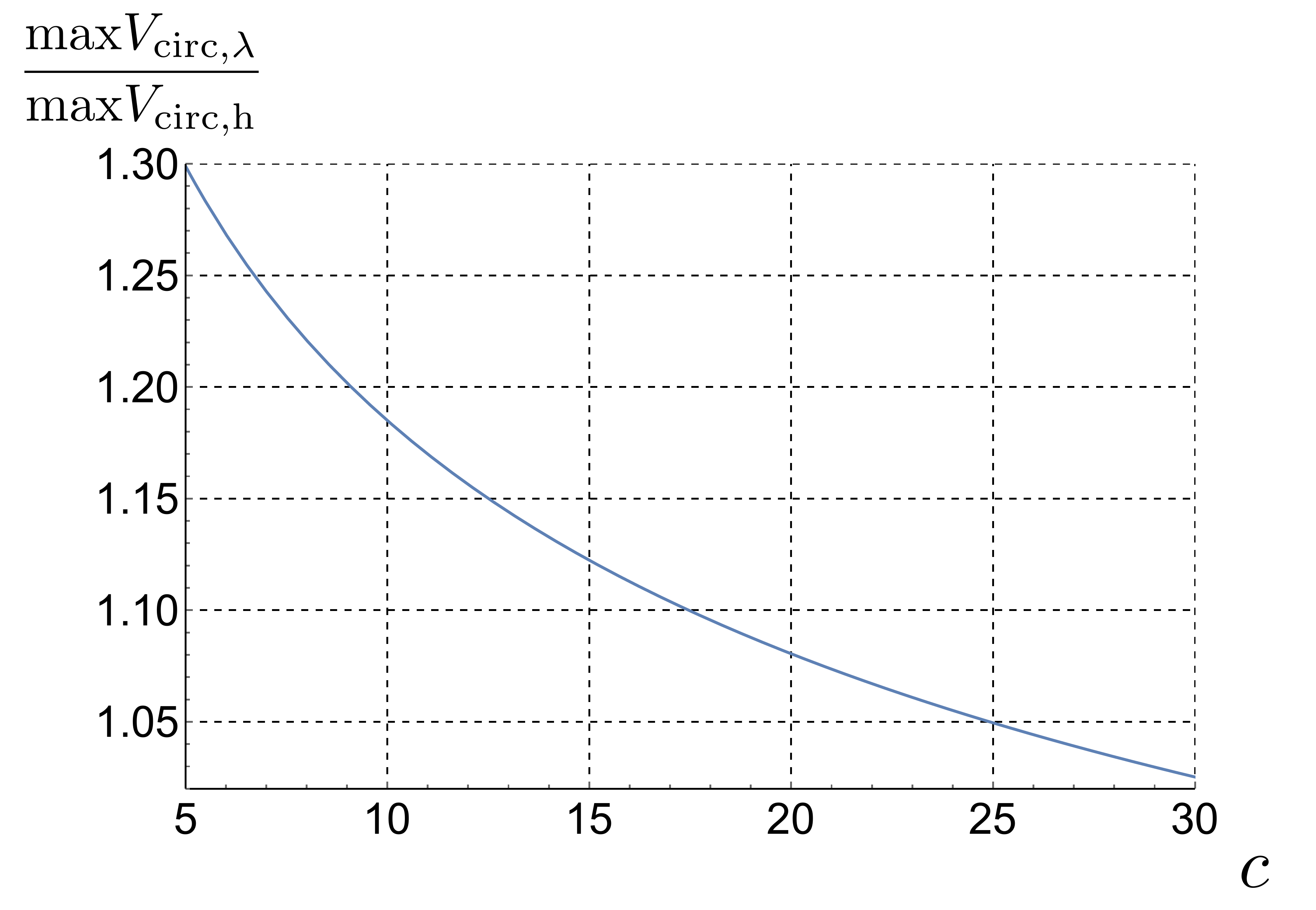}
\caption{Ratio between halo and soliton peak circular
  velocities as a function of the halo concentration.
}\label{fig:ctilderoot}
\end{figure}

As anticipated in the beginning of this section, 
Eq.~(\ref{eq:E2MsolE2Mh}) predicts approximately equal peak circular
velocities for the inner soliton component and for the host halo, 
\be\label{eq:vv}\frac{{\max}V_{\rm circ,\lambda}}{{\max}V_{\rm circ,h}}&\approx&1.1\left(\frac{\tilde c}{0.4}\right)^{\frac{1}{2}},\ee
independent of the particle mass $m$, independent of the halo mass
$M_{200}$, and only weakly dependent on the details of the halo via
the factor $(\tilde c/0.4)^{\frac{1}{2}}$. Eq.~(\ref{eq:vv}) 
is plotted in
Fig.~\ref{fig:ctilderoot} as function of the concentration parameter. 

While ${\max}V_{\rm circ,\lambda}$ and the approximate equality
Eq.~(\ref{eq:vv}) are $m$-independent, the soliton peak velocity
occurs in an $m$-dependent location, 
\be\label{eq:xpeakVch} x_{\rm peak,\lambda}
&\approx&184\left(\frac{10^{-22}\,\rm
    eV}{m}\right)\left(\frac{{\max}V_{\rm circ,\lambda}}{200~\rm
    km/s}\right)^{-1}~{\rm pc}.
\ee

\begin{figure}[hbp!]
\centering
\includegraphics[width=0.475\textwidth]{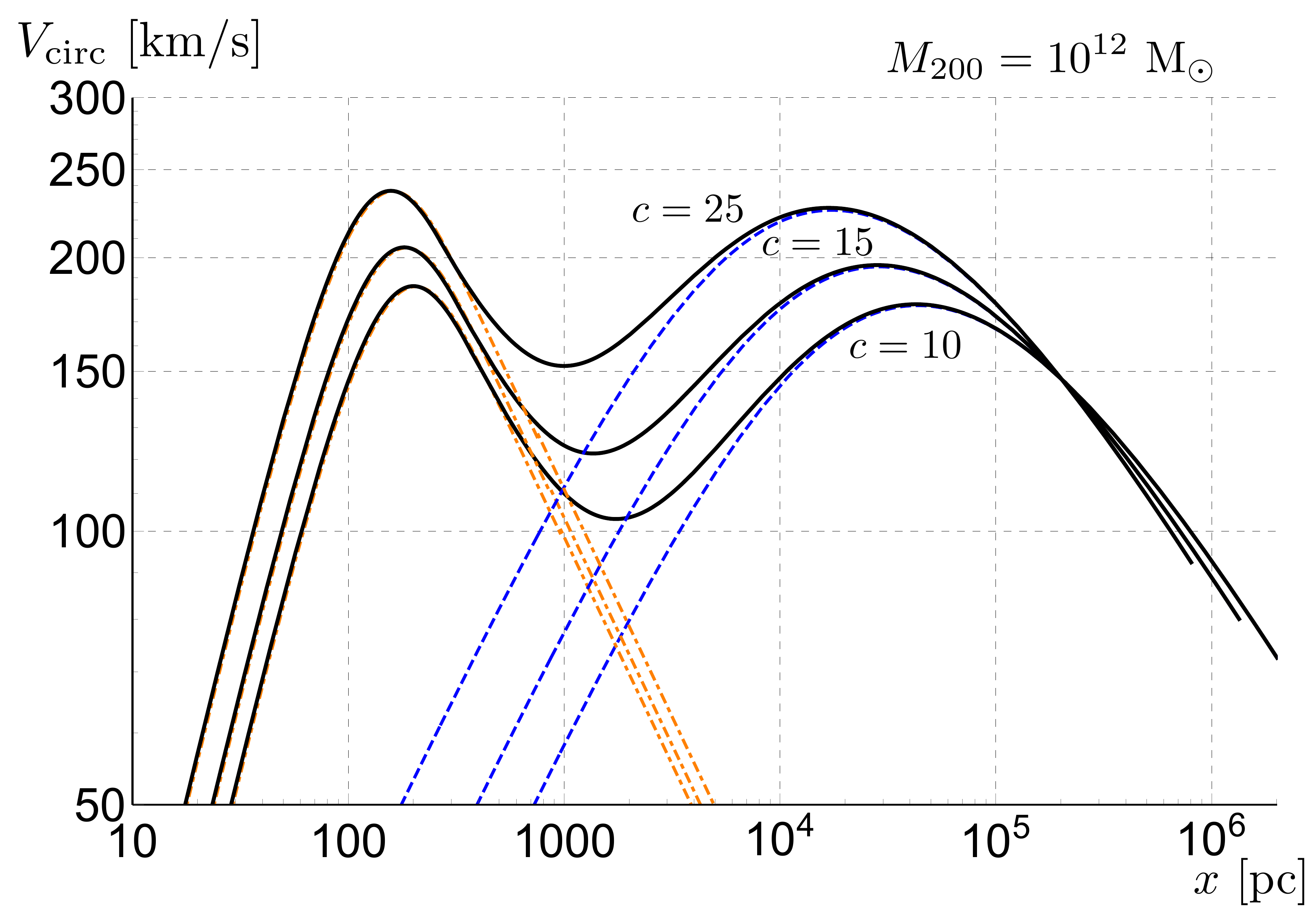}
\includegraphics[width=0.475\textwidth]{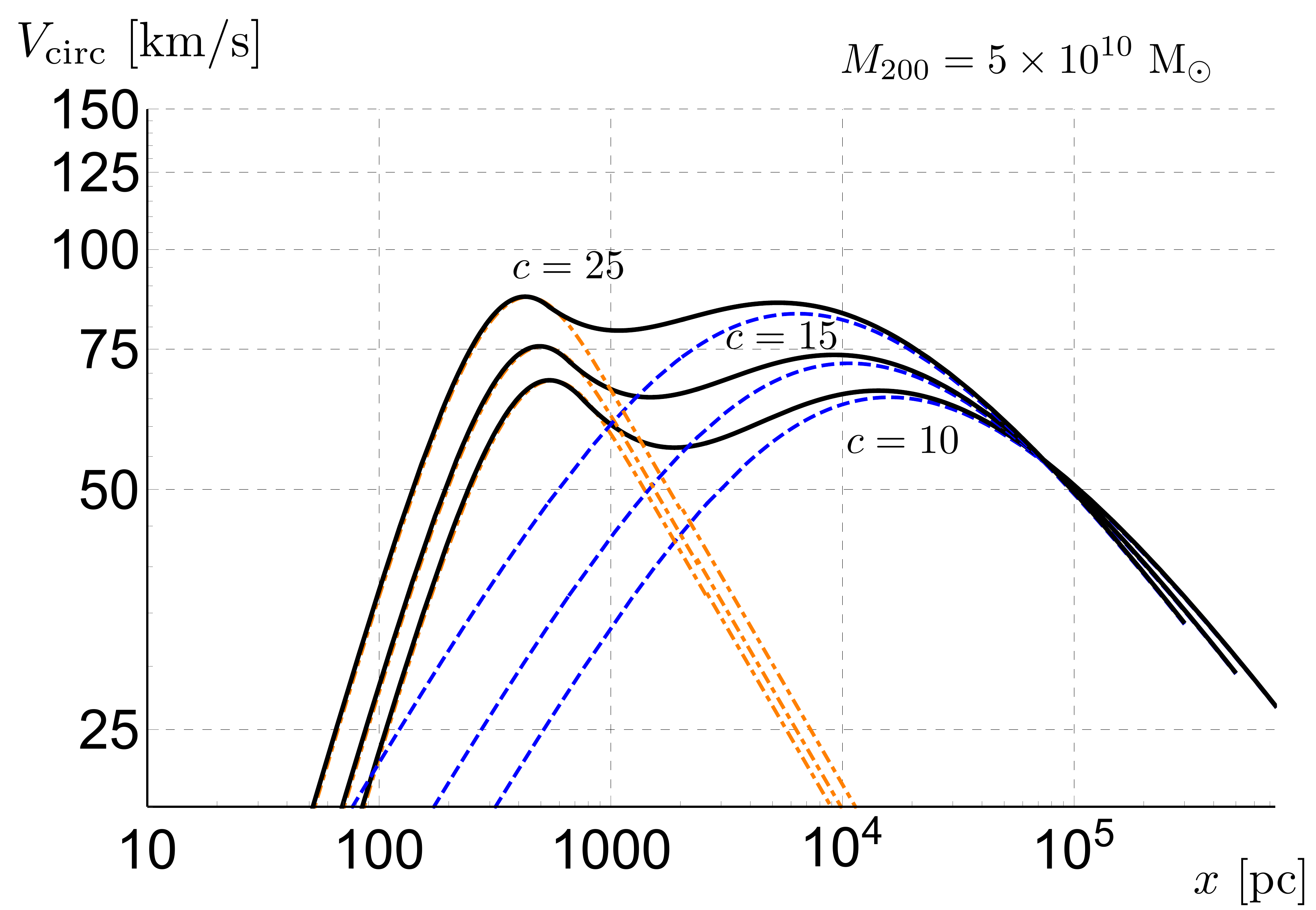}
\caption{Rotation curves for the ULDM soliton+halo system, obtained
  for a DM-only NFW halo using the soliton--halo relation
  Eq.~(\ref{eq:E2MsolE2Mh}) with $m=10^{-22}$~eV. Solid black, 
    dot-dashed orange, and dashed blue show $V_{\rm circ}$ due to the
  total soliton+halo system, the soliton only, and the halo
  only. Results are shown for NFW concentration parameter
  $c=10,\,15,\,25$, with $M_{200}=10^{12}$~M$_\odot$ and
  $5\times10^{10}$~M$_\odot$ on the upper and lower panels,
  respectively.  
}\label{fig:VcircM12}
\end{figure}

Fig.~\ref{fig:VcircM12} shows the circular velocity curve for the NFW
halo+soliton system, following from Eq.~(\ref{eq:E2MsolE2Mh}), with
ULDM particle mass $m=10^{-22}$~eV. The solid black, dot-dashed
orange, and dashed blue lines show the contributions to $V_{\rm circ}$
due to the total system, the soliton only, and the halo only. Results
are shown for three different values of the NFW concentration
parameter, $c=10,\,15,\,25$, with $M_{200}=10^{12}$~M$_\odot$ and
$5\times10^{10}$~M$_\odot$ on the top and bottom panels,
respectively. For larger $m>10^{-22}$~eV, the soliton bump in the
rotation curve would shift to smaller $x$ according to
Eq.~(\ref{eq:xpeakVch}), but would maintain its height. 

In Fig.~\ref{fig:VcircM12}, to define the rotation velocity for the total system, we set the ULDM mass density for the total system to be $\rho(x)={\rm max}\left\{\rho_\lambda(x),\,\rho_{NFW}(x)\right\}$, calculate the resulting mass profile $M(x)$, and use spherical symmetry to find $V_{\rm circ}(x)=\sqrt{G\,M(x)/x}$. This prescription for matching between the soliton and NFW parts is ad-hoc and only roughly consistent with the simulations of~\cite{Schive:2014hza,Schive:2014dra}. The true transition region between the NFW part and the soliton part probably deviates from the pure NFW form. Ref.~\cite{Vicens:2018kdk} considered this transition region and concluded that the density profile in this region should follow approximately $\rho\sim x^{-\frac{5}{3}}$, steeper than the usual inner NFW form $\rho\sim x^{-1}$. This would affect the detailed shape of the rotation curve in the intermediate region between the two peaks, but not our general results.

Eq.~(\ref{eq:vv}) was derived for an NFW host halo, but it is the manifestation of  Eq.~(\ref{eq:E2MsolE2Mh}) that is not tied to a particular parametrisation of the halo profile. Building on Eq.~(\ref{eq:E2MsolE2Mh}), we expect in general that for DM-dominated galaxies, the soliton peak circular velocity should roughly equal the peak circular velocity in the host halo. The NFW example demonstrates that details of the host halo profile affect this result at the 10\% level or so. 

In the rest of this paper, when we refer to Eq.~(\ref{eq:vv}), we set the RHS to unity.  
Approximating the RHS of Eq.~(\ref{eq:vv}) by unity, and replacing ${\max}V_{\rm circ,h}$ instead of ${\max}V_{\rm circ,\lambda}$ in Eq.~(\ref{eq:maxVc}), the peak circular velocity of a host halo allows to  predict the scale parameter $\lambda$ and thus the soliton relevant for that host halo.

\subsection{Comparison to numerical simulations}\label{ss:compsim}
\begin{figure}[hbp!]
\centering
\includegraphics[width=0.45\textwidth]{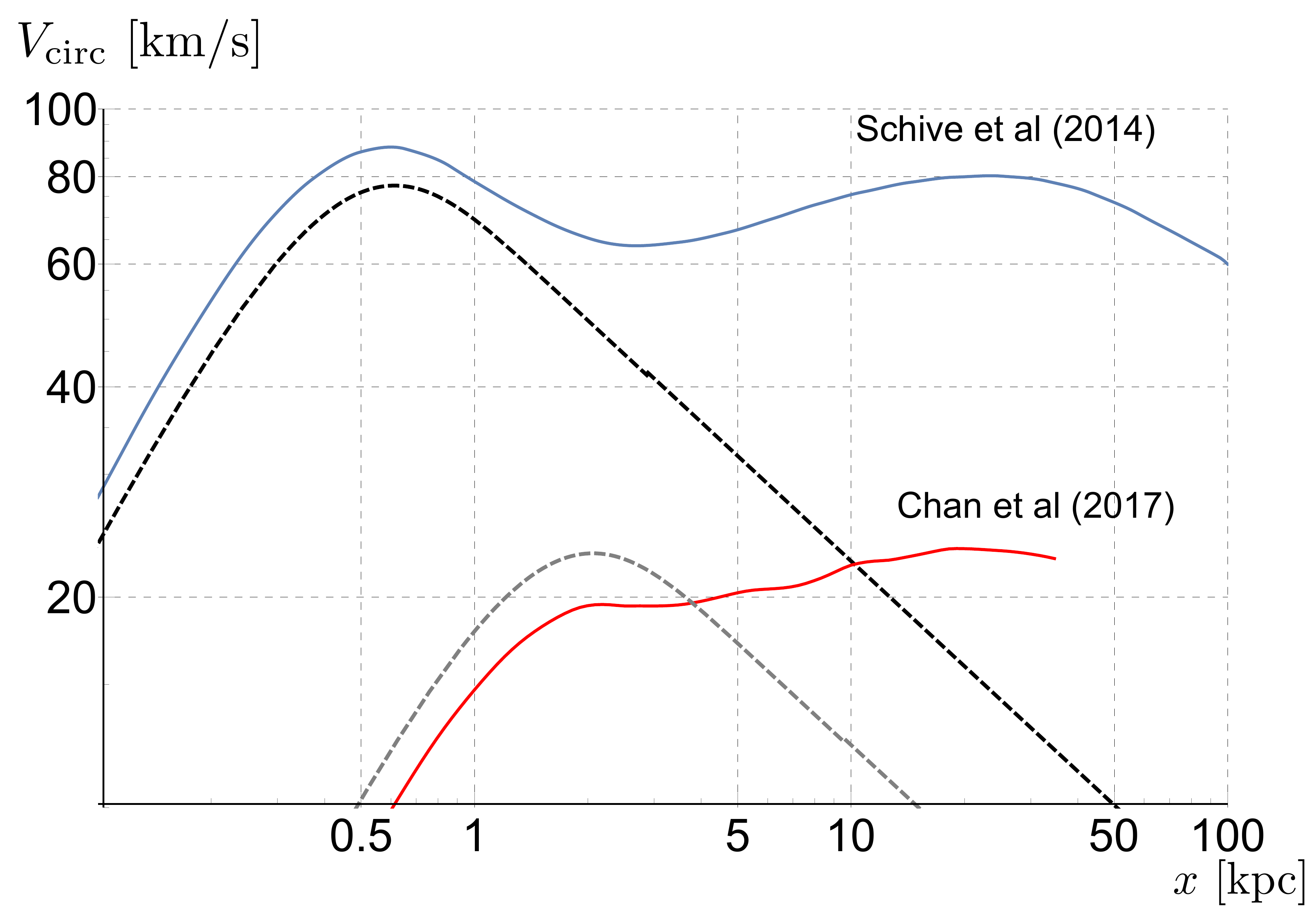}
\caption{Comparison of the prediction of Eq.~(\ref{eq:vv}) (dashed
  lines) to the numerical simulation results (solid lines) of
  Schive et al (2014)~\cite{Schive:2014dra} and
Chan et al (2017)~\cite{2017arXiv171201947C}. 
}\label{fig:Vcirc1406.6586}
\end{figure}

In Fig.~\ref{fig:Vcirc1406.6586} we compare our results to two soliton+halo configurations from the simulations of~\cite{Schive:2014dra} and~\cite{2017arXiv171201947C} (for~\cite{Schive:2014dra}, we take the largest halo, and for~\cite{2017arXiv171201947C} we take the initial state of Case C). To calculate the soliton, we read ${\max}V_{\rm circ,h}$ from the large-scale peak (at $x\sim20$~kpc) of the numerically extracted halo rotation curves (solid lines). Following Eq.~(\ref{eq:vv}), we use ${\max}V_{\rm circ,h}$ instead of ${\max}V_{\rm circ,\lambda}$ in Eq.~(\ref{eq:maxVc}), and read off the value of $\lambda$. %
The predicted soliton bump is shown in dashed lines. It gives the correct soliton peak rotation velocity to $\sim20$\% accuracy in both cases. 

\begin{figure}[hbp!]
\centering
\includegraphics[width=0.45\textwidth]{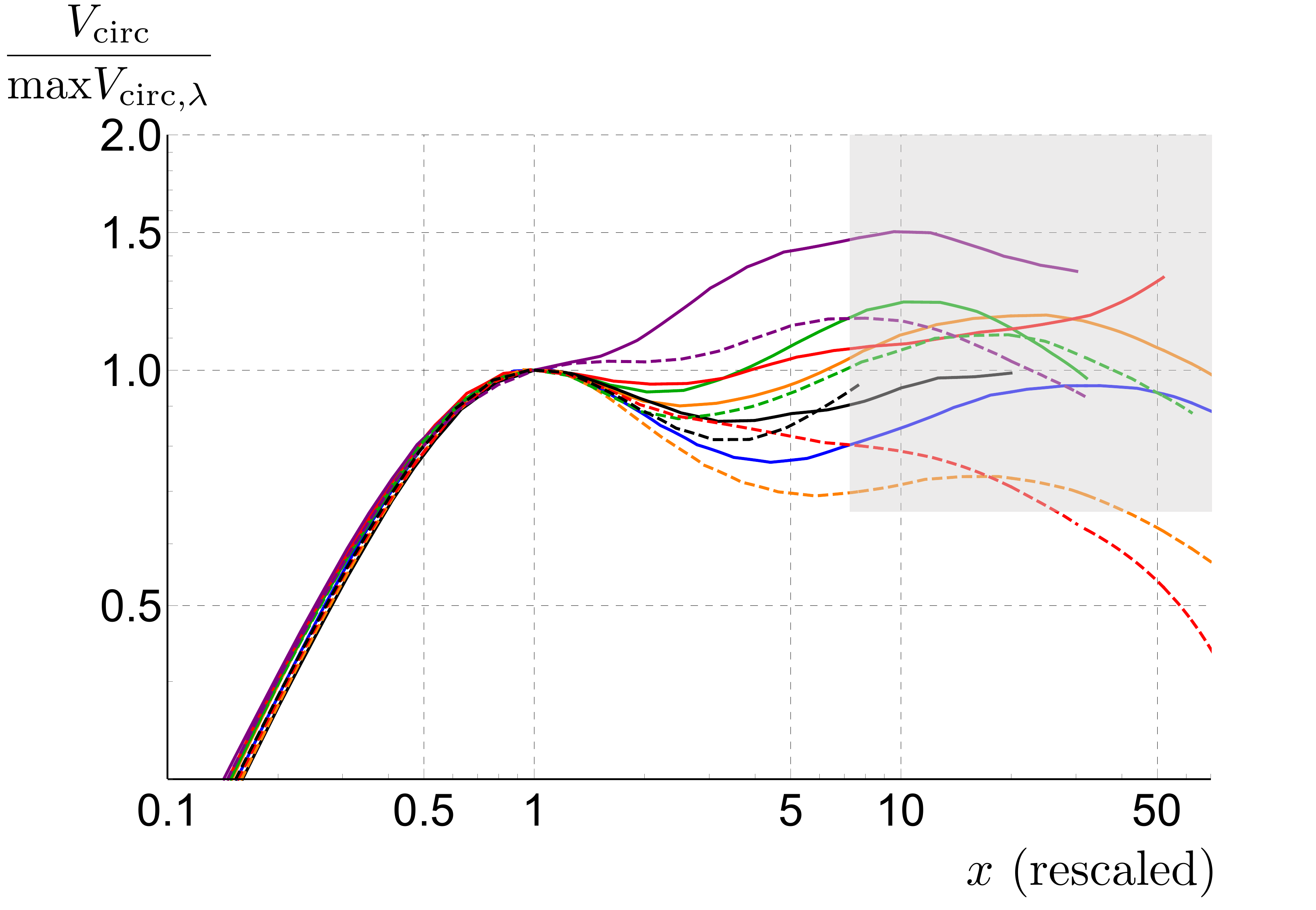}
\caption{Circular velocity profiles for halos, calculated for simulation results published in~\cite{Schive:2014hza} (solid lines) and~\cite{Schive:2014dra} (dashed lines), scaled and normalised such that $x_{\rm peak,\lambda}=1$ and ${\max}V_{\rm circ,\lambda}=1$. Shaded band highlights the range $0.5<\frac{{\max}V_{\rm circ,\lambda}}{{\max}V_{\rm circ,h}}<1.5$.
}\label{fig:Vcsim}
\end{figure}

In Fig.~\ref{fig:Vcsim} we show the velocity profiles for 11 simulated
halos, calculated for 6 halos from~\cite{Schive:2014hza} (solid lines)
and 5 halos from~\cite{Schive:2014dra} (dashed lines). The rotation
curves are scaled and normalised such that $x_{\rm peak,\lambda}=1$
and ${\max}V_{\rm circ,\lambda}=1$ in each case. All of the halos
satisfy $0.65<\frac{{\max}V_{\rm circ,\lambda}}{{\max}V_{\rm
    circ,h}}<1.4$;  
for later reference, the shaded band highlights the range
$0.5<\frac{{\max}V_{\rm circ,\lambda}}{{\max}V_{\rm circ,h}}<1.5$. 

\subsection{Comparison to real galaxies}\label{ss:compdat}

We now consider some observational consequences of our analysis. We
choose to do so by examining the rotation curves of nearby disc galaxies
with halo masses in the range covered by the simulations
of~\cite{Schive:2014hza,Schive:2014dra}, and above the minimal mass of
an ULDM halo, see Eq.~(\ref{eq:minMh}). We divide our analysis into two parts. First, in Sec.~\ref{sssec:few}, we concentrate on a small set of galaxies for which high-resolution kinematical and photometric data is available. For this small set of galaxies, we present detailed rotation curve data and compare them to the ULDM prediction. We find clear tension between the data and the ULDM prediction. Next, in Sec.~\ref{sssec:many}, we extend the analysis to include a large sample of galaxies, confirming and reinforcing the tension between the ULDM prediction and the rotation curve data.

\subsubsection{A few specific examples}\label{sssec:few}
In our first analysis of the data, we study a set of four representative rotation curves from Ref.~\cite{deBlok:2002vgq} (see Ref.~\cite{Lelli:2016zqa} for a recent
rendering of these and many other rotation curves), for which
high-resolution kinematical data is available. Our goal is to check if the ULDM prediction, that we 
demonstrated in the previous section using simulated galaxies, actually holds in real observed galaxies. The key prediction we check is summarised by Eq.~(\ref{eq:vv}), and was demonstrated for simulated galaxies in Figs.~\ref{fig:Vcirc1406.6586} and ~\ref{fig:Vcsim}: this prediction states that given a measurement of the large-radius halo rotation curve, ULDM prescribes a soliton-induced peak in the inner part of the halo, with height specified by Eq.~(\ref{eq:vv}), at a location specified by Eq.~(\ref{eq:xpeakVch}).

\begin{figure}[hbp!]
\centering
\includegraphics[width=0.45\textwidth]{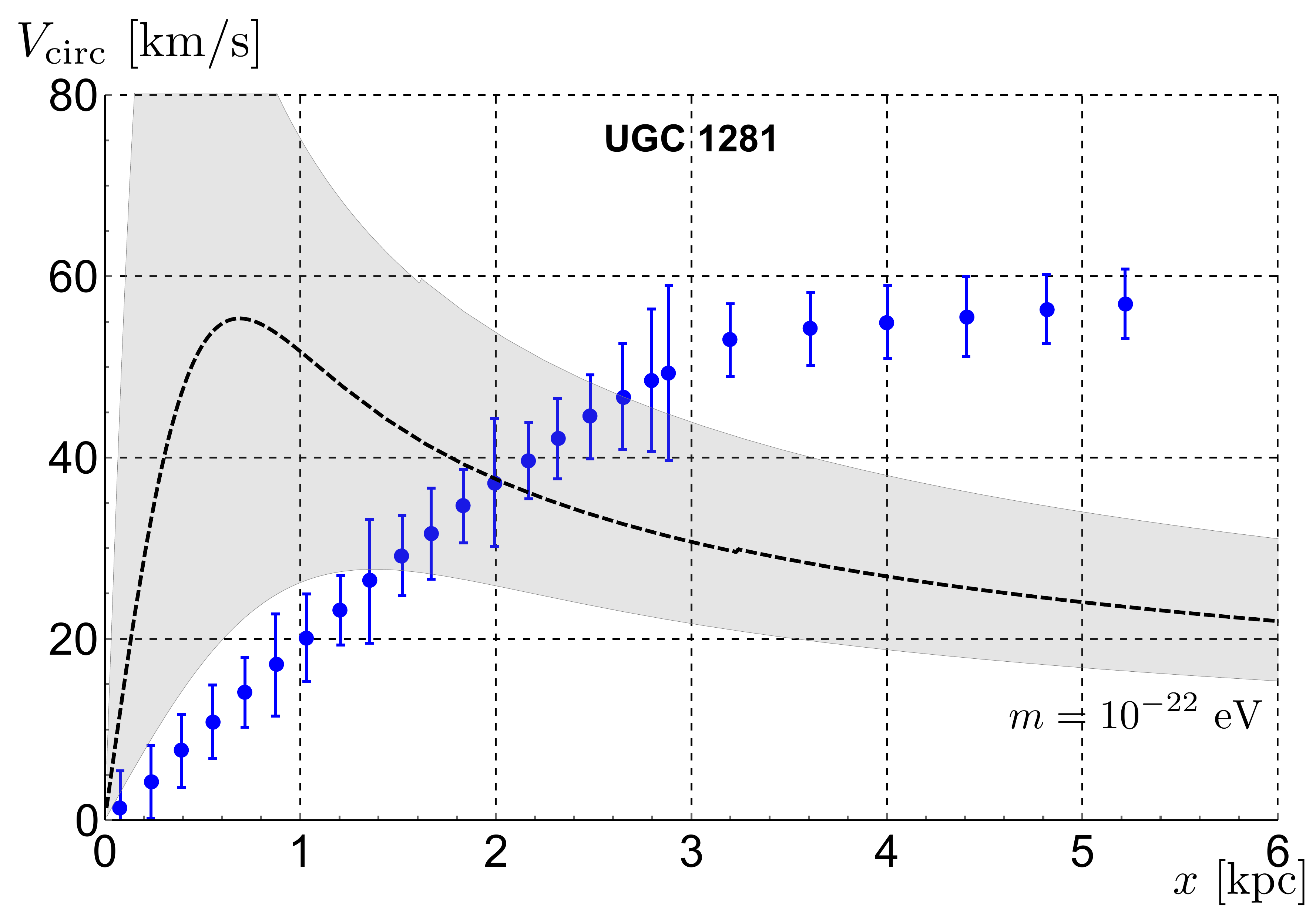}
\includegraphics[width=0.45\textwidth]{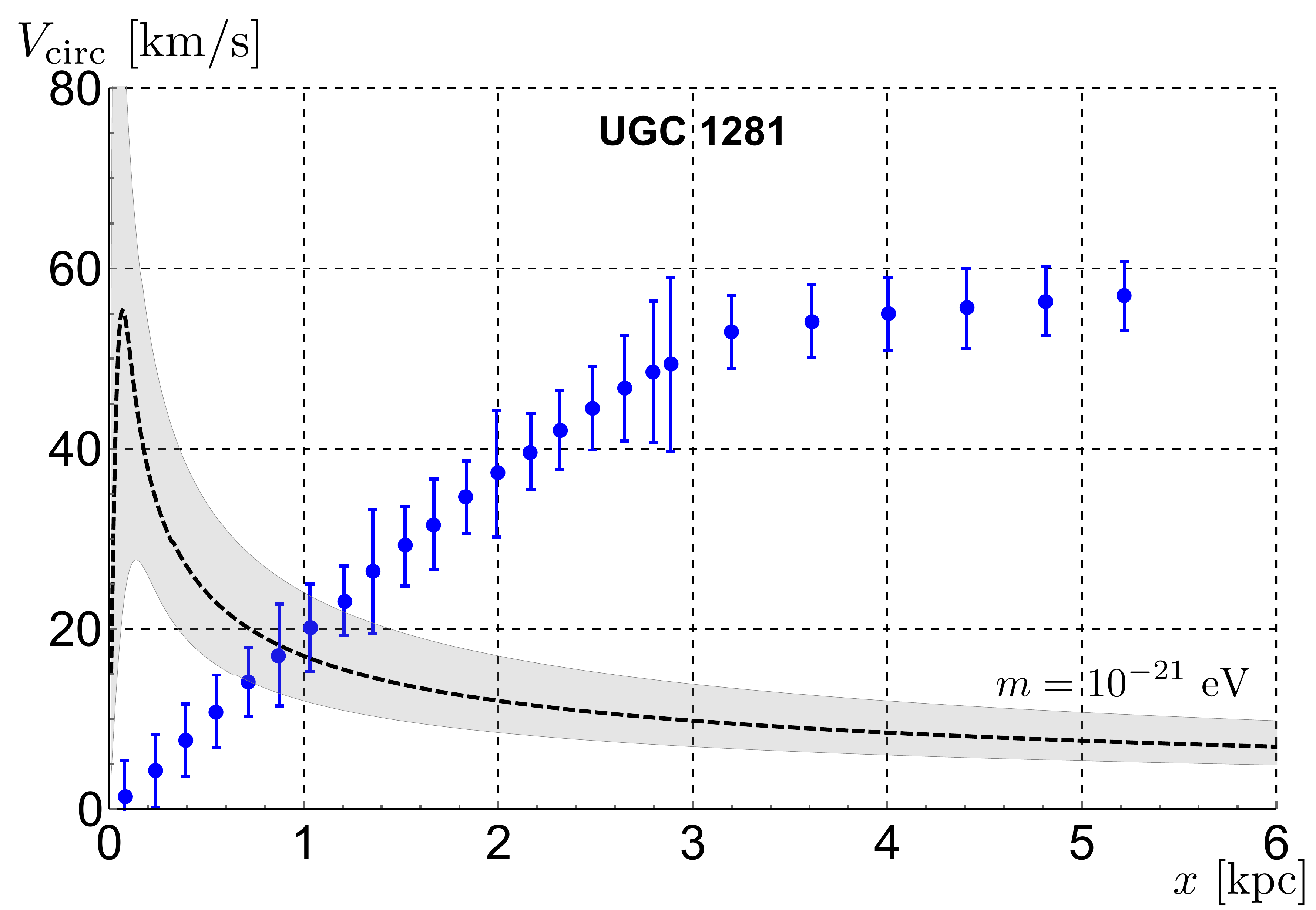}
\caption{Measured rotation curve of UGC 1281 superimposed on the
  prediction from Eq.~(\ref{eq:vv}) following from the soliton--host
  halo relation. 
The ULDM mass is $m=10^{-22}$~eV (upper panel) and
  $m=10^{-21}$~eV (lower panel). 
The shaded band accounts for the intrinsic scatter of
  the soliton--host 
  halo relation. 
}\label{fig:manygal1}
\end{figure}

\begin{figure}[hbp!]
\centering
\includegraphics[width=0.45\textwidth]{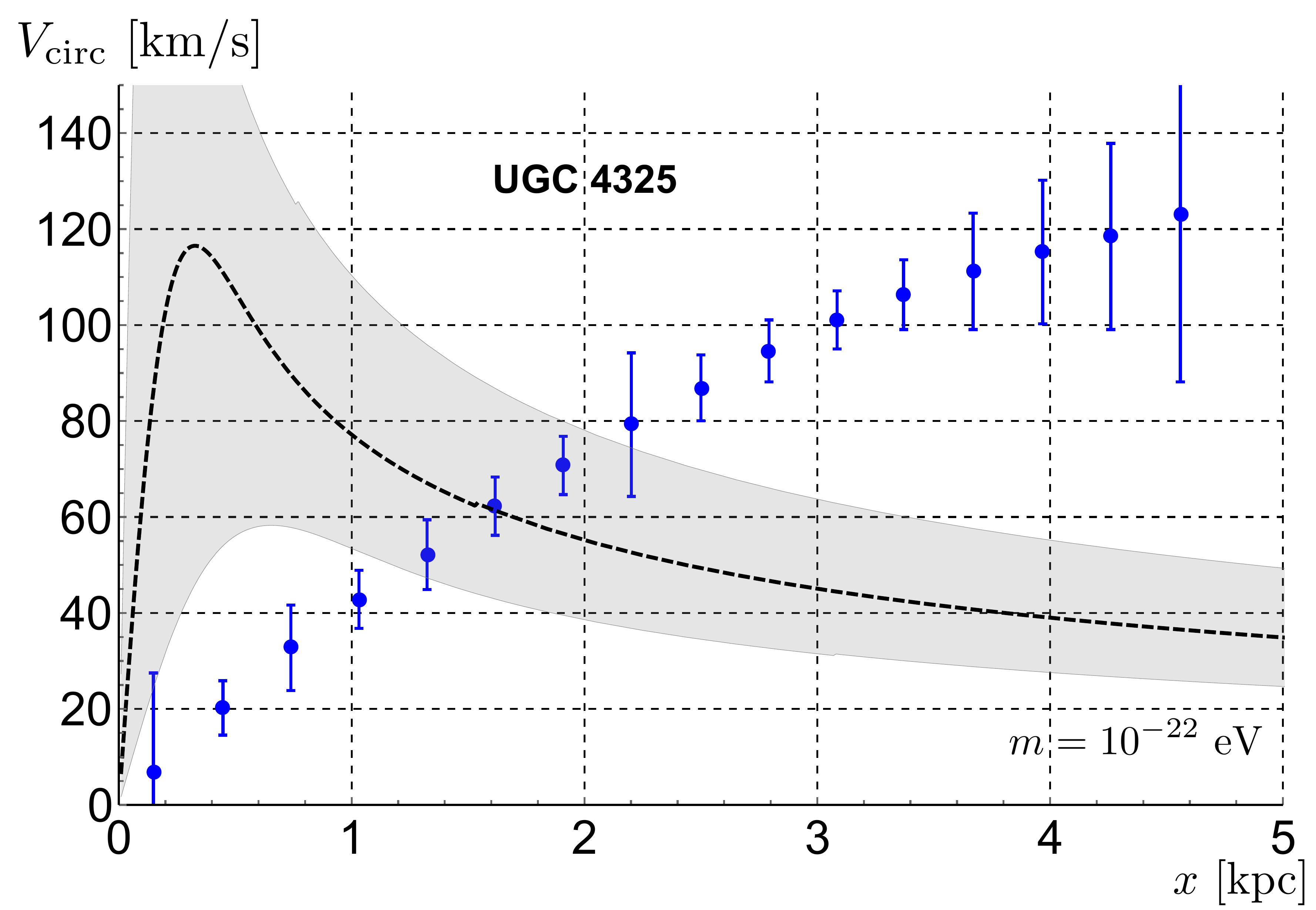}
\includegraphics[width=0.45\textwidth]{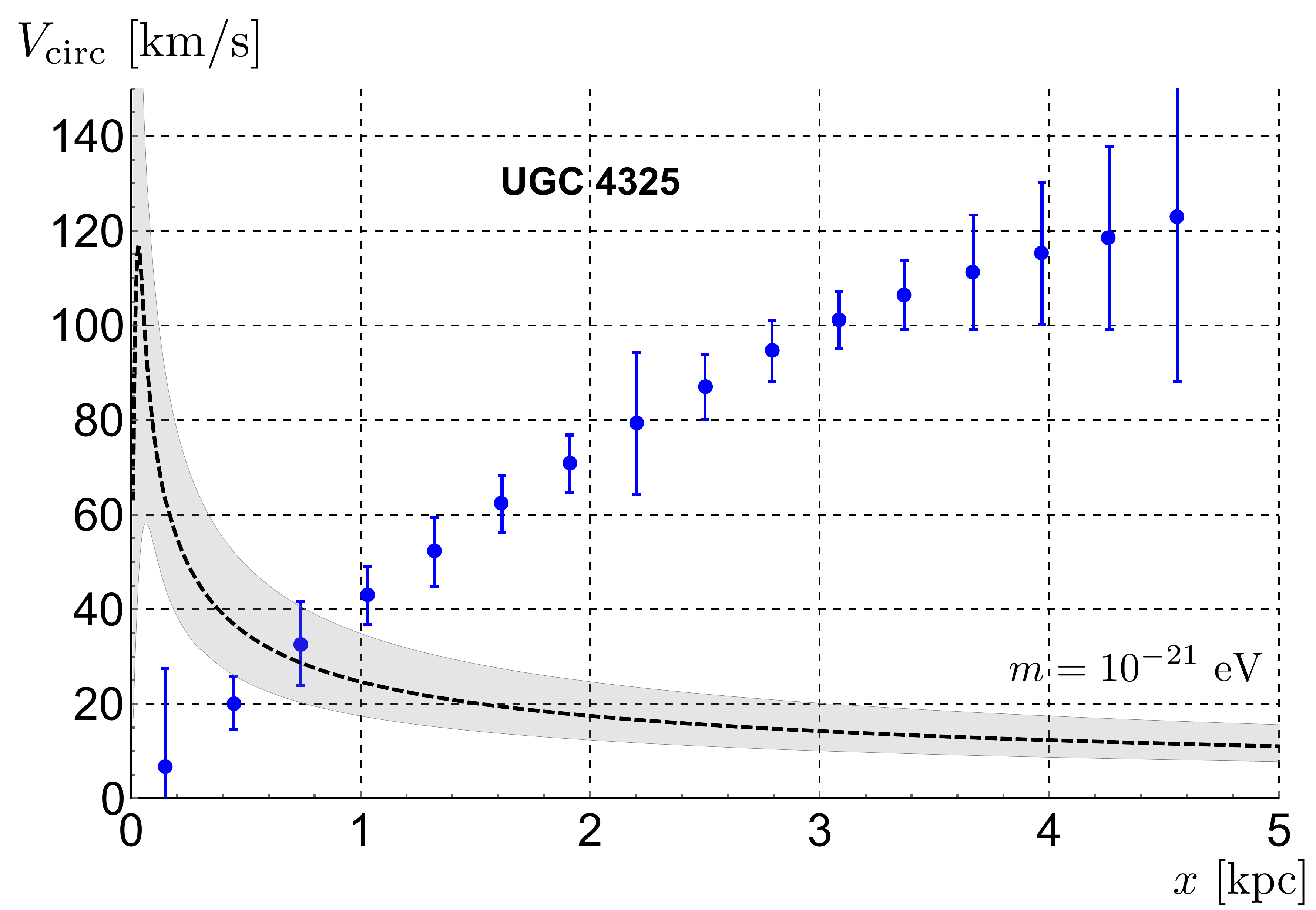}
\caption{Same as Fig.~\ref{fig:manygal1} for UGC 4325.
}\label{fig:manygal2}
\end{figure}

\begin{figure}[hbp!]
\centering
\includegraphics[width=0.45\textwidth]{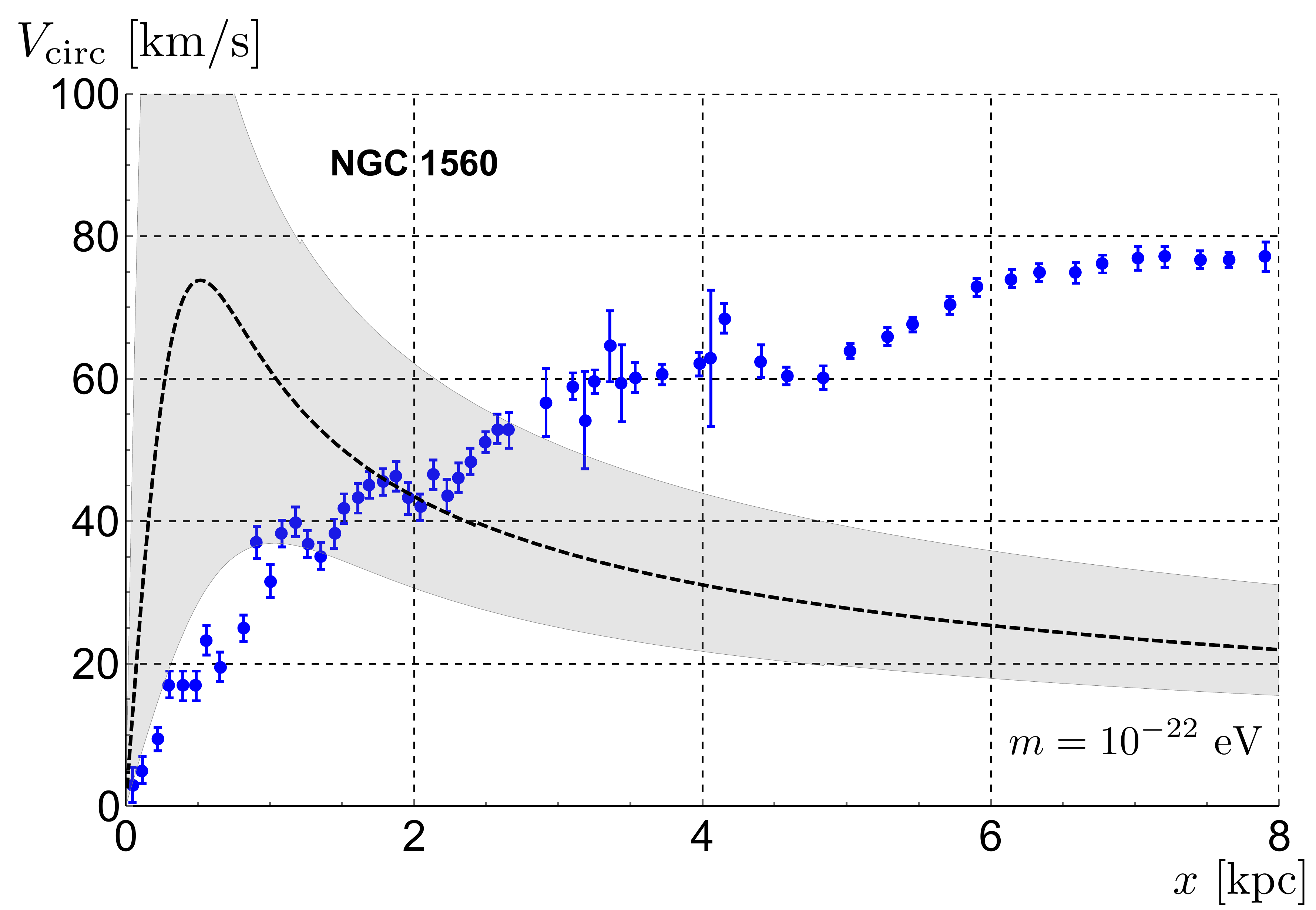}
\includegraphics[width=0.45\textwidth]{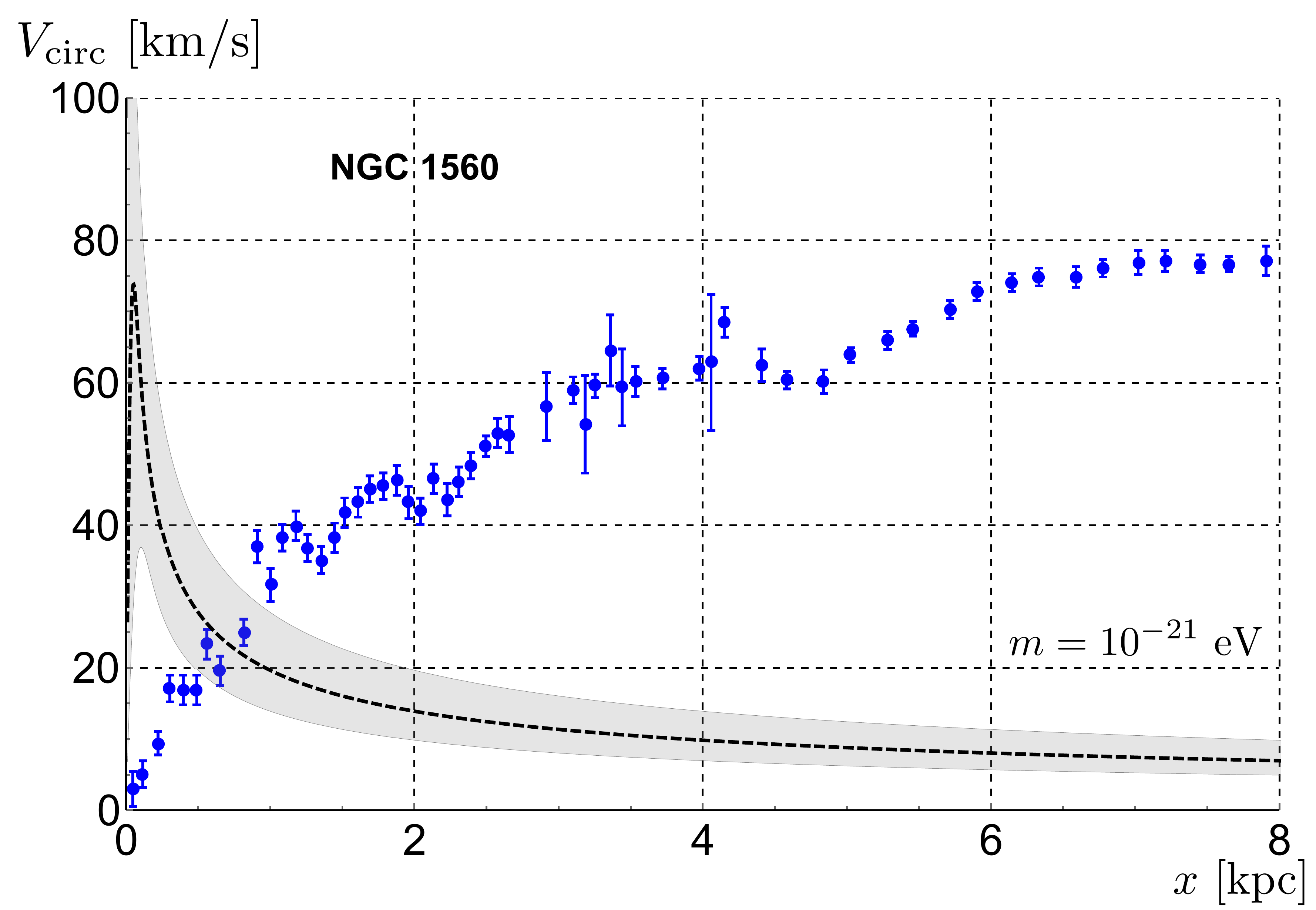}
\caption{Same as Fig.~\ref{fig:manygal1} for NGC 1560.
}\label{fig:manygal3}
\end{figure}

\begin{figure}[hbp!]
\centering
\includegraphics[width=0.45\textwidth]{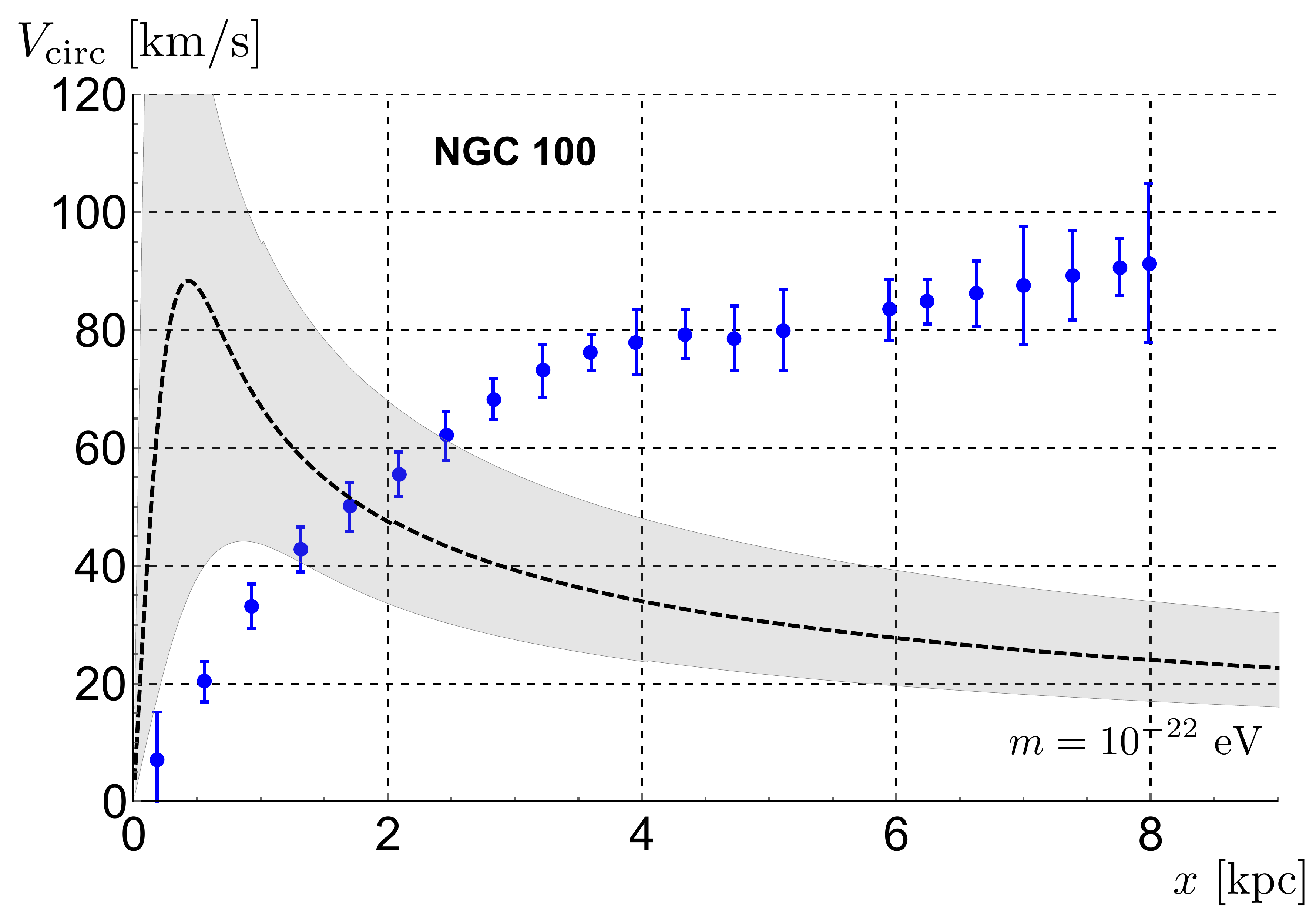}
\includegraphics[width=0.45\textwidth]{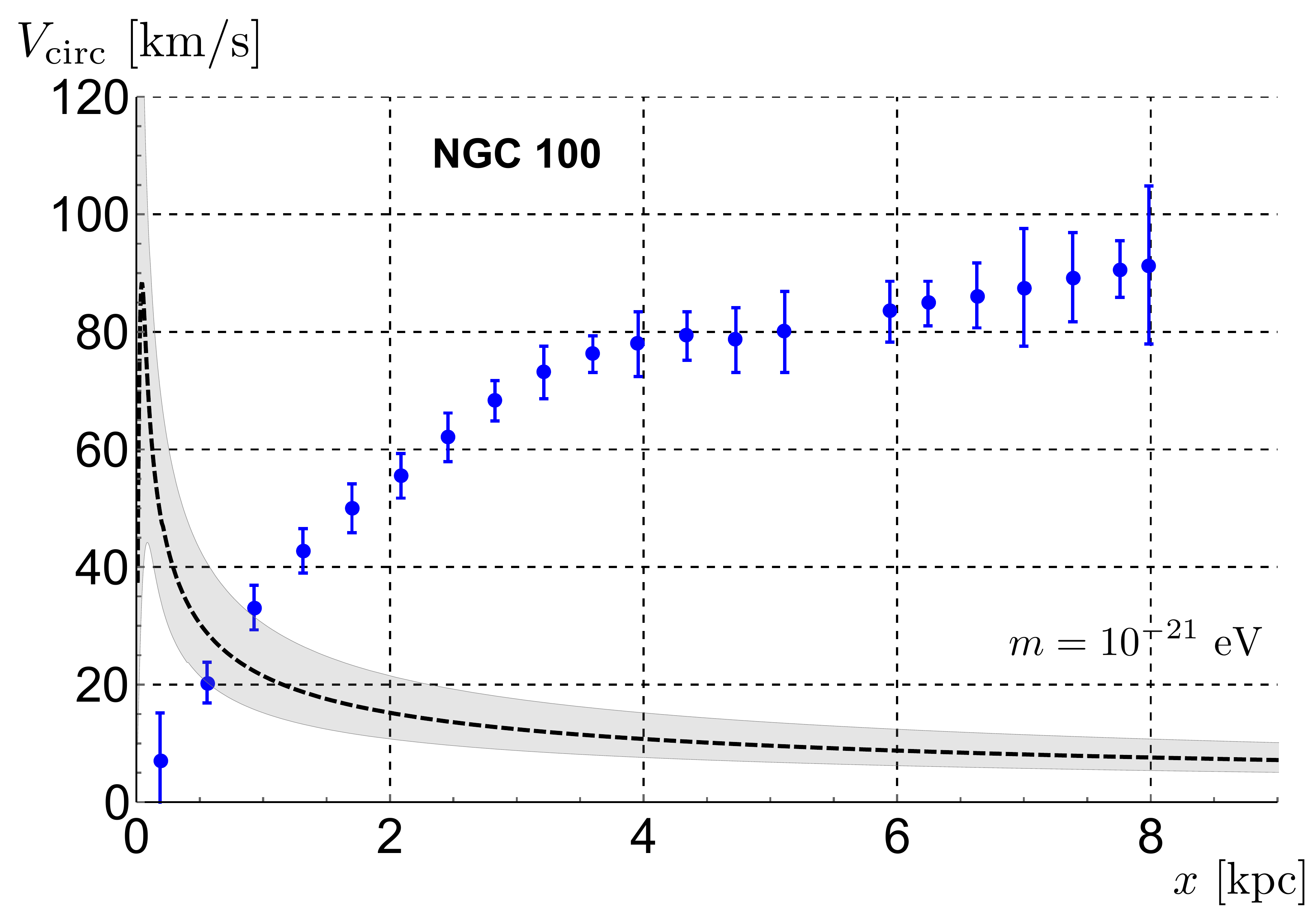}
\caption{Same as Fig.~\ref{fig:manygal1} for NGC 100.
}\label{fig:manygal4}
\end{figure}

The four measured rotation curves are shown in Figs.~\ref{fig:manygal1}-\ref{fig:manygal4}, as blue markers. For each of these rotation curves, we use the measured circular velocity at the farthest radius, to serve as an input for ${\rm max}V_{\rm circ,h}$ in Eq.~(\ref{eq:vv}). In turn, Eq.~(\ref{eq:vv}) gives as output the predicted soliton-induced peak rotation velocity, ${\rm max}V_{\rm circ,\lambda}$. Given the soliton-induced peak rotation velocity, the soliton $\lambda$ parameter is fixed by Eq.~(\ref{eq:maxVc}) and with it, the full soliton-induced rotation curve. The result is plotted as dashed line in Figs.~\ref{fig:manygal1}-\ref{fig:manygal4}; the upper panels show the result for $m=10^{-22}$~eV and the lower panels show it for $m=10^{-21}$~eV.

We find that the ULDM soliton--host halo relation  
significantly overestimates the rotation velocity in the inner part of
all of the galaxies in Figs.~\ref{fig:manygal1}-\ref{fig:manygal4}.  This puts the predictions of ULDM in the mass range $m\sim (10^{-22}\div 10^{-21})$ eV in tension with the data.

We emphasize that in using Eq.~(\ref{eq:vv}) to predict the soliton,
we set the RHS of that equation to unity, and thus we ignore any
details of the shape of the host halo. We also neglect corrections due to baryons. 
On the one hand, as we have learned from the NFW
analysis, this prescription for deriving the soliton profile would
suffer $\mathcal{O}(10\%)$ corrections from the detailed halo shape. 
The baryonic (stellar and gas) contribution to the gravitational potential affects the halo velocity at a similar level: in Ref.~\cite{Lelli:2016zqa}, the baryonic contribution to the halo peak rotation velocity of UGC 1281, UGC 4325, and NGC 100 was estimated from photometric data  to be well bellow the DM contribution\footnote{For this estimate we use 3.6$\mu$m mass-to-light ratio $\Upsilon_*=0.5~{\rm M_\odot/L_\odot}$.}, $V_{\rm circ,h}^{(\rm bar)}/V_{\rm circ,h}^{(\rm obs)}<0.5$. This means that the observed velocity $V_{\rm circ,h}^{(\rm obs)}$ is equal to the DM-induced velocity $V_{\rm circ,h}^{(\rm DM)}$ to better than 15\%.
On the other hand, this simple procedure relieves us from the need to fit for the virial mass or  other details of the host halo. All that is needed is the peak halo 
rotation velocity, a directly observable quantity\footnote{\label{fn}The
  rotation curves in Figs.~\ref{fig:manygal1}-\ref{fig:manygal4} do
  not show a clear peak within the range of the measurement; this
  means that our soliton bump, derived from the maximal velocity 
  seen in the data, underestimates the true predicted soliton and is
  thus conservative.}. For completeness, we will return to the issue of baryonic effects and deal with it more systematically in the second part of this section, considering SPARC data~\cite{Lelli:2016zqa}. 
Eq.~(\ref{eq:vv}) (represented by the dashed line in
Figs.~\ref{fig:manygal1}-\ref{fig:manygal4}) corresponds to the central value
of the soliton--host halo
relation. Ref.~\cite{Schive:2014hza,Schive:2014dra} showed a scatter
of about a factor of two around Eq.~(\ref{eq:McEh}) between
simulated halos. This translates to a factor of two scatter in the
soliton $\lambda$ parameter (we have illustrated this scatter for a sub-sample of simulated halos in Fig.~\ref{fig:Vcsim}).  
In Figs.~\ref{fig:manygal1}-\ref{fig:manygal4} we represent this
 scatter by a shaded band, showing the results
when the $\lambda$ parameter inferred from Eq.~(\ref{eq:vv}) is
changed by a factor of 2. 

\subsubsection{A large sample of galaxies: SPARC}\label{sssec:many}

It is important to check if scatter between different
galaxies could explain the discrepancy, with the four galaxies in
Figs.~\ref{fig:manygal1}-\ref{fig:manygal4} being accidental outliers.  
To address this question, we analyse the 175 rotation curves contained
in the SPARC data base~\cite{Lelli:2016zqa}.  
This sample includes, in particular, the galaxies UGC 1281, UGC 4325, and NGC 100, shown in
Figs.~\ref{fig:manygal1}-\ref{fig:manygal4}.  

Our SPARC analysis is as follows. For each galaxy, we make a crude
estimate of the halo mass contained within the observed rotation curve
profile by $M_{\rm gal}\sim RV^2/G$, where $R$ is the radial distance
of the last data point in the rotation curve, and
$V$ the corresponding velocity. We keep only galaxies with
$5\times10^{11}~{\rm M}_{\odot}>M_{\rm gal}>5\times10^8\left(m/10^{-22}~{\rm eV}\right)^{-3/2}$~M$_\odot$. We
do this in order to limit ourselves to galaxy masses 
that are comfortably above the minimal halo mass (\ref{eq:minMh}), and not above the range simulated in~\cite{Schive:2014hza,Schive:2014dra}.
Our results are not sensitive
to the details of this mass cut.

Next, for each galaxy we determine the observed maximal halo rotation
velocity ${\rm max} V_{\rm circ,h}$, and use it to compute the soliton
prediction from Eq.~(\ref{eq:vv}). To avoid confusion between halo
peak velocity and soliton peak velocity, we search for the halo peak
velocity restricting to radial distance $x>3\left(m/10^{-22}~{\rm
    eV}\right)^{-1}$~kpc. Galaxies with no data above
$x=3\left(m/10^{-22}~{\rm eV}\right)^{-1}$~kpc are discarded. Our
results are not sensitive to this criterion; defining the halo cut anywhere at
$\gtrsim 1\,\left(m/10^{-22}~{\rm eV}\right)^{-1}$~kpc guarantees that
such confusion is avoided.  

SPARC galaxies come with photometric data, allowing to model the baryonic contribution to the gravitational potential~\cite{Lelli:2016zqa}. We use this information to limit baryonic effects on our analysis, and to explore the sensitivity of our results to baryonic corrections. For each galaxy, we estimate the baryonic contribution to the observed rotation velocity using the mass models of~\cite{Lelli:2016zqa} with 3.6$\mu$m mass-to-light ratio $\Upsilon_*=0.5~{\rm M_\odot/L_\odot}$. Setting $V_{\rm circ,h}^{(\rm bar),2}+V_{\rm circ,h}^{(\rm DM),2}=V_{\rm circ,h}^{(\rm obs),2}$, we calculate the ratio $f_{\rm bar2DM}=V_{\rm circ,h}^{(\rm bar)}/V_{\rm circ,h}^{(\rm DM)}$. We present results when cutting on different values of $f_{\rm bar2DM}<1,\,0.5,\,0.33$. 

Our first pass on the data includes only galaxies for which the
predicted soliton is resolved, namely, $x_{\rm peak,\lambda}$ from
Eq.~(\ref{eq:xpeakVch}), with ${\rm max} V_{\rm circ,\lambda}={\rm
  max} V_{\rm circ,h}^{\rm (DM)}$, lies within the rotation curve data. For these
galaxies, we compute from data the ratio 
\be\frac{V_{\rm circ,\,obs}(x_{\rm peak,\lambda})}{{\rm max} V_{\rm circ,h}^{\rm (DM)}}.\ee
Here, $V_{\rm circ,\,obs}(x_{\rm peak,\lambda})$ is the measured
velocity at the expected soliton peak position. 

\begin{figure}[hbp!]
\centering
\includegraphics[width=0.5\textwidth]{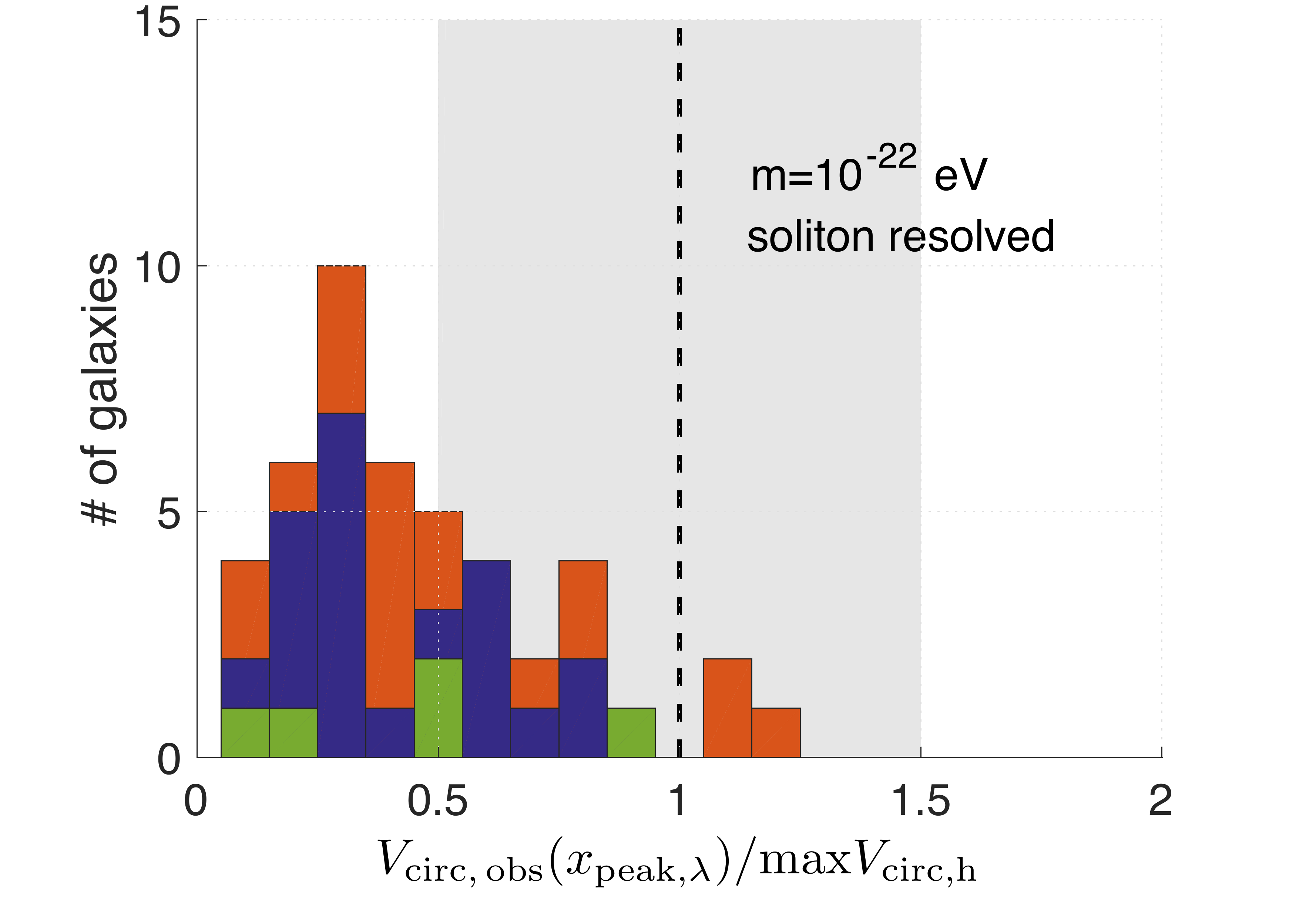}
\includegraphics[width=0.5\textwidth]{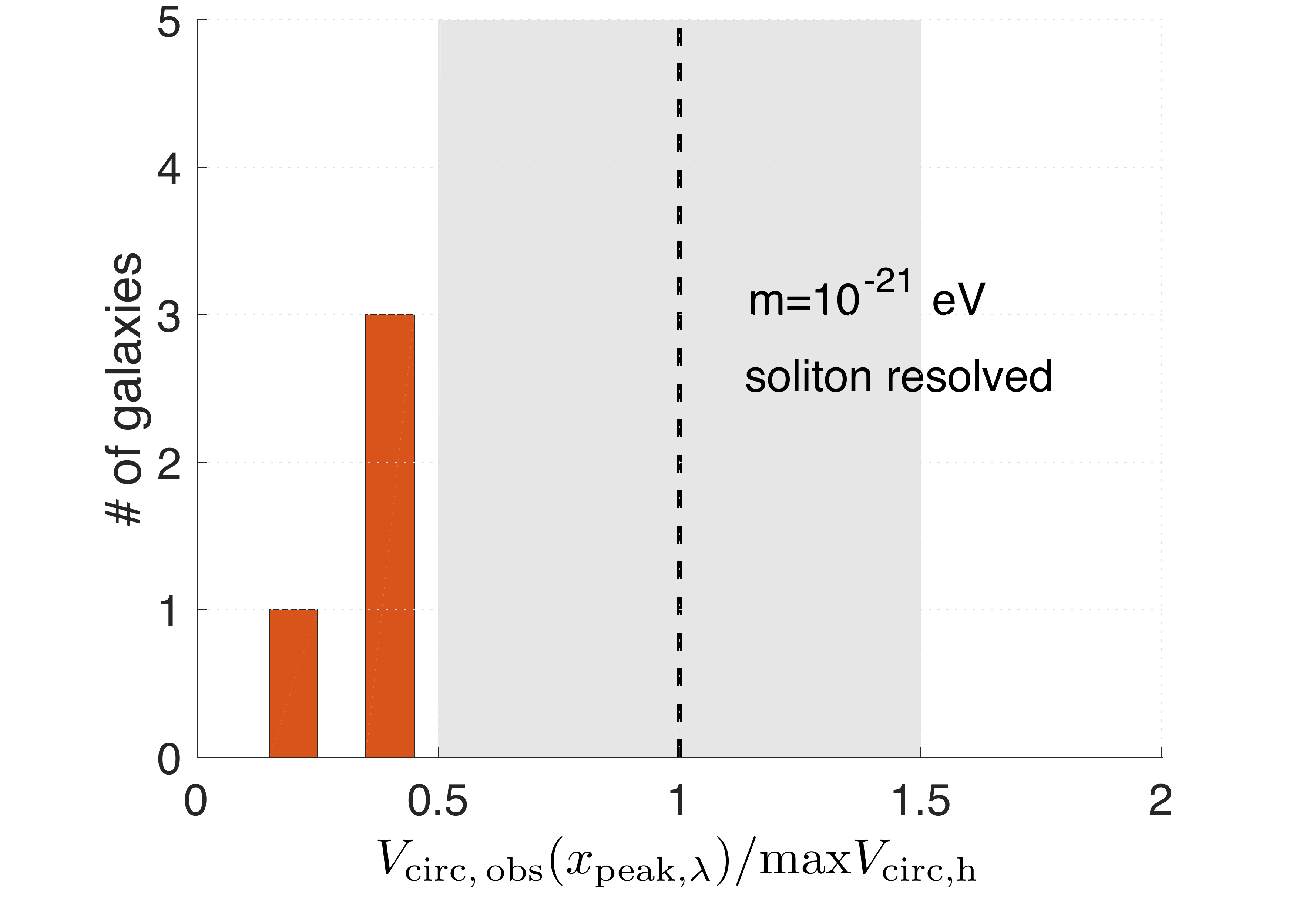}
\caption{Distribution of SPARC galaxies~\cite{Lelli:2016zqa} with
  respect to the ratio of observed circular velocity at the soliton
  peak to the maximal circular velocity of the halo. The vertical
  dashed line shows the prediction for the mean implied by
  the soliton-host halo relation and the shaded region accounts for
  the intrinsic scatter in this relation. The ULDM mass is
  $m=10^{-22}$~eV (upper panel) and $m=10^{-21}$~eV (lower panel).
Red, blue, green histograms correspond to the cuts $f_{\rm bar2DM} <1,
~0.5,~ 0.33$, respectively.
Only rotation curves with resolved solitons are included (see the main
text for details).
}\label{fig:sparc1}
\end{figure}

The results of this first pass on the data are shown in
Fig.~\ref{fig:sparc1}. Red, blue, and green histograms show the result when imposing $f_{\rm bar2DM}<1,\,0.5,\,0.33$, respectively. For $m=10^{-22}$~eV, we find 45, 26, and 5 galaxies that pass the resolved soliton cut
for $f_{\rm bar2DM}<1,\,0.5,\,0.33$. For $m=10^{-21}$~eV, only 4 galaxies pass the resolved soliton cut for $f_{\rm bar2DM}<1$, and none for $f_{\rm bar2DM}<0.5,\,0.33$.

\begin{figure}[hbp!]
\centering
\includegraphics[width=0.5\textwidth]{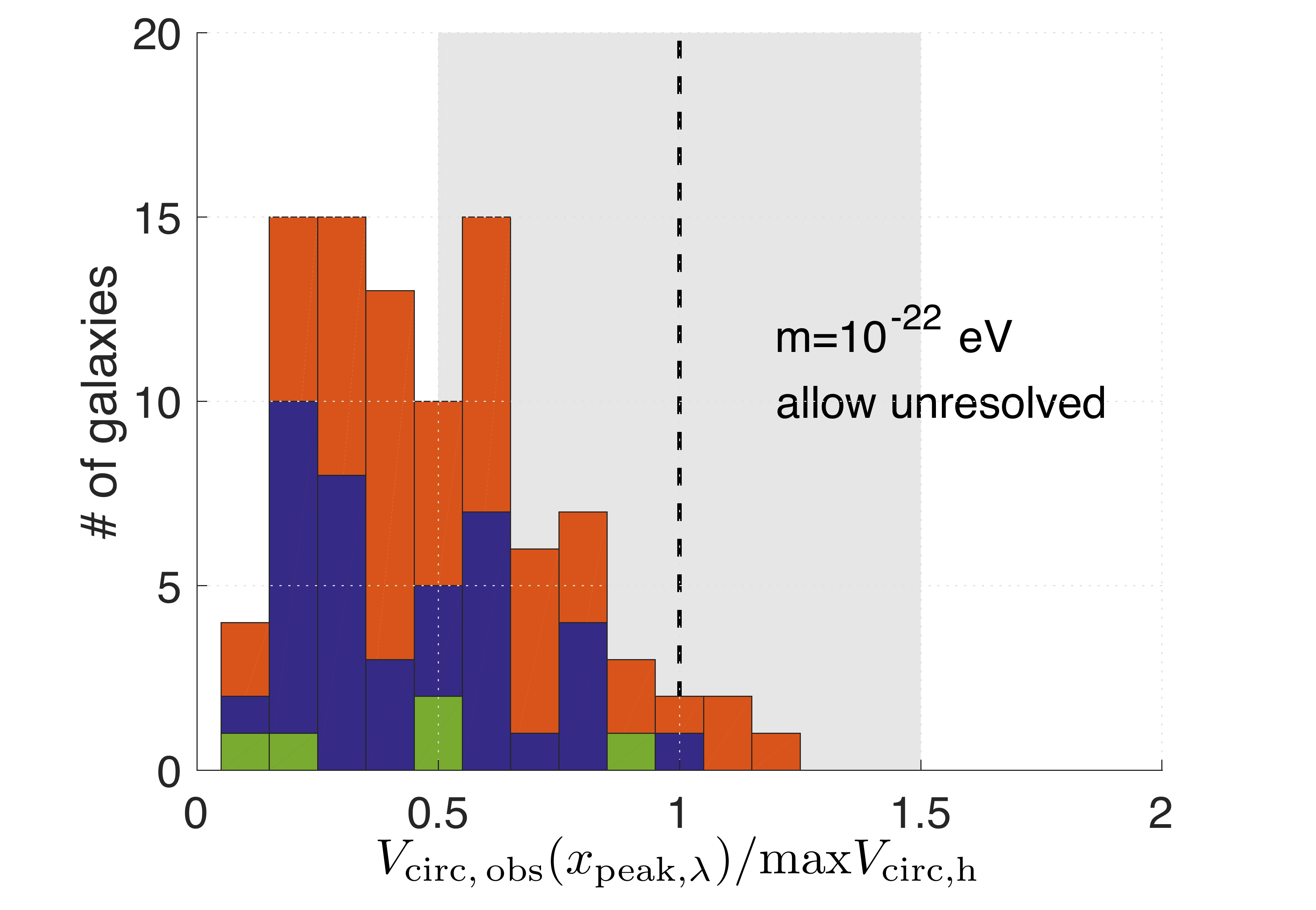}
\includegraphics[width=0.5\textwidth]{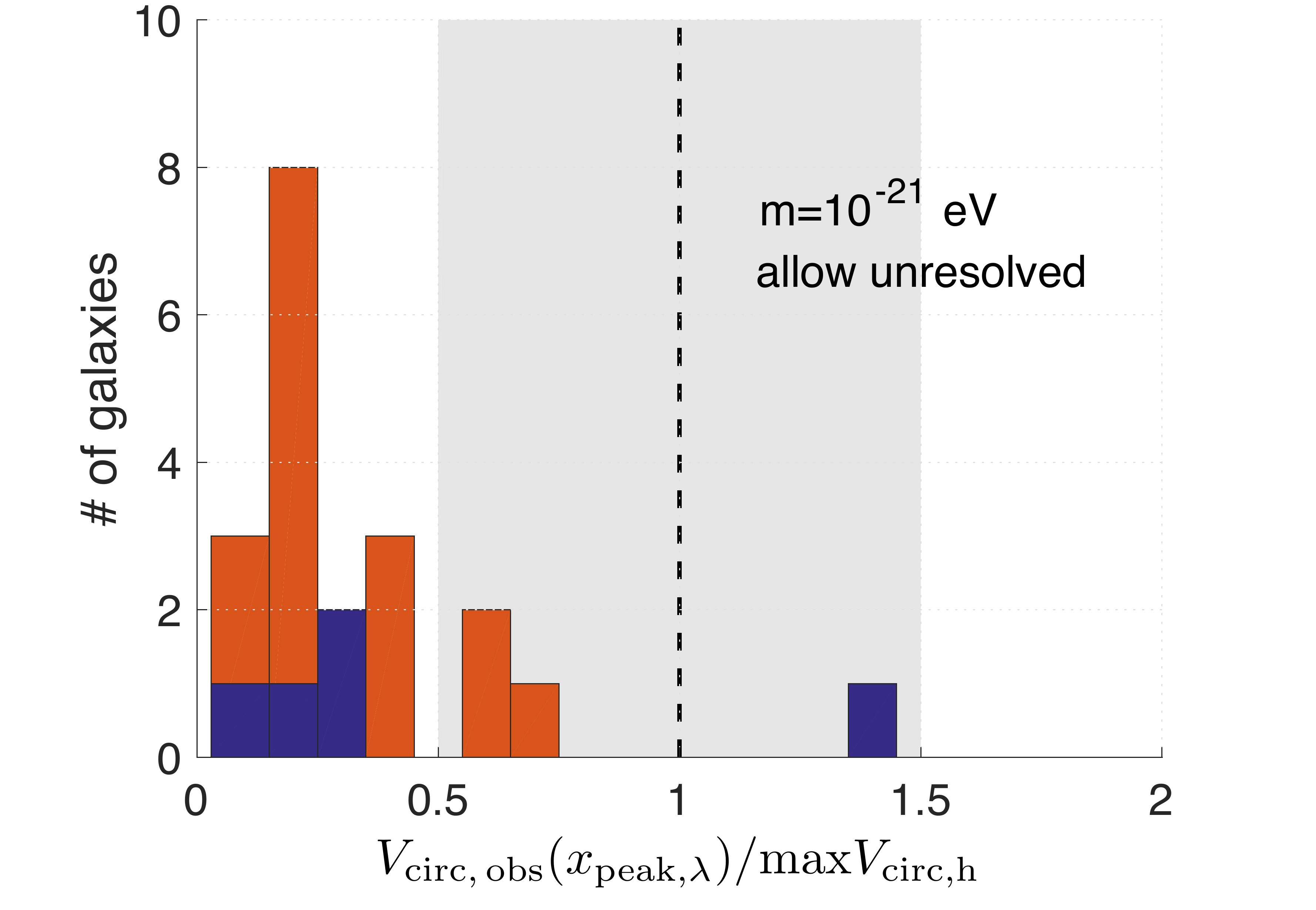}
\caption{Same as Fig.~\ref{fig:sparc1}, including galaxies with 
unresolved solitons (see the main text).
}\label{fig:sparc2}
\end{figure}

Including only galaxies with a resolved soliton causes us to loose
many rotation curves with discriminatory power. For example,
UGC 4325 drops out of the analysis for $m=10^{-21}$~eV, though it
clearly constrains the model, as seen from the lower panel
of Fig.~\ref{fig:manygal2}.
To
overcome this without complicating the analysis, we
perform a second 
pass on the data. Here, we allow galaxies with unresolved soliton, as
long as the innermost data point is located not farther than $3\times
x_{\rm peak,\lambda}$. We need to correct for the fact that the
soliton peak velocity is outside of the measurement resolution. To do
this, we modify our observable as 
\be\frac{V_{\rm circ,\,obs}(x_{\rm peak,\lambda})}{{\rm max} V_{\rm circ,h}^{\rm (DM)}}\to \frac{V_{\rm circ,\,obs}(x_{\rm min,data})}{{\rm max} V_{\rm circ,h}^{\rm (DM)}}\times\sqrt{\frac{x_{\rm min,data}}{x_{\rm peak,\lambda}}},\no\\\ee
where $x_{\rm min,data}$ is the radius of the first data point. This
correction is conservative, because it takes the fall-off of the
soliton gravitational potential at
at $x>x_{\rm peak,\lambda}$ to be the same as for a point mass. 
In reality, the potential decays slower and
the soliton-induced velocity decreases slower. Keeping this caveat in
mind, Fig.~\ref{fig:sparc2} presents our results including unresolved
solitons. 
For $m=10^{-22}$~eV, with $f_{\rm bar2DM}<1,\,0.5$, we find 48 and 16 galaxies with unresolved soliton, that can be added to the sample of Fig.~\ref{fig:sparc1}. No galaxy is added for $f_{\rm bar2DM}<0.33$. For $m=10^{-21}$~eV, 16 and 5 galaxies are added with $f_{\rm bar2DM}<1, 0.5$, and none for $f_{\rm bar2DM}<0.33$.

In Figs.~\ref{fig:sparc1}-\ref{fig:sparc2}, vertical dashed line indicates the soliton--host halo prediction. The shaded region shows the range of the prediction, modifying the RHS of Eq.~(\ref{eq:vv}) between $0.5-1.5$, consistent with the scatter seen in the simulations. 

We conclude that the four galaxies in
Figs.~\ref{fig:manygal1}-\ref{fig:manygal4} are not outliers: they are
representative of a systematic discrepancy, that would be difficult to
attribute to the scatter seen in the simulations.  
If the soliton-host halo relation
of~\cite{Schive:2014hza,Schive:2014dra} is correct, then ULDM in the
mass range $m\sim10^{-22}$~eV to $m\sim10^{-21}$~eV is in tension with the data.  

We have limited our attention to the range $m=(10^{-22}\div
10^{-21})$~eV, for which we believe the results are clear. We leave a
detailed study of the precise exclusion range to future work. We note
that for lower particle mass, $m\lesssim10^{-23}$~eV, the soliton
contribution extends over much of the velocity profile of many of the
SPARC galaxies, leaving little room for a host halo. This limit, where
the galaxies are essentially composed of a single giant soliton, was
considered in other works. We do not pursue it further, one reason
being that this range of small $m$ is in significant tension with
Ly-$\alpha$ data~\cite{Armengaud:2017nkf,Bozek:2014uqa}.  

For higher particle mass, $m\gtrsim10^{-21}$~eV, the soliton peak is
pushed deep into the inner 100~pc of the rotation curve. Although high
resolution data (e.g. NGC 1560,
Fig.~\ref{fig:manygal3}) is sensitive to and disfavours this
situation, a more careful analysis would be needed to draw a
definitive conclusion.
Note that in this range of $m$, ULDM ceases to offer a solution to the
small-scale puzzles of $\Lambda$CDM (see, e.g., review
in~\cite{Hui:2016ltb}).

\section{Baryonic effects}\label{s:bareff}

\subsection{A fixed distribution of baryons}\label{s:fixbar}
We now consider the soliton solution, and some aspects of the
soliton-host halo relation, with a coexisting distribution of baryonic
mass. In this section we consider a smooth mass distribution,
deferring the analysis of the effect of a super-massive black hole
(SMBH) to App.~\ref{ss:smbh}. 

The analysis of baryonic effects is qualitatively important, and
quantitatively relevant to ULDM in, e.g., MW-like galaxies. 
We will see that in order to make a significant impact, baryons need to constitute an
$\mathcal{O}(1)$ fraction of the total mass in the soliton
region. For many of the galaxies that we analysed in Sec.~\ref{ss:compdat}, the predicted soliton contribution to the rotation 
velocity exceeds the
observed velocities by a factor of 3-5, implying that the soliton
over-predicts the mass in the central region of the galaxy by about an
order of magnitude. 
This leaves little room for baryons (and, we think, baryonic feedback) to significantly affect the dynamics of the inner ULDM halo. 

Proceeding to the analysis, we denote the gravitational potential due
to baryons by $\Phi_b(r)$. We assume that the distribution of
baryonic mass is spherically symmetric\footnote{The assumption of
  spherical symmetry is not realistic; some of the galaxies
  considered in Sec.~\ref{ss:compdat} may in fact exhibit maximal
  discs. Nevertheless, we expect our simplified analysis to give us
  correct order of magnitude estimates, and defer the analysis of
  non-spherical baryonic+soliton+halo systems to future work.} and
dies off at infinity sufficiently fast, so that
$\Phi_b(r\to\infty)=-GM_b/r$, where $M_b$ is the total baryonic mass. 
Adding $\Phi_b(r)$ changes Eqs.~(\ref{eq:SPselfX}-\ref{eq:SPselfV}) into~\cite{Arbey:2001qi}
\be\partial_r^2\left(r\chi\right)&=&2r\left(\Phi+\Phi_b\left(r\right)-\gamma\right)\chi,\label{eq:SPselfXbar}\\
\partial_r^2\left(r\Phi\right)&=&r\chi^2\label{eq:SPselfVbar}.\ee
It remains convenient to solve the problem using boundary conditions
with $\chi(0)=1$.  
Let us denote this solution (satisfying $\chi(0)=1$) by
$\chi_1(r;\Phi_b)$, accompanied by the soliton potential
$\Phi_1(r;\Phi_b)$. The general solution $\chi_\lambda(r;\Phi_b)$
satisfying Eqs.~(\ref{eq:SPselfXbar}-\ref{eq:SPselfVbar}) with
boundary condition $\chi_\lambda(0;\Phi_b)=\lambda^2$ is given by  
\be
\chi_\lambda(r;\Phi_b)&=&\lambda^2\chi_1\left(\lambda r;\lambda^{-2}\Phi_b\left(\lambda^{-1}r\right)\right), \label{eq:solbar1}\\
\Phi_\lambda(r;\Phi_b)&=&\lambda^2\Phi_1\left(\lambda r;\lambda^{-2}\Phi_b\left(\lambda^{-1}r\right)\right).\label{eq:solbar2}\ee
Defining the soliton mass and energy in the presence of the baryons by $M_\lambda(\Phi_b),\,E_\lambda(\Phi_b)$, we have
\be\label{eq:MlamBar} M_\lambda(\Phi_b)&=&\lambda M_1\left(\lambda^{-2}\Phi_b\left(\lambda^{-1} r\right)\right),\\
E_\lambda(\Phi_b)&=&\lambda^3E_1\left(\lambda^{-2}\Phi_b\left(\lambda^{-1} r\right)\right),\ee
where the ULDM energy is 
\be \label{eq:Egenbar}E(\Phi_b)&=&\int
d^3x\left(\frac{\left|\nabla\psi\right|^2}{2m^2}
+\left(\frac{\Phi}{2}+\Phi_b(x)\right)\left|\psi\right|^2\right).\no\\&&\ee
Given the function $\Phi_b$, and plotting $M_\lambda(\Phi_b)$ and
$E_\lambda(\Phi_b)$ vs. $\lambda$, we can find the value of $\lambda$
and hence the profile for a soliton solution of any desired mass or
energy. 

\begin{figure}[hbp!]
\centering
\includegraphics[width=0.45\textwidth]{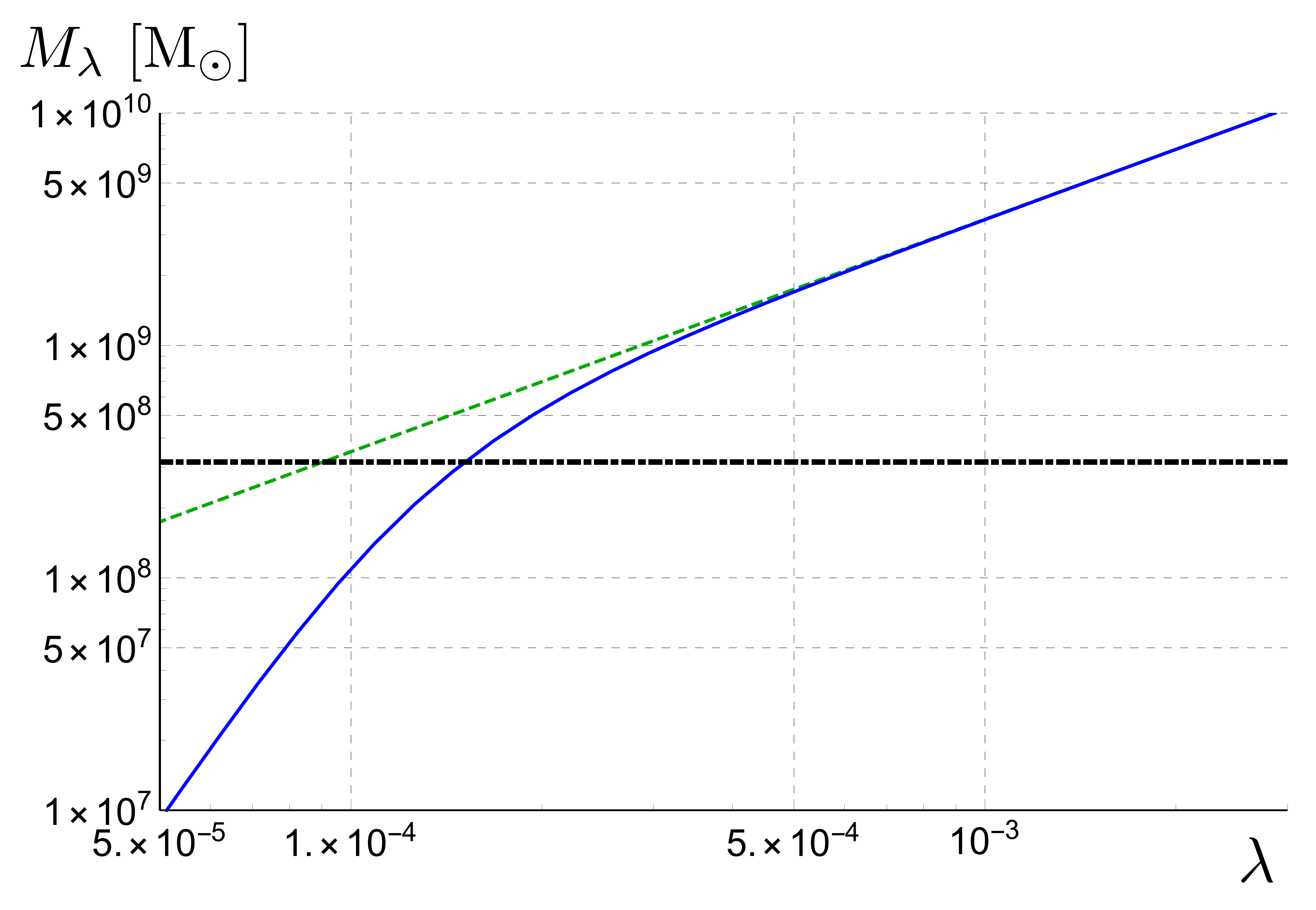}\\
\vspace{10pt}
\includegraphics[width=0.45\textwidth]{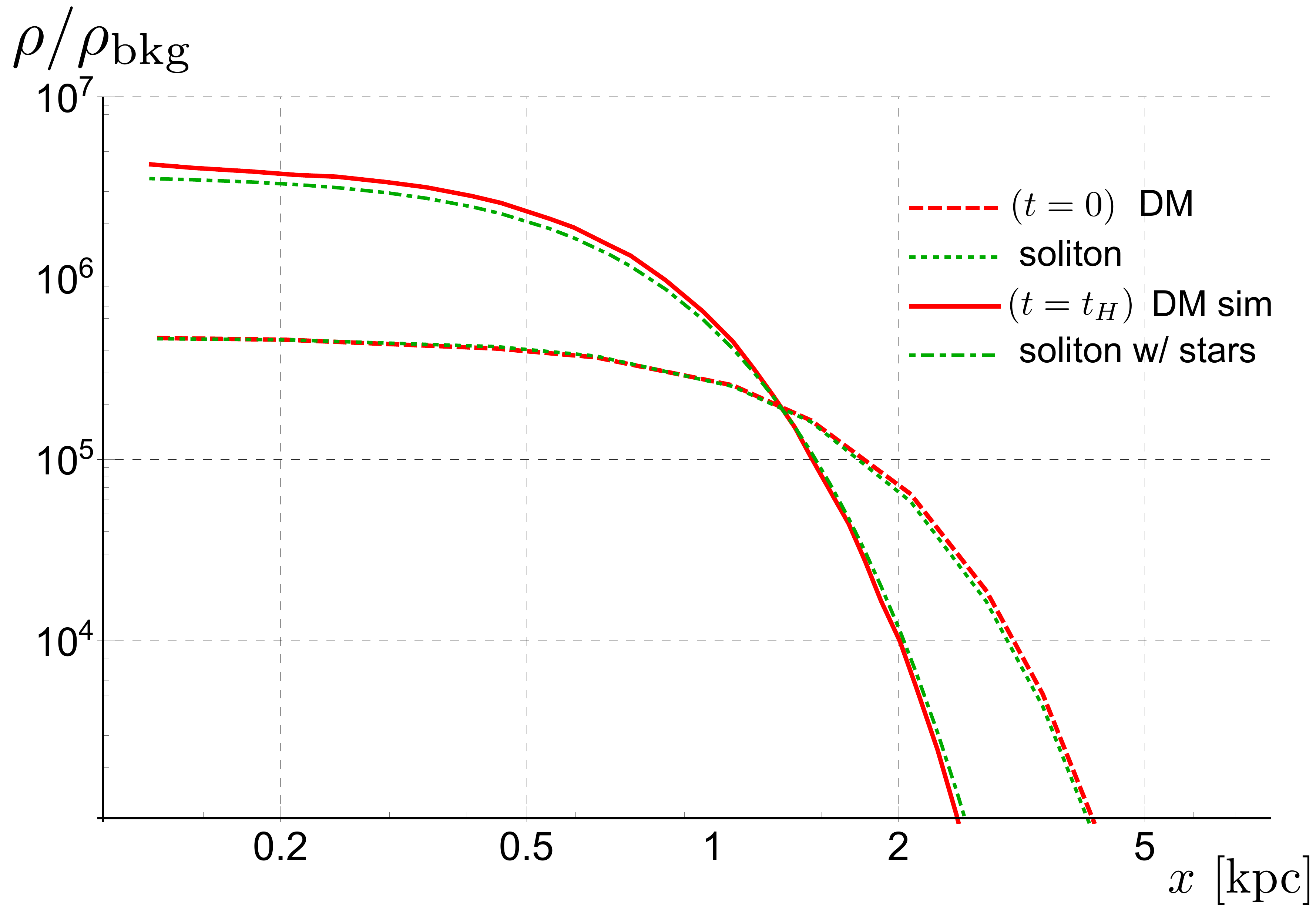}
\caption{
{\bf Top:} Soliton mass vs. $\lambda$ parameter, determined using
Eq.~(\ref{eq:MlamBar}) with 
the stellar gravitational potential as in the numerical simulation
shown in Fig.~1 of Ref.~\cite{2017arXiv171201947C}. 
Blue solid (green dashed) lines show the
$M_\lambda$ vs. $\lambda$ with (without) the stellar
potential. Black horizontal (dot-dashed) 
line denotes the mass $M$ chosen in the
simulation. {\bf Bottom:} Soliton profiles with and without stars,
describing the results of numerical simulation shown in Fig.~1 of
Ref.~\cite{2017arXiv171201947C} ($\rho_{\rm bkg}$ is the cosmological background
matter density).  
}\label{fig:DMstarsMlambda}
\end{figure}

Solutions of Eqs.~(\ref{eq:solbar1}-\ref{eq:solbar2}) can be compared
with the results of numerical simulations.  
Ref.~\cite{2017arXiv171201947C} added a distribution of mass in the form of point particles (``stars'') to ULDM simulations 
with $m=0.8\times10^{-22}$~eV. 
The first toy model they studied included an isolated soliton (without
an ULDM host halo) of mass $M=3.3\times10^8$~M$_\odot$, to which a
stellar distribution was added and evolved to a virialised state.  
From Fig.~1 of Ref.~\cite{2017arXiv171201947C} we derive the baryonic
contribution to the gravitational potential, and solve for the
distorted soliton at the stated ULDM mass. In the top panel of
Fig.~\ref{fig:DMstarsMlambda} we show $M_\lambda$ vs. $\lambda$,
derived from Eq.~(\ref{eq:MlamBar}). Blue solid (green dashed) lines
show $M_\lambda$ with (without) the stellar potential. Black
horizontal line denotes the value of $M$ that was fixed in the
simulation: the intersection of the black and the blue lines gives the
parameter $\lambda$ describing the distorted soliton. In the bottom
panel we show the density profiles. Red lines are taken
from~\cite{2017arXiv171201947C}, while green lines show the analytic
soliton solutions. Clearly, the numerical profile
from~\cite{2017arXiv171201947C} closely matches the
analytically-derived distorted soliton of Eq.~(\ref{eq:solbar1}).  

Ref.~\cite{2017arXiv171201947C} also simulated the
soliton in an ULDM halo, the output of a DM-only cosmological
simulation. Then, a stellar mass distribution was introduced as before
and the system allowed to evolve.  

\begin{figure}[hbp!]
\centering
\includegraphics[width=0.45\textwidth]{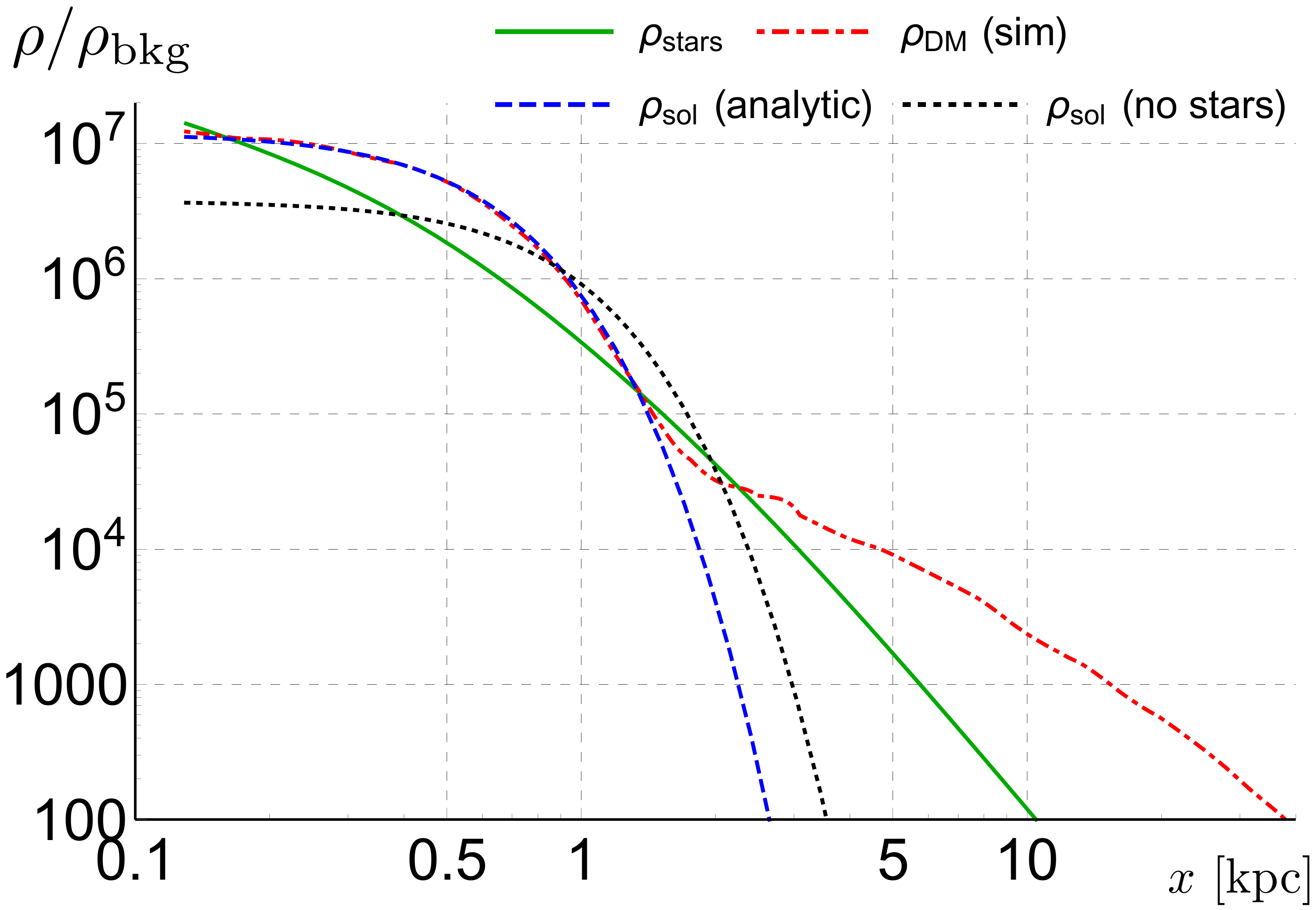}
\\
\vspace{10pt}
\includegraphics[width=0.45\textwidth]{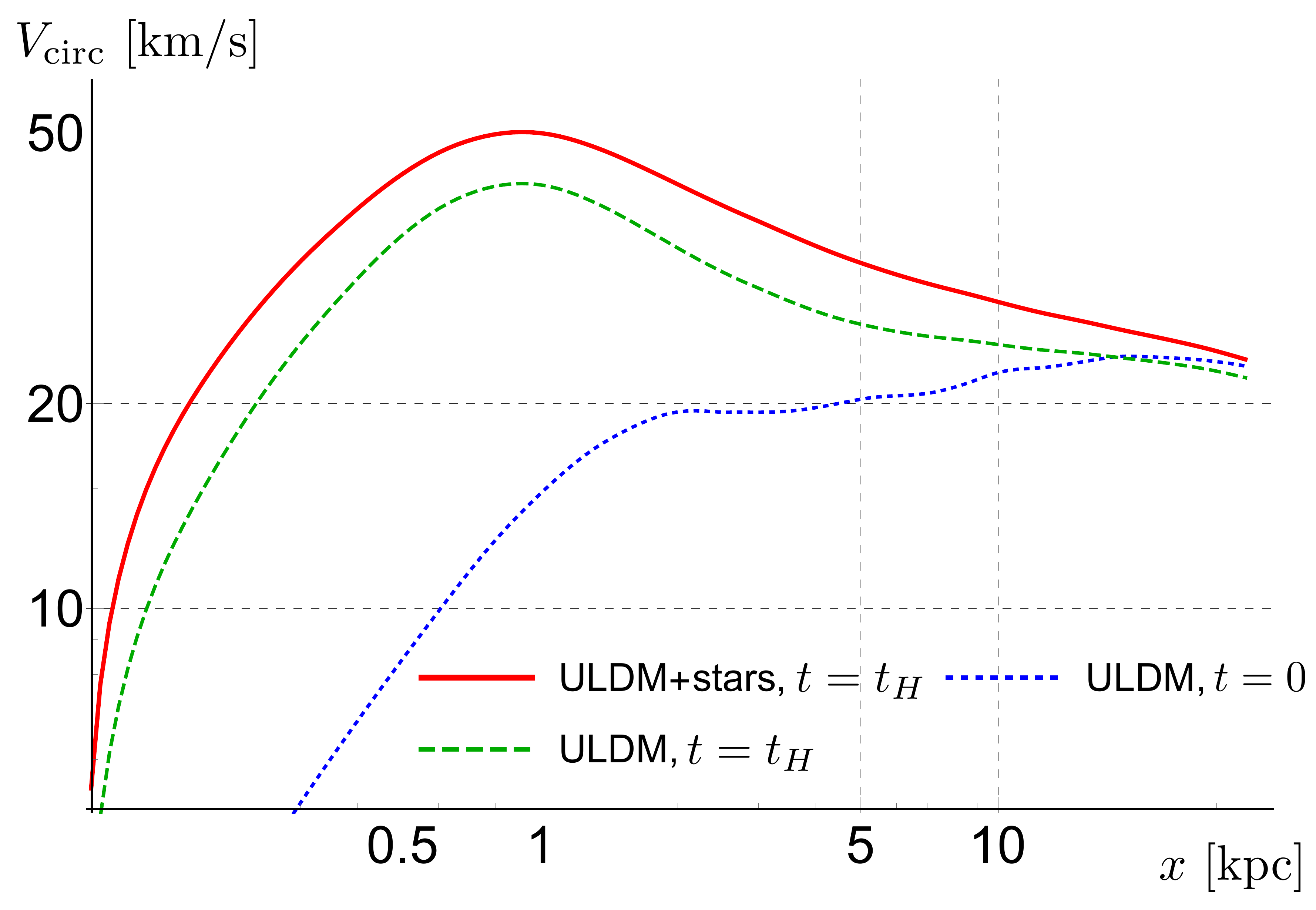}
\caption{
{\bf Top:} Soliton profiles with and without stars, describing the
inner part of the halo obtained in the numerical simulation (Case C
in Fig.~5 of 
Ref.~\cite{2017arXiv171201947C}). Green and red solid lines show the
stellar and ULDM densities, respectively, taken from the
simulation. Blue dashed show a distorted soliton solution, computed
including the stellar potential. Dotted black shows an un-distorted
soliton with the same mass. {\bf Bottom:} Rotation curve for the total
ULDM+stellar system (red solid) and including only the ULDM contribution (green dashed), at $t=t_H$. We also show the initial ULDM only
distribution at $t=0$ (blue dotted). 
}\label{fig:DMstarsCaseC}
\end{figure}

When embedded in a halo, the mass of the soliton is not constant
anymore but can grow by absorbing mass from the halo. However,
regardless of the large-scale halo, in the core region the perturbed
soliton profile is still fixed by the SP equations up to the ambiguity
of $\lambda$. We checked that in all of the virialised
soliton+halo+stellar mass simulations, presented in
Ref.~\cite{2017arXiv171201947C}, the soliton profile matches that of
the analytic distorted soliton, once accounting for the stellar
potential.  
In the top panel of Fig.~\ref{fig:DMstarsCaseC} we illustrate this
point for the $t=t_H$ snapshot of Case C, described in Fig.~5 of
Ref.~\cite{2017arXiv171201947C} 

With stars included, the total $t=t_H$ mass distribution, given by
summing the red (DM) and green (stars) solid lines in the upper panel
of Fig.~\ref{fig:DMstarsCaseC}, leads to the rotation curve shown by
the red solid line in the lower panel of the same figure. This
rotation curve is peaked at small radius due to the distorted soliton,
that is more massive than the soliton-halo prediction of the
DM-only simulations. The inner velocity exceeds the prediction of
Eq.~(\ref{eq:vv}) by a factor of two. We also show, by green dashed line, the rotation velocity at $t=t_H$ due to ULDM alone. To compare, the blue dotted
curve shows the rotation curve of the initial ULDM system, extracted
from the cosmological simulations. This is the same curve we showed in
Fig.~\ref{fig:Vcirc1406.6586}; as we discussed,
it satisfies Eq.~(\ref{eq:vv}) to $20\%$.

These numerical results suggest that the presence of baryonic matter
tends to make the soliton somewhat more compact and massive than
predicted by the pure ULDM soliton--host halo relation
(\ref{eq:Mc}). This, in turn, increases the peak in the rotation curve
due to the soliton, suggesting that our constraints
on ULDM obtained in Sec.~\ref{ss:compdat} are robust with respect to inclusion
of baryons. We now consider an example, where baryonic
effects are expected to be important: the Milky Way.

\subsection{The Milky Way: Nuclear Bulge vs. Soliton}\label{ss:MW}

Refs.~\cite{Schive:2014hza,Schive:2014dra} commented that for a
MW-like galaxy, Eq.~(\ref{eq:Mc}) predicts that ULDM with
$m\sim10^{-22}$~eV should produce a bump in the rotation curve at a
radial distance $\sim200$~pc, consistent at face value with a feature
observed in the MW.  Our results in
Sec.~\ref{ss:compdat} show that this value of $m$ is in tension with
observations of small disc galaxies. Nevertheless, the MW provides an
interesting example for studying the impact of  baryons and of a
SMBH on an ULDM soliton, and developing intuition as to what extent
this effect can be important. Our goal in this section is to provide a
preliminary study along these lines, using photometric baryonic mass
estimates. As an interesting outcome, we find that precision
kinematical studies of the MW inner bulge could in principle be
sensitive to ULDM with $10^{-21}~{\rm eV}\lesssim m \lesssim 10^{-19}$~eV, 
where the analysis along the lines of Sec.~\ref{ss:compdat}
may become challenging. 

\begin{figure}[hbp]
\centering
\includegraphics[width=0.5\textwidth]{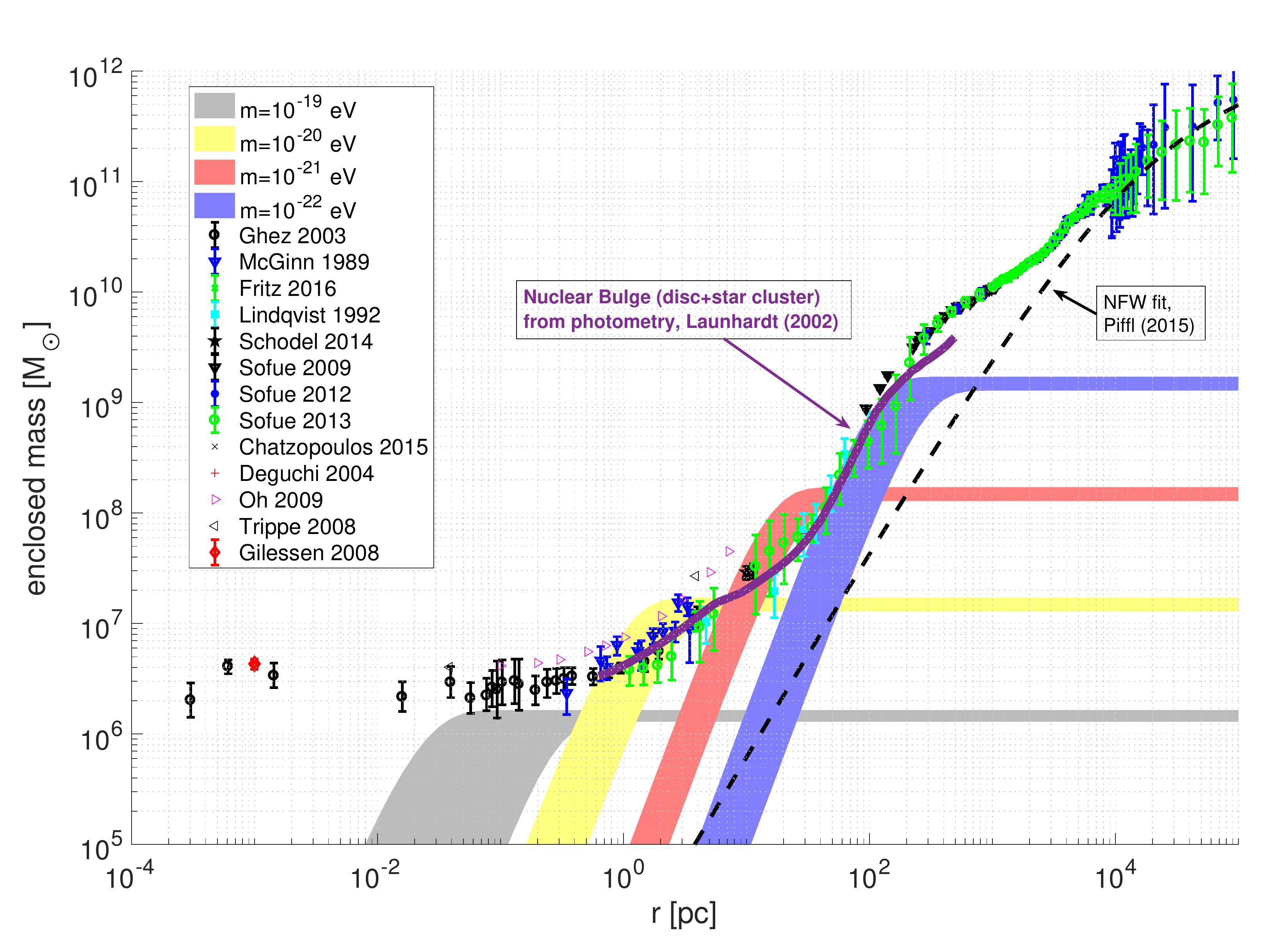}
\caption{
Spherically-averaged mass profile in the Milky Way, vs. ULDM soliton contributions. 
See text for details. 
}\label{fig:obs}
\end{figure}

Fig.~\ref{fig:obs} shows a spherically-averaged enclosed mass profile,
derived for the MW via various dynamical
tracers~\cite{Ghez2003,Feldmeier-Krause2017,McGinn1989,Lindqvist1992,Deguchi2004a,Trippe2008,Oh2009,Schoedel2014,Chatzopoulos2014,Fritz2016,Sofue2009,Sofue2012,Sofue2013,Chemin2015b}. 
Flattening of the enclosed mass at small radii reflects the
contribution of the SMBH. 
From the large radius ($r\gtrsim10$~kpc) part of the rotation curve, the parameters of the large-scale DM host halo can be extracted by standard analysis. We demonstrate this in Fig.~\ref{fig:obs} by a black dashed line, showing an NFW profile fitted in
Ref.~\cite{Piffl:2014mfa} to $r\gtrsim10$~kpc SDSS data. The large-scale analysis yields for the MW an host halo mass roughly in the range $M_h=(0.8\div 2)\times10^{12}$~M$_\odot$.

Given the host halo mass $M_h$, the soliton--host halo relation, Eq.~(\ref{eq:Mc}), predicts the soliton mass $M$. This, in turn, allows us to extract the soliton scaling parameter $\lambda$ (from Eq.~(\ref{eq:lamM}), with $M_\lambda=M$) and with it the entire predicted soliton contribution to the enclosed mass. In Fig.~\ref{fig:obs}, shaded bands show the resulting soliton-induced rotation curves for $m$ in the range
$(10^{-22}\div 10^{-19})$~eV. The width of the shaded bands comes from varying the host halo mass $M_h$ in the range $(0.8\div 2)\times10^{12}$~M$_\odot$. The approximate effects of baryon mass contributions, which cannot be neglected in the inner MW and which distort the soliton profile, are included and discussed below. 

We stress that the purpose of Fig.~\ref{fig:obs} is to illustrate the
possible signature of ULDM in the inner MW, and not for statistical
analyses of the MW mass distribution. Modeling the inner MW is a
complicated task. The measurement of inner kinematics of the galaxy, 
below a few
kpc, is subject to large systematic uncertainties due, among other
issues, to the effects of the Galactic bar and spiral arm
structures~\cite{Chemin2015b}, which impact tangent-point velocity
measurements like those utilised
in~\cite{Sofue:2013kja,Sofue:2015xpa}. Our simplified derivation of
the spherically-averaged mass profile in Fig.~\ref{fig:obs} combines
many tracers with different systematics, and accounts for none of
these subtleties. 

Ref.~\cite{Portail:2016vei} analysed the MW central gravitational
potential using a large set of observational constraints. In addition
to the classical bulge and disc, Ref.~\cite{Portail:2016vei} found
dynamical evidence for the presence of a mass component of
$\sim2\times10^9$~M$_\odot$ extending to $\sim250$~pc. This mass
component is visible as a mass bump in Fig.~\ref{fig:obs} (see,
e.g. green data points extracted
from~\cite{Sofue:2013kja}). Consistent with comments
in~\cite{Schive:2014hza,Schive:2014dra}, the bump is in tantalising
agreement with the soliton prediction of Eq.~(\ref{eq:Mc}) for
$m=10^{-22}$~eV (blue shaded band).  

Unfortunately, there are about a billion stars in there, too: the bump
in the mass profile at $r\sim200$~pc has been associated in the
literature with the nuclear bulge (NB). 
Ref.~\cite{Launhardt:2002tx}
fitted the NB mass and light by a dense disc of stars, with mass
density $\rho_*\sim200$~M$_\odot$/pc$^3$, scale height $\lesssim45$~pc
and
scale radius $\sim230$~pc. In all, the NB is thought to contain
$(1.4\pm0.6)\times10^9$~M$_\odot$ in stars, roughly enough to match
the dynamically inferred mass. Subsequent kinematic detection
supporting the stellar mass and disc-like morphology of this component
was given in~\cite{2041-8205-812-2-L21}. Microlensing
analyses~\cite{2017ApJ...843L...5W} lend further support to the
results of~\cite{Portail:2016vei,2041-8205-812-2-L21,Launhardt:2002tx}
down to $r\gtrsim220$~pc. 

The photometrically-derived NB mass model of~\cite{Launhardt:2002tx}
is superimposed as purple line in Fig.~\ref{fig:obs}.  We stress that
the photometric derivation is subject to large uncertainties due to
the need to correct for very strong extinction and due to unknown
stellar mass-to-light ratios. What we learn from this photometric mass
model, therefore, is that stars {\it could} plausibly account for all
of the kinematically inferred mass in this region.   

Assuming that the NB is due to stars, we now use a toy model of this
mass distribution to see its effect on an ULDM soliton. We replace
the disc-like morphology of the NB in~\cite{Launhardt:2002tx} by a
spherical model with the same radially averaged mass. The nominal
model, containing the NB and additional subleading components
described in~\cite{Launhardt:2002tx}, contains
$\sim1.7\times10^9$~M$_\odot$ in stars inside of $r=300$~pc. Adding a
SMBH of $M_{BH}=4.3\times10^6$~M$_\odot$~\cite{Ghez2003}, we calculate
soliton solutions in this baryonic potential.  

\begin{figure}[hbp]
\centering
\includegraphics[width=0.45\textwidth]{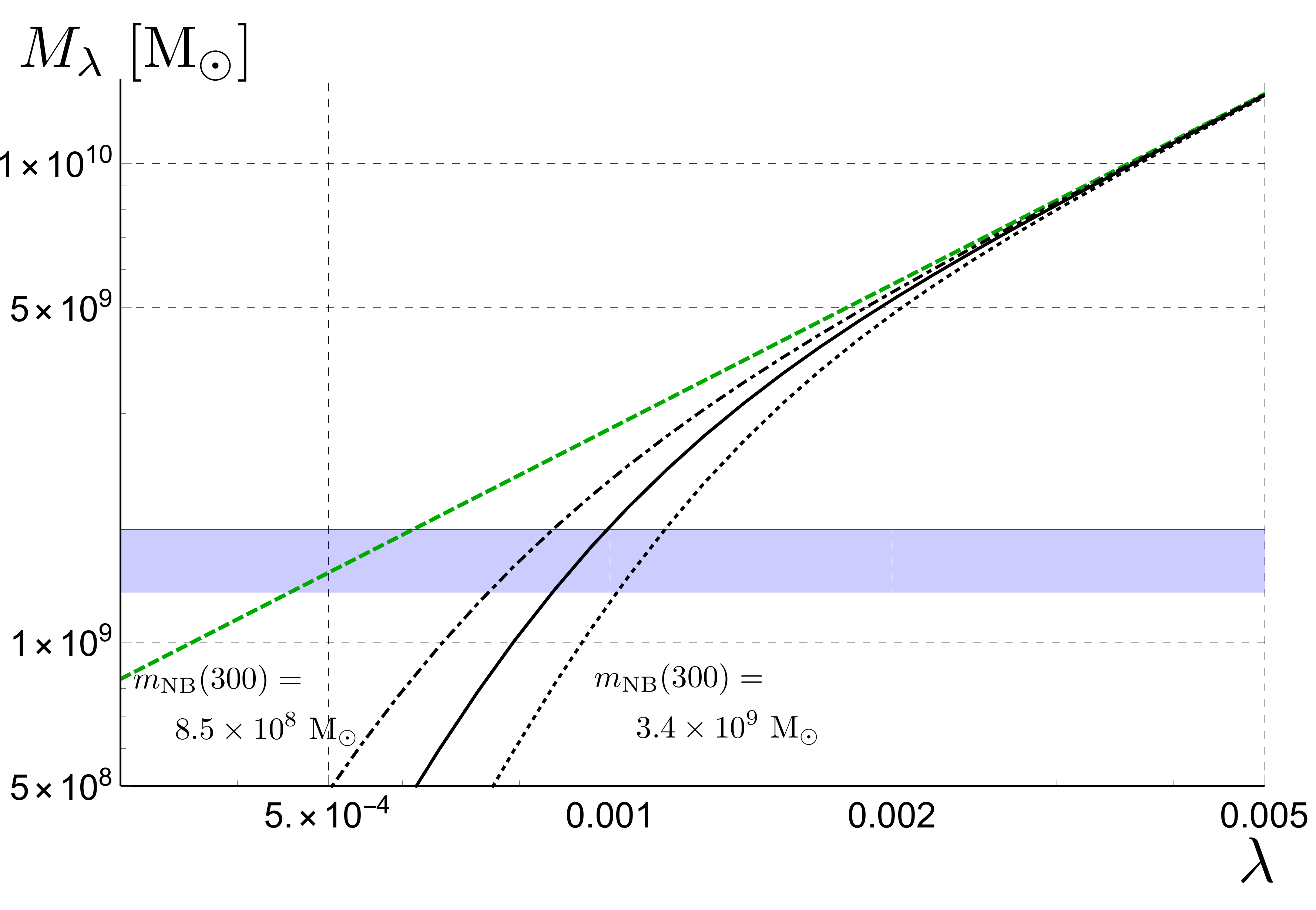}
\caption{
Soliton mass vs. $\lambda$ parameter, accounting for the
spherically-averaged gravitational potential due to
stars~\cite{Launhardt:2002tx}. The ULDM particle mass is
$m=10^{-22}$~eV. See text for more details. 
}\label{fig:MWMvslam}
\end{figure}

Fig.~\ref{fig:MWMvslam} shows the soliton mass as function of the
$\lambda$ parameter, for $m=10^{-22}$~eV. Green dashed line shows the
unperturbed $M_\lambda$ vs. $\lambda$ relation. Solid, dashed, and dotted
black lines show the relation for the nominal NB model and for two
other models, obtained by scaling the NB mass density by an over-all
factor of 0.5 and 2, respectively. For orientation, shaded blue band
shows the soliton mass predicted by Eq.~(\ref{eq:Mc}) for a host halo
with mass $M_h=(0.8-2)\times10^{12}$~M$_\odot$. 

For $M_\lambda\gtrsim3\,M_{NB}\sim5\times10^9$~M$_\odot$, the NB makes
a negligible impact on the soliton. 
For larger ratio of the stellar to ULDM mass, $M_\lambda\lesssim
M_{NB}$, the NB becomes important, contracting the soliton
profile. For the MW, this is the parametric region predicted by
Eq.~(\ref{eq:Mc}), implying that the solitons would receive
significant distortion. In Fig.~\ref{fig:obs} we illustrated this
effect by presenting, in shaded bands, the soliton mass profiles
computed accounting for the nominal NB model. 
We observe that the solitons for $m$ in the range
$(10^{-22}\div 10^{-19})$~eV 
are expected
to affect the potential at an order unity level.
Thus, a dedicated analysis using our formalism to calculate
steady state soliton profiles 
consistent with a given baryonic mass model could be sensitive to ULDM
solitons all the way up to $m\sim 10^{-19}$~eV.
For larger $m$, the expected soliton becomes
subdominant with respect to the SMBH. 

Another limitation comes from
absorption of the ULDM from the soliton by SMBH.
As discusses in App.~\ref{ss:absorption}, the accretion rate is
negligible as long as $m\lesssim 5\times 10^{-20}$~eV. However, for
higher ULDM masses the time scale of accretion can become shorter than the
age of the universe and our steady-state Newtonian analysis in
this section may not apply.

\section{Discussion}\label{sec:discussion}

\subsection{Comparison to previous work}\label{ss:prev}
An earlier analysis of ULDM halos including the effect of baryons was
done in~\cite{Arbey:2001qi} (see also~\cite{Lesgourgues:2002hk}). This
work attempted to fit the rotation curves of spiral galaxies
from~\cite{Persic:1995ru}, assuming that the entire ULDM halo is
contained in a giant soliton (see
also~\cite{Fernandez-Hernandez:2017rds} for a similar approach in
modeling low surface brightness galaxies). This exercise
led~\cite{Arbey:2001qi} to report $m<10^{-23}$~eV, in tension with
Ly-$\alpha$ data. This exercise is different from the current work. As suggested by numerical simulations, we
expect heavier ULDM particles with $m\gtrsim10^{-22}$~eV to produce
galaxies with a large scale host halo following roughly the usual NFW
profile, in which the soliton affects the rotation curve only in the
inner part of the halo. 

Refs.~\cite{Schive:2014dra,Marsh:2015wka} performed a Jeans analysis,
fitting a soliton+halo configuration to a small sample of
dispersion-dominated MW-satellite dwarf spheroidal galaxies
(dSph). They found that the modelling of stellar kinematics in the
dSph is consistent with a cored inner region, as expected in ULDM,
with good fits to the data for $m\lesssim10^{-22}$~eV. The fitted core
found in~\cite{Schive:2014dra} was consistent with expectations from
the soliton--host halo relation. In~\cite{Marsh:2015wka}, the
transition between an assumed host halo NFW profile and the soliton
was modelled using corresponding free parameters. To conclude, these
analyses considered a small sample of dispersion-dominated galaxies,
as opposed to our larger sample of rotation-dominated galaxies. Their
best-fit ULDM mass $m\lesssim 10^{-22}$~eV is disfavoured by our findings.

Ref.~\cite{Bernal:2017oih} fitted ULDM soliton+halo profiles
to SPARC~\cite{Lelli:2016zqa} galaxies. Without a soliton--host
halo relation, Ref.~\cite{Bernal:2017oih} assigned separate
free parameters to the NFW halo and to the soliton, for each galaxy. As a result of
this freedom, it is difficult to understand from that analysis if ULDM
is ruled out or not in the mass range
$m=(10^{-22}\div 10^{-21})$~eV. Nevertheless, Ref.~\cite{Bernal:2017oih} did
state a best fit of $m\approx0.5\times10^{-23}$~eV, and noted the
tension with Ly-$\alpha$ data, in agreement with our findings.  
Our work here shows that the soliton--host halo relation reduces the modelling freedom, and leads to disagreement with rotation curve data.   

Recently, Ref.~\cite{Deng:2018jjz} addressed the problem from a different
angle. It builds on the results of Ref.~\cite{Rodrigues:2017vto} that
fitted rotations curves of a sample of galaxies assuming that their
halos are described by the Burkert profile. Ref.~\cite{Deng:2018jjz}
points out a correlation between the parameters of the fitted 
profiles --- the central dark matter density $\rho_c$ and the size of
the core $R_c$,
\be
\rho_c\propto R_c^{-\beta}
\ee
with $\beta \sim 1$. This is different from $\beta=4$ predicted by
ULDM if the cores are identified with the solitons, see Eq.~(\ref{eq:rhosol}). Thus, Ref.~\cite{Deng:2018jjz}
concludes that ULDM cannot explain the origin of the halo cores, and
cannot solve the core--cusp problem of $\Lambda$CDM. Our results are
consistent with these findings and strengthen them by showing that the
soliton--host halo relation implies a tension between
predictions of ULDM and the data for $m\sim(10^{-22}\div
10^{-21})$~eV, disfavouring this mass range altogether.  

\subsection{Caveats}

\subsubsection{Soliton formation time}
Our analysis has been based on the assumption that ULDM solitons
satisfying the soliton--host halo relation exist in the centres of all
(or most)
galactic halos. Formation of such solitons within the lifetime of the
universe has been demonstrated in numerical simulations~\cite{Schive:2014dra,Veltmaat:2018dfz} for ULDM masses $m\sim
10^{-22}$~eV and halos with mass $(10^9\div
10^{11})\,M_\odot$. Here we discuss whether such solitons have enough
time to form for higher values of $m$ and larger halos.

Recently, Ref.~\cite{Levkov:2018kau} considered the scenario, where
solitons form as a result of gravitational relaxation of the inner
part of the halo. In this case the relaxation time can be estimated
as,
\be
\label{eq:trelax}
\tau_{\rm relax}=\frac{b\sqrt{2}\,m^3 v^6}{12\pi^3G^2\rho^2\Lambda}\;,
\ee 
where $v$ and $\rho$ are the mean velocity and density in the
inner part of the halo, $\Lambda=\log(mv R)$ is the Coulomb logarithm depending
on the size $R$ of the relaxation region, and $b$ is a numerical
coefficient close to one. The formula has been confirmed in
\cite{Levkov:2018kau} by numerical simulations that found $b$ to range
between $0.7\div 0.9$ depending on the precise form of the initial
conditions. Substituting into Eq.~(\ref{eq:trelax}) the values for a
typical halo and taking $\Lambda\sim 4$, we obtain,
\be
\label{eq:trelaxnum}
\begin{split}
\tau_{\rm relax}\sim\, &7\times 10^8
\left(\frac{m}{10^{-22}{\rm eV}}\right)^3\\
&\times\left(\frac{v}{100\,{\rm km}/{\rm s}}\right)^6
\left(\frac{\rho}{0.1 M_\odot/{{\rm pc}^3}}\right)^{-2}
{\rm yr}\;.
\end{split}
\ee
We observe that the relaxation time has strong dependence on the
velocity and the density in the central region of the halo, that have
large uncertainties. Fixing the fiducial values of $v=100\,{\rm
  km}/{\rm s}$ and $\rho=0.1M_\odot/{\rm pc}^3$ one would conclude
that the relaxation takes longer than the age of the universe if
$m\gtrsim 3\times 10^{-22}$~eV.

However, it can be incorrect to use Eq.~(\ref{eq:trelax})
for the estimate of the soliton formation 
time inside ULDM halos in the cosmological
context. Indeed, this equation has been derived and tested in the
kinetic regime corresponding to large velocity 
dispersion of ULDM particles, such 
that their de Broglie wavelength is much
shorter than the size of the system. 
Starting from initial conditions of a gas of wavepackets with large
velocity dispersion in a box, 
Ref.~\cite{Levkov:2018kau}
observed a fast formation of a virialised halo, followed by 
a long period of kinetic evolution populating the low-lying
energy levels, until the system was able to form the soliton through
Bose--Einstein condensation. The time scale 
(\ref{eq:trelax}) refers to the long second stage of the process.
On the other hand, the cosmological initial conditions
prior to virialisation are characterised by very low velocities, 
where the kinetic description
does not apply. In this situation a long relaxation seems unnecessary and
the formation of the soliton can
proceed much faster, on the scale of the halo free-fall time, see
discussion in \cite{Levkov:2018kau}.
This is supported by the simulations with cosmological
initial conditions which appear to imply formation of the soliton in every halo
at the moment of virialisation
\cite{Schive:2014dra,Veltmaat:2018dfz}. We believe that the issue of
the soliton formation time requires further investigation, including 
possible effects of baryonic matter.

\subsubsection{Non-gravitational interactions}

In our analysis we have neglected any non-gravitational
interactions of the ULDM particles. Let us discuss under which
conditions this approximation is justified. As an example we consider
quartic self-interaction of the scalar field $\phi$ with coupling
constant $\kappa$,
\be
\label{eq:selfint}
\delta V(\phi)=\frac{\kappa\phi^4}{4}\;.
\ee
The self-interaction is negligible as long as the corresponding 
potential energy is much smaller than the 
gradient energy of the field,
\be
\frac{|\kappa|\phi^4}{4}\ll\frac{(\nabla\phi)^2}{2}\;.
\ee 
For a soliton with the scale parameter $\lambda$, this translates into a bound,
\be
|\kappa|<\frac{2m^2}{x_{c\lambda}^2\rho_{c\lambda}}\;.
\ee
Using the soliton--host halo relation in the form of Eqs.~(\ref{eq:rc}),
(\ref{eq:rhosol}) we obtain the condition,
\be
\label{eq:kappabound}
|\kappa|<4\times 10^{-93}\left(\frac{m}{10^{-22}{\rm eV}}\right)^2
\left(\frac{M_h}{10^{12}M_\odot}\right)^{-\frac{2}{3}}.
\ee 
If this condition is satisfied, our analysis is applicable. Otherwise,
the effects of the self-interaction on the soliton properties must be
taken into account.

While the condition (\ref{eq:kappabound}) appears stringent, it is
naturally fulfilled in a broad class of theories containing 
axion-like particles. These typically have a periodic potential of the form,
\be
\label{eq:Vcos}
V(\phi)=m^2f^2\big(1-\cos(\phi/f)\big)\;.
\ee
For the masses of interest $m\sim (10^{-22}\div 10^{-18})$~eV, the
axion periodicity should be in the range
$f\sim (10^{16}\div 10^{17})$~GeV to produce the right dark matter
abundance via the misalignment mechanism (see e.g.
\cite{Hui:2016ltb}). Expanding the potential (\ref{eq:Vcos}) in powers
of $\phi$ we obtain the value of the coupling,
\be
\begin{split}
\kappa=&-\frac{m^2}{6f^2}\\
\approx&\,-1.7\times 10^{-97}\left(\frac{m}{10^{-22}{\rm eV}}\right)^2
\left(\frac{f}{10^{17}{\rm GeV}}\right)^{-2}.
\end{split}
\ee
We see that (\ref{eq:kappabound}) is indeed satisfied for all galactic
halos considered in this paper, independently of the ULDM mass.

Similar analysis can be performed in the case of non-gravitational
interactions between ULDM and baryonic matter, though in this case it
will be strongly model-dependent.

\section{Summary}\label{s:sum}

Bosonic ultra-light dark matter (ULDM) is an interesting paradigm for
dark matter. For particle mass $m\sim10^{-22}$~eV, ULDM would form
solitonic cores in the center of galaxies, and it has been suggested
that this could solve several puzzles of $\Lambda$CDM on small scales.  

We analysed the results of numerical simulations of ULDM, which have
found 
scaling relations between the mass of the central soliton and the mass or
energy of the host halo. Simulations by different groups converge on a
soliton profile in good agreement with the analytical solution of the
Schroedinger-Poisson (SP) equation, admitting important analytic insight
into the numerical results. 

We showed that the simulations of Ref.~\cite{Mocz:2017wlg} contain a
central soliton that dominates the total energy of the entire
soliton+halo system. This situation is unlikely to describe realistic
galactic halos\footnote{This does not invalidate the beautiful
  simulations of~\cite{Mocz:2017wlg}. We only think that their setup
  for initial conditions
  does not match real galaxies.} 
 with mass above $\sim5\times
10^7\left(m/10^{-22}{\rm eV}\right)^{-3/2}
{\rm M}_\odot$.

We have demonstrated 
that the soliton--host halo relation found in the simulations of
Refs.~\cite{Schive:2014hza,Schive:2014dra} can be summarised by the
statement, that $(E/M)|_{\rm soliton}=(E/M)|_{\rm halo}$. The
simulations of~\cite{Schive:2014hza} show a small spread, less than a
factor of two, around this relation. This $E/M$ relation could apply
to real galaxies.  

The $E/M$ soliton--host halo relation implies that the peak circular
velocity of the large-scale halo should approximately reproduce itself
in the inner soliton region. This can be tested against observations
without free parameters. Contrasting this prediction with high
resolution disc galaxy data, we showed that ULDM in the mass range
$m\sim (10^{-22}\div 10^{-21})$~eV is disfavoured by observations. Given
that smaller $m$ is in tension with cosmological measurements, this disfavours ULDM
as a solution to the small scale puzzles of $\Lambda$CDM. 

We also analysed the effect 
of a fixed background distribution of baryons and of a super-massive
black hole (SMBH) on the soliton 
solution. We showed that this analysis would be important towards
using high-resolution kinematical data in the Milky Way or Milky
Way-like galaxies. Mapping the Milky Way gravitational potential down
to $r\sim 10^{-2}$~pc may allow to probe ULDM with
mass up to $m\sim10^{-19}$~eV.

It should be stressed that the $E/M$ soliton--host halo relation
(\ref{eq:E2MsolE2Mh}) is an empirical result, deduced from numerical
simulations without baryons and
tested in a limited range of ULDM and halo masses. The importance of
its phenomenological implications motivates further investigation
to better understand the physics underlying this relation 
and map out its domain of
validity. 

Throughout the paper we have assumed that ULDM comprises the total
amount of dark matter. We leave for future the study of scenarios
where only a fraction of dark matter is in the form of ULDM.

\acknowledgments
We thank James Hung-Hsu Chan, Josh Eby, Lam Hui, Dmitry Levkov, Moti Milgrom, Eran Ofek,
Alexander Panin, and Igor Tkachev for discussions, and Yossi Nir, Scott Tremaine, and Eli Waxman for comments on an early version of the draft. KB thanks the participants of the VBSM Polynesia workshop, where part of this work was conducted, for useful discussions. 
KB is incumbent of the Dewey David Stone
and Harry Levine career development chair. The work of KB and NB is
supported by grant 1937/12 from the I-CORE program of the Planning and
Budgeting Committee and the Israel Science Foundation and by grant
1507/16 from the Israel Science Foundation. SS is grateful to Columbia
University for hospitality during completion of this work. 

\begin{appendix}

\section{Soliton mass--host halo mass relation, $M_h$ vs. $M_{200}$}\label{app:MhM200}

Ref.~\cite{Schive:2014hza} defined the halo virial mass $M_h$ and radius $R_v$,
\be M_h&=&\frac{4\pi}{3}R_v^3\,\zeta(z)\,\rho_{m0},\ee
where $\rho_{m0}$ is the present matter density and
$\zeta(z)=(18\pi^2-82(1-\Omega_m(z))-39(1-\Omega_m(z))^2)/\Omega_m(z)$. 
The halo energy per unit mass was estimated as $E_h/M_h=-3GM_h/(10R_v)=-G\frac{3}{10}\left(\frac{4\pi}{3}\zeta\,\Omega_{m0}\right)^{\frac{1}{3}}\rho_c^{\frac{1}{3}}\,M_h^{\frac{2}{3}}$. 
In Sec.~\ref{s:bhconsp}, we prefer to work with $M_{200}$ and
$R_{200}$ and have calculated the halo energy per unit mass, using
Eq.~(\ref{eq:E2MNFW}), as $E_{200}/M_{200}=-G\frac{\tilde c}{4}\left(\frac{4\pi\delta_c}{\left(\ln(1+c)-\frac{c}{1+c}\right)^2}\right)^{\frac{1}{3}}\rho_c^{\frac{1}{3}}\,M_{200}^{\frac{2}{3}}$. 

To express the soliton mass--host halo mass relation,
Eq.~(\ref{eq:Mhostan}), in terms of $M_h$ of
Ref.~\cite{Schive:2014hza}, we need to do so at fixed
$E_{200}/M_{200}=E_h/M_h\equiv E/M$. This gives: 
\be\label{eq:M200toMh}
\left(\frac{M_{200}}{M_h}\right)^{\frac{1}{3}}\bigg|_{E/M}&=&\left(\frac{\frac{3}{10}\left(\frac{4\pi}{3}\zeta\,\Omega_{m0}\right)^{\frac{1}{3}}}{\frac{\tilde c}{4}\left(\frac{4\pi\delta_c}{\left(\ln(1+c)-\frac{c}{1+c}\right)^2}\right)^{\frac{1}{3}}}\right)^{\frac{1}{2}}.
\ee 
For $c$ in the range $5\div 30$, the RHS of Eq.~(\ref{eq:M200toMh}) varies
between 0.58 and 0.47. Plugging this result into
Eq.~(\ref{eq:Mhostan}) shows that it agrees with Eq.~(\ref{eq:Mc})
of~\cite{Schive:2014hza} to 20\%.

\section{Adding a super-massive black hole}\label{ss:smbh}
Most galaxies, if not all, host a SMBH, and we should check how it
affects the ULDM soliton. 
This is an important point to verify and we do this analysis here in
some detail. 
The upshot of our discussion is that a SMBH would not affect our
  results for small disc galaxies, whereas in the MW it becomes relevant
  at the upper end of considered ULDM masses, $m\gtrsim 5\times
  10^{-20}$~eV.  

We start with a Newtonian analysis, dealing with the ULDM configuration
far from the BH Schwarzschild radius. Then we will consider the
limitations of this analysis imposed by absorption of ULDM on SMBH
that leads to the decay of the soliton+SMBH configuration.

\subsection{Soliton shape in the presence of SMBH }\label{ss:Newtonian}

Adding a SMBH, coincident with the soliton center of mass, changes Eqs.~(\ref{eq:SPselfX}-\ref{eq:SPselfV}) into
\be\partial_r^2\left(r\chi\right)&=&2r\left(\Phi-\frac{A}{r}-\gamma\right)\chi,\label{eq:SPselfXbh}\\
\partial_r^2\left(r\Phi\right)&=&r\chi^2\label{eq:SPselfVbh},\ee
where
\be A&=&GM_{BH}m\\
&\approx&3\times10^{-6}\left(\frac{M_{BH}}{4\times10^6~\rm M_\odot}\right)\left(\frac{m}{10^{-22}~\rm eV}\right).\no\ee
We chose the reference value for $M_{BH}$ to represent the case of the MW.

It remains convenient to solve the problem using boundary conditions
with $\chi(0)=1$.  
Let us denote this solution (satisfying $\chi(0)=1$) by $\chi_1(r;A)$, accompanied by the potential $\Phi_1(r;A)$. Having found $\chi_1(r;A)$ and $\Phi_1(r;A)$ for any value of $A$, the physical solution $\chi_\lambda(r;A)$ satisfying Eqs.~(\ref{eq:SPselfXbh}-\ref{eq:SPselfVbh}) with boundary condition $\chi_\lambda(0;A)=\lambda^2$ is given by 
\be
\chi_\lambda(r;A)&=&\lambda^2\chi_1(\lambda r;A/\lambda), 
\label{eq:solbh1}\\
\Phi_\lambda(r;A)&=&\lambda^2\Phi_1(\lambda r;A/\lambda).
\label{eq:solbh2}
\ee
Defining $M_\lambda(A)$, with obvious notation, the soliton mass is
\be M_\lambda(A)&=&\lambda M_1(A/\lambda).\ee

\begin{figure}[hbp!]
\centering
\includegraphics[width=0.45\textwidth]{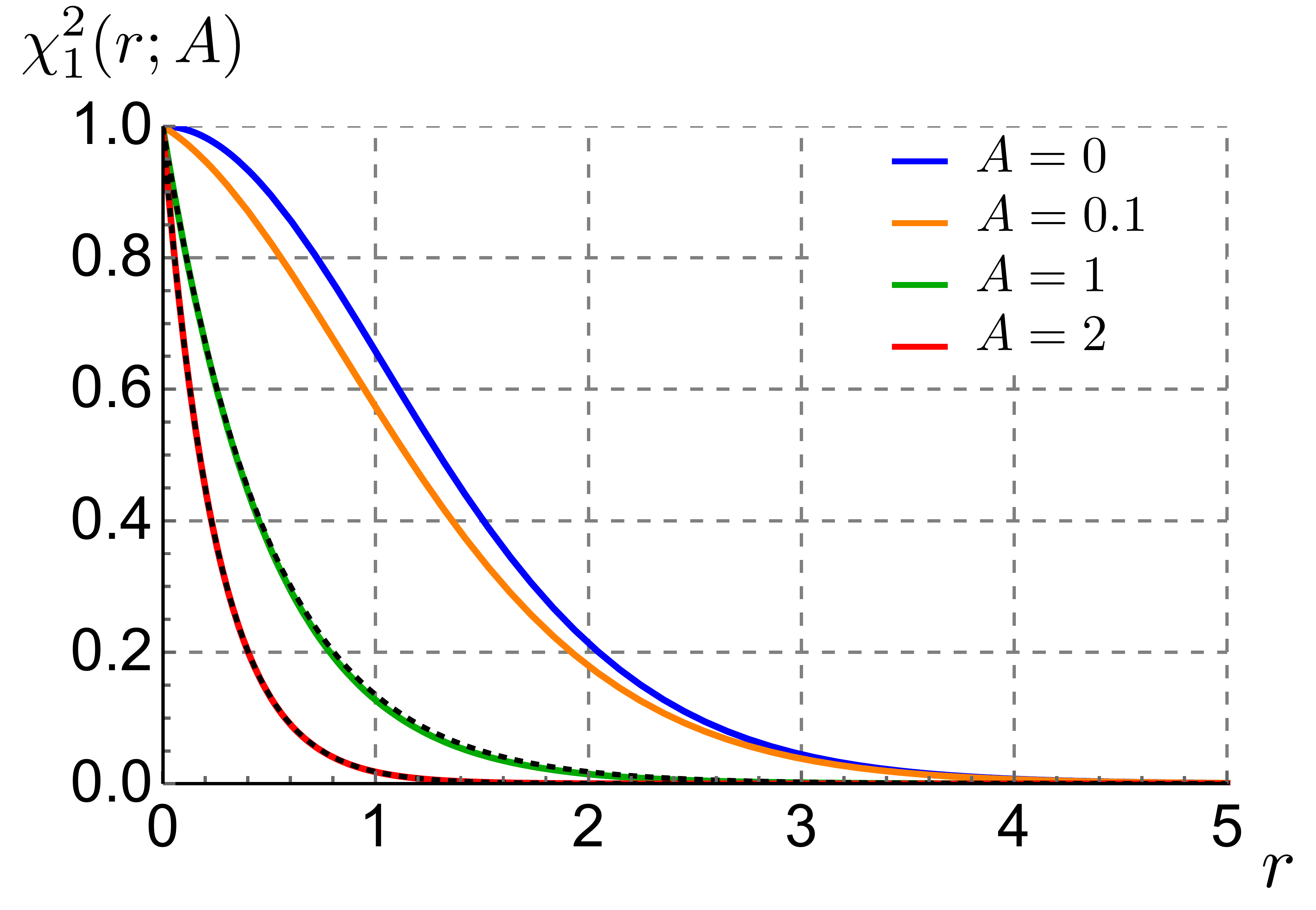}
\caption{
$\chi_1^2(A)$ profiles, with $A=0,0.1,1,2$ shown in blue, orange, green and red, respectively. For the $A=1$ and $A=2$ profiles, the large-$A$ analytic solution is shown in dotted black. 
}\label{fig:solSMBH}
\end{figure}

\begin{figure}[hbp!]
\centering
\includegraphics[width=0.4\textwidth]{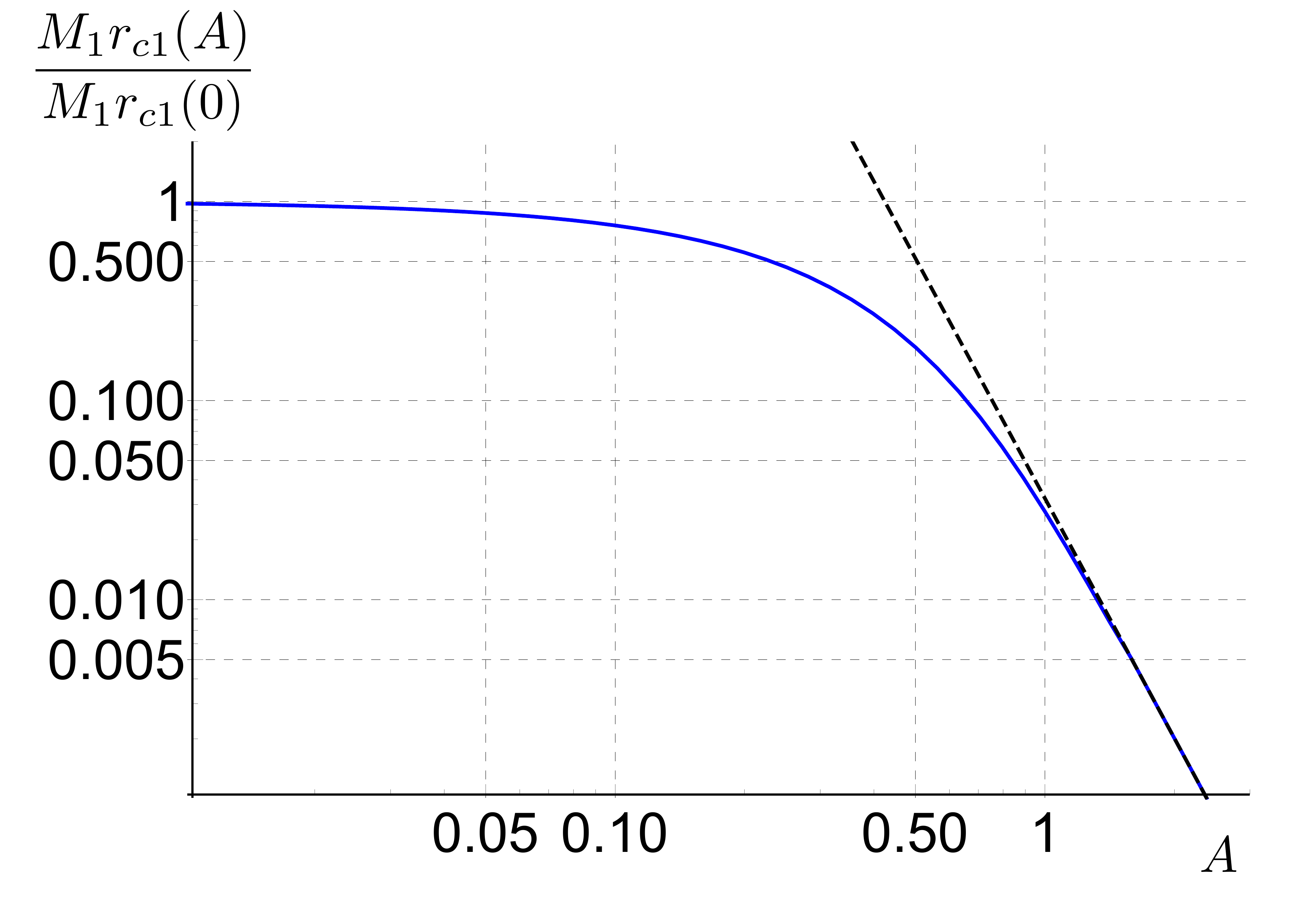}
\includegraphics[width=0.4\textwidth]{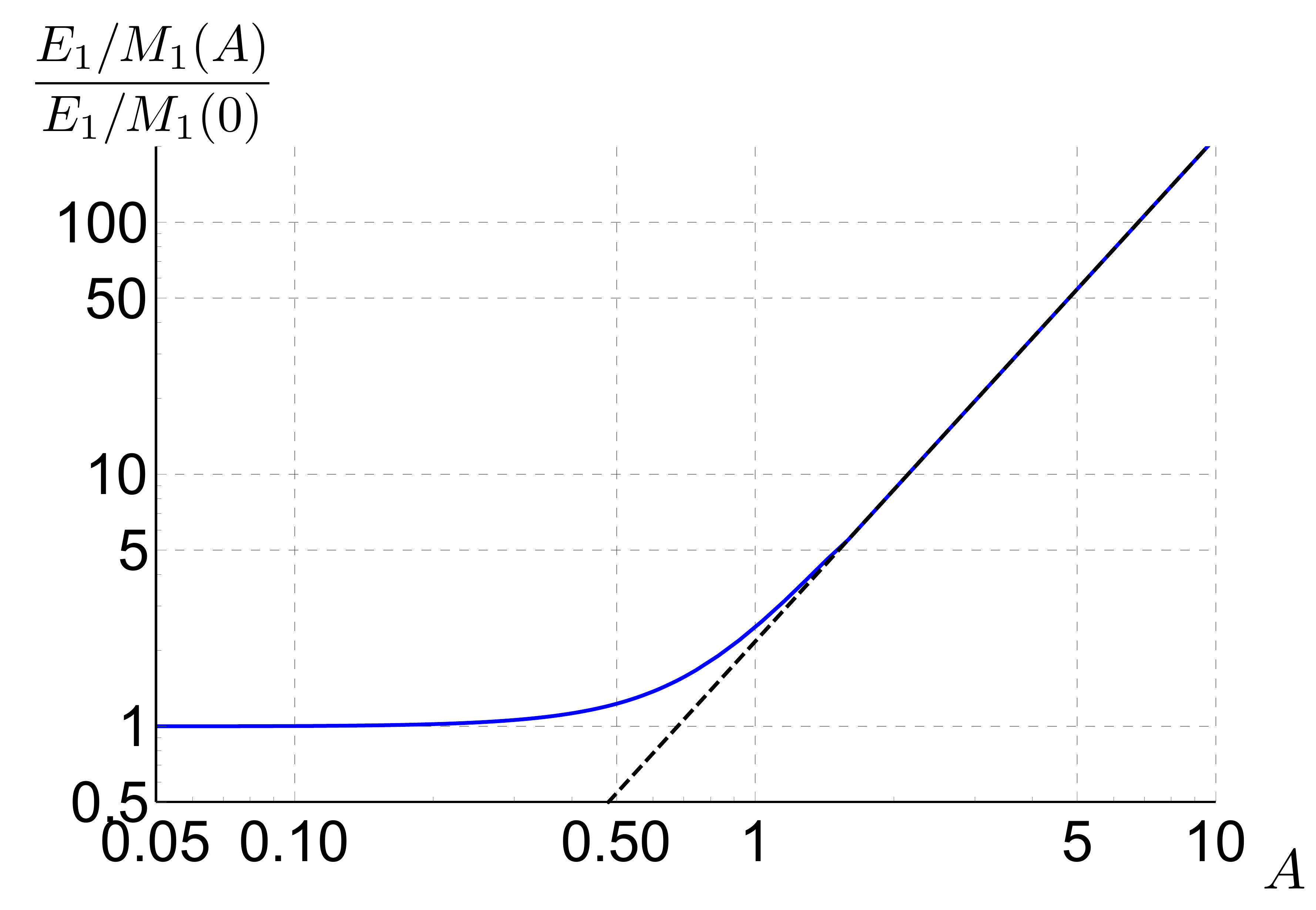}
\caption{
Mass-core radius relation ({\bf top panel}) and $E/M$ ({\bf bottom panel}) for the $\chi_1(A)$ soliton, as function of $A$. The large-$A$ analytic solution is shown in dashed black.}\label{fig:Mrc}
\end{figure}

Fig.~\ref{fig:solSMBH} shows the density profile $\chi^2_1(r;A)$ for
different values of $A$. For $A\lesssim10^{-2}$ the solution converges
to the unperturbed $\chi_1$. For $A\gtrsim1$ the solution approaches
$\chi_1(r;A\to\infty)\to e^{-Ar}$ which is nothing, but the
wavefunction of the ground state in the Coulomb potential. 
This means that for
$A/\lambda\gtrsim1$, solitons with different values of $\lambda$ have
the same profile up to over-all normalization:
$\chi_\lambda(r;A\gtrsim\lambda)\to\lambda^2e^{-Ar}$. In this limit
the soliton core radius is $r_{c\lambda}(A)\to\ln(2)/(2A)$, and the
mass is
$M_\lambda(A)\to\frac{\lambda^4}{4A^3}\frac{M_{pl}^2}{m}$. More
generally, for any $A$ the mass-radius relation is 
\be M_\lambda(A)\,r_{c\lambda}(A)&=&M_1(A/\lambda)\,r_{c1}(A/\lambda).\ee
The mass-radius product is shown by the solid blue curve in
Fig.~\ref{fig:Mrc} (top panel). The large-$A$ analytic solution,
$M_1(A)\,r_{c1}(A)\huge|_{A\to\infty}\to
\frac{\ln(2)}{8A^4}\frac{M_{pl}^2}{m}$ (recall $r=m\,x$), is
shown in dotted black, normalised to the $A=0$ result. 

As discussed in Sec.~\ref{ss:Schive}, Eqs.~(\ref{eq:Mc})
and~(\ref{eq:E2MsolE2Mh}) are equivalent when the ULDM soliton is
self-gravitating, but this equivalence is broken by the SMBH
gravitational potential. It is unlikely that Eq.~(\ref{eq:Mc}) remains
realistic when the SMBH becomes important. Instead,
Eq.~(\ref{eq:E2MsolE2Mh}) may still be valid.  
In computing the ULDM energy we need to account for the SMBH contribution,
\be \label{eq:EgenBH}E(A)&=&\int d^3x
\left(\frac{\left|\nabla\psi\right|^2}{2m^2}+\left(\frac{\Phi}{2}
-\frac{A}{m\,x}\right)\left|\psi\right|^2\right),\no\\
&&\ee
which means that
\be E_\lambda(A)&=&\lambda^3E_1(A/\lambda).\ee
The large $A$ limit (attained for a $\chi_\lambda(r;A)$ soliton at
$A\gtrsim\lambda$) is
$E_\lambda(A\to\infty)\to-\frac{\lambda^4}{8A}\frac{M_{pl}^2}{m}$. In
this limit $E_\lambda/M_\lambda$ tends to a $\lambda$-independent
constant, $E_\lambda/M_\lambda(A\to\infty)\to-A^2/2$, because the
soliton is not held together by self gravity, but by the SMBH
potential.  
The $E/M$ relation for the $\chi_1(A)$ soliton is shown in the bottom
panel of Fig.~\ref{fig:Mrc}.  

\begin{figure}[hbp!]
\centering
\includegraphics[width=0.45\textwidth]{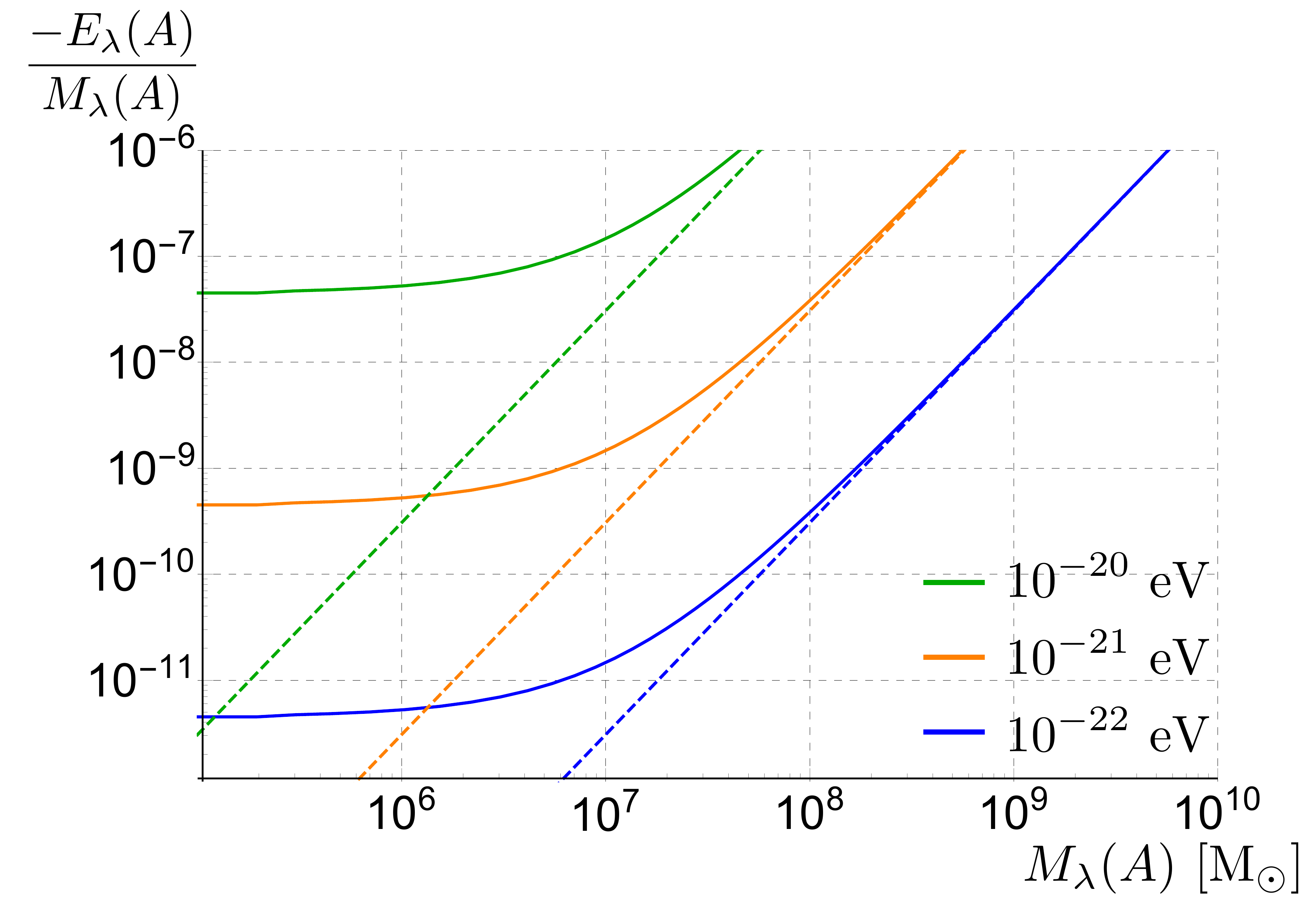}
\caption{
Solid lines: energy per unit mass vs. mass for a soliton, including the effect of a SMBH with $M_{BH}=4\times10^6$~M$_\odot$. Dashed lines: switching off the SMBH.
}\label{fig:E2MvsM}
\end{figure}

In Fig.~\ref{fig:E2MvsM}, solid lines show $E/M$ for the soliton with
different values of particle mass $m$, in the presence of a SMBH with
$M_{BH}=4\times10^6$~M$_\odot$. The dashed lines show the undistorted
result, obtained from Eq.~(\ref{eq:MlamSchieve}).  
The large $A$ limit with constant $E_\lambda/M_\lambda$ is attained
when the soliton mass becomes smaller than the SMBH mass. 

To further
clarify this
point we relate $A/\lambda$ to the
unperturbed soliton mass using Eq.~(\ref{eq:Mlambda}),  
\be \label{eq:effA}A/\lambda&\approx&2.06\left(\frac{M_{BH}}{M_\lambda(A=0)}
\right)
\\
&\approx&0.8\left(\frac{M_{BH}}{4\times10^6~\rm M_\odot}\right)
\left(\frac{M_\lambda(A=0)}{10^{7}~\rm M_\odot}\right)^{-1}.\no\ee
Next, we see from Eqs.~(\ref{eq:solbh1}),
(\ref{eq:solbh2}) that $A/\lambda$ gives an effective $A$-parameter
that must be substituted in the $\chi_1$-soliton to obtain the actual
solution. Thus, whether the effects of SMBH are important or not can be
read from the ratio of the SMBH mass to the mass of the undistorted
soliton. 

\begin{figure}[htbp!]
\centering
\includegraphics[width=0.45\textwidth]{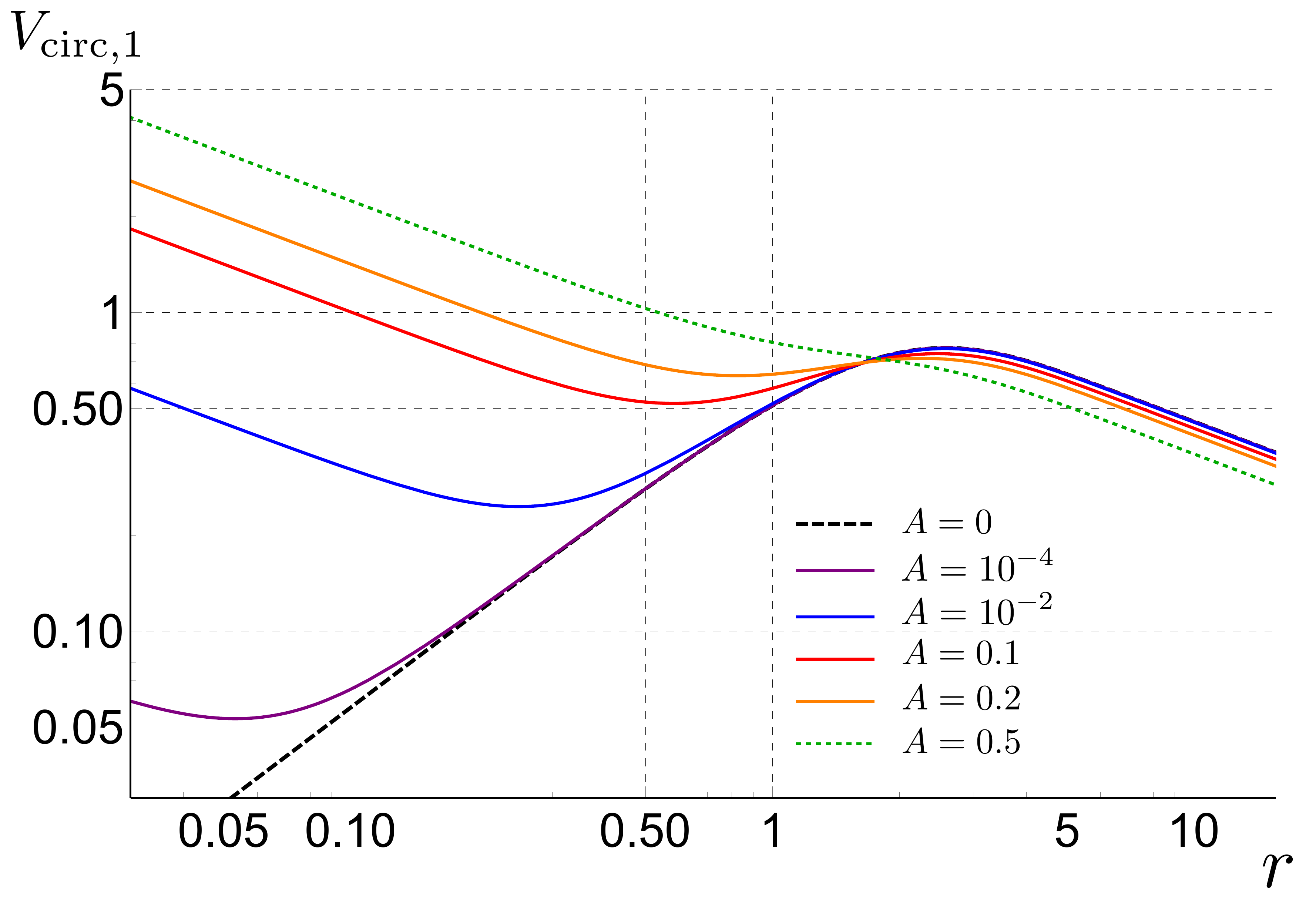}
\caption{
Rotation curve for the $\chi_1(A)$ soliton.
}\label{fig:VcircBH}
\end{figure}

To analyse the rotation curve for the soliton+SMBH system, it is useful to view $V_{\rm circ}$ for the $\chi_1(A)$ soliton, obtained for different values of $A$. This is shown in Fig.~\ref{fig:VcircBH}. Because  $\chi_\lambda(A)$ solitons are a scale transformation of $\chi_1(A/\lambda)$, plotting $\chi_1(A)$ with different values of $A$ is equivalent to plotting the rescaled velocity curve of solitons at different values of $M_{BH}/M_\lambda$. For example, using Eq.~(\ref{eq:effA}), the $A=0.2$ curve in Fig.~\ref{fig:VcircBH} corresponds to a soliton with $M_\lambda\approx 10\,M_{BH}$; while the $A=10^{-2}$ corresponds to $M_\lambda\approx 200\,M_{BH}$. 

For $M_\lambda\gtrsim10\,M_{BH}$, the soliton's circular velocity peak and the peak location are affected very little by the SMBH. However, at small $x$ approaching the SMBH, the velocity curve turns back up  reflecting the gravitational potential of the SMBH itself. This implies that the rotation curve of a galaxy hosting a SMBH may not decrease appreciably below $x_{\rm peak,\lambda}$, even when $M_{BH}$ is much smaller than $M_\lambda$. For example, even for $M_{BH}=0.05\,M_\lambda$ ($A=0.1$ in Fig.~\ref{fig:VcircBH}), the circular velocity decreases by only $\sim25\%$ at $x<x_{\rm peak,\lambda}$ before it goes back up due to the BH. For the same parameters, Fig.~\ref{fig:Mrc} shows that $E_\lambda/M_\lambda$ is essentially unperturbed, meaning that the soliton-host halo relation of Eq.~(\ref{eq:E2MsolE2Mh}) and the rotation curve properties discussed in Sec.~\ref{s:bhconsp} remain unaffected. Such a galaxy would have an approximately flat rotation curve all the way in to the region of SMBH dominance.

The impact of a SMBH on the analysis of Sec.~\ref{s:bhconsp} can be
concluded as follows. For $M_\lambda\gtrsim4M_{BH}$ (corresponding, by
using Eq.~(\ref{eq:effA}), to effective $A/\lambda\approx0.5$),
Fig.~\ref{fig:Mrc} shows that $E_\lambda/M_\lambda$ is corrected by
$\lesssim25\%$ compared to its unperturbed
value. Eqs.~(\ref{eq:maxVc}) and~(\ref{eq:E2Mmatch}) then imply that
$\lambda$ and ${\rm max}V_{\rm circ,\lambda}$ are corrected by an
insignificant $\sim12\%$.  For a large SMBH with $M_{BH}\gtrsim
M_\lambda$, on the other hand, the SMBH gravitational potential itself
must produce high rotation velocity at $x_{\rm peak,\lambda}$,
comparable or higher than what the soliton itself would provide. As a
result, besides from the fact that the rotation curve may not decrease
towards lower $x$ below the soliton peak, our analysis in and
observational limits from Sec.~\ref{s:bhconsp} should be robust
against the effect of a SMBH\footnote{This conclusion seems in
  contradiction with the claim of Ref.~\cite{Lee:2015yws}, who
  proposed that ULDM solitons distorted by SMBHs could explain the
  $M_{BH}-\sigma$ correlation, observed between SMBHs and the stellar
  bulge of the galaxies hosting them.}. 

\subsection{Absorption of soliton by SMBH}\label{ss:absorption}

We now discuss accretion of ULDM from the
soliton onto SMBH. To estimate the lifetime of the soliton+SMBH
configuration we follow the approach of \cite{Hui:2016ltb} (see also
\cite{Burt:2011pv,Barranco:2012qs} for the study of scalar bound
states in external Schwarzschild metric and \cite{Barranco:2017aes}
for a numerical 
analysis of the lifetime of soliton+BH
system). Unruh~\cite{Unruh:1976fm} has derived the cross section for
absorption of a scalar particle with mass $m$ and momentum $k$ by a
Schwarzschild BH, whose size is much smaller than the 
Compton wavelength of the particle. 
In the non-relativistic limit, $k\ll m$, it has the form,
\be
\sigma=\frac{32\pi^2(GM_{BH})^3m^3}{k^2(1-e^{-\zeta})}\;,
\ee
where 
\be
\zeta=2\pi GM_{BH}m^2/k\;.
\ee
As the cross section is s-wave dominated, it is appropriate for
calculation of accretion from a spherically symmetric scalar field
configuration. 

The
growth of BH mass due to an infalling flux of particles with density
$\rho$ is,
\be
\frac{dM_{BH}}{dt}=\frac{32\pi^2 (GM_{BH})^3m^2\rho}{k(1-e^{-\zeta})}\;.
\ee
In the case of the soliton we can estimate the absorption
rate by substituting into this expression the central density of the
soliton and the characteristic momentum of particles in the soliton
that can be read from the
soliton profile (\ref{eq:rhosol}),
\be
k\sim 0.3\,x_c^{-1}\;.
\ee
This approximation is justified as long as the mass of the soliton is
bigger than the BH mass, so that the effect of BH on the soliton shape
can be neglected. 

The expression for the accretion rate simplifies in the limiting cases
when the parameter $\zeta$ is small or large. 
Using Eq.~(\ref{eq:rc}) (which is
consistent with the soliton--host halo relation) we obtain,
\begin{equation}
\zeta=0.16\left(\frac{m}{10^{-22}{\rm
      eV}}\right)\left(\frac{M_{BH}}{4\times 10^6 M_{\odot}}\right)
\left(\frac{M_h}{10^{12} M_{\odot}}\right)^{-\frac{1}{3}}\;.\no
\end{equation}
If $\zeta\ll 1$, the characteristic accretion time, during which the
BH mass grows by a factor 2 is,
\begin{align}
\label{eq:tau1}
\tau_{\zeta\ll 1}\sim &\, 2.4\times 10^{17}
\left(\frac{m}{10^{-22}{\rm eV}}\right)^{-2}\\ 
&\times\left(\frac{M_{BH}}{4\times 10^6 M_{\odot}}\right)^{-1}
\left(\frac{M_h}{10^{12} M_{\odot}}\right)^{-\frac{4}{3}}{\rm
  yr}\;.\no
\end{align}
In the opposite regime, $\zeta\gg 1$, the characteristic time is, 
\begin{align}
\label{eq:tau2} 
\tau_{\zeta\gg 1}\sim &\, 1.5\times 10^{18}
\left(\frac{m}{10^{-22}{\rm eV}}\right)^{-3}\\ 
&\times\left(\frac{M_{BH}}{4\times 10^6 M_{\odot}}\right)^{-2}
\left(\frac{M_h}{10^{12} M_{\odot}}\right)^{-1}{\rm yr}\;.\no
\end{align}
In deriving these expressions we made use of
Eqs.~(\ref{eq:rc},\ref{eq:rhosol}). 
Both expressions give the accretion time longer than the age of the
universe for MW's SMBH with
$M_{BH}\approx4.3\times10^6$~M$_\odot$ and
$m\lesssim5\times10^{-20}$~eV. Our Newtonian analysis, addressing
stationary ULDM soliton solutions, is adequate in this case. For
galaxies with a more massive SMBH, e.g. M31 with
$M_{BH}\sim10^8$~M$_\odot$~\cite{Bender:2005rq},
Eq.~(\ref{eq:tau2}) implies that absorption of the ULDM field
into the SMBH becomes important already for
$m\gtrsim 6\times10^{-21}$~eV, and our stationary Newtonian analysis is
not expected to capture the physics correctly for such values of
$m$. Finally, for the reference value of $m=10^{-22}$~eV, our analysis
should be valid as long as $M_{BH}\lesssim10^{10}$~M$_\odot$.

\end{appendix}

\vspace{6 pt}

\bibliography{ref}
\bibliographystyle{utphys}

\end{document}